\documentclass[fleqn,usenatbib]{mnras}

\usepackage{natbib}
\usepackage[caption=false]{subfig}    % don't load caption or captcont
\usepackage{graphicx}
\usepackage{appendix}
\usepackage{amsmath}
\usepackage{amssymb}

\newcommand{\eg}{e.g.,}
\newcommand{\ie}{i.e.,}

\newcommand{\etal}{et~al.}

\newcommand{\logten}{\ensuremath{\log_{10}\,}}
\usepackage{cleveref}
\crefname{section}{\S}{\S\S}
\Crefname{section}{\S}{\S\S}

\title[Close Pairs in the CANDELS fields and the SDSS.]{Major Merging History in CANDELS. I. Evolution of the Incidence of Massive Galaxy-Galaxy Pairs from $z=3$ to $z\sim0$}
\author[K. B. Mantha et al.]{Kameswara Bharadwaj Mantha$^{1}$\thanks{E-mail: km4n6@mail.umkc.edu}, Daniel H. McIntosh$^{1}$, Ryan Brennan$^{2}$, 
	\newauthor	
	Henry C. Ferguson$^{3}$, Dritan Kodra$^{4}$, Jeffrey A. Newman$^{4}$, Marc Rafelski$^{3}$, 
	\newauthor
	Rachel S. Somerville$^{2}$, Christopher J. Conselice$^{5}$, Joshua S. Cook$^{1}$, 
	\newauthor
	 Nimish P. Hathi$^{3}$, David C. Koo$^{6}$, Jennifer M. Lotz$^{3}$, Brooke D. Simmons$^{7}$,
	\newauthor
	Amber N. Straughn$^{8}$, Gregory F. Synder$^{3}$, Stijn Wuyts$^{9}$, Eric F. Bell$^{10}$,
	\newauthor
	Avishai Dekel$^{11}$, Jeyhan Kartaltepe$^{12}$, Dale D. Kocevski$^{13}$, Anton M. Koekemoer$^{3}$,
	\newauthor
	Seong-Kook Lee$^{14}$, Ray A. Lucas$^{3}$, Camilla Pacifici$^{15}$, Michael A. Peth$^{16}$,
	\newauthor
	Guillermo Barro$^{17}$, Tomas Dahlen$^{3}$, Steven L. Finkelstein$^{18}$, Adriano Fontana$^{19}$,
	\newauthor
	 Audrey Galametz$^{20}$, Norman A. Grogin$^{3}$, Yicheng Guo$^{6}$, Bahram Mobasher$^{21}$,
	\newauthor
	 Hooshang Nayyeri$^{21}$, Pablo G. P\'erez-Gonz\'alez$^{22}$, Janine Pforr$^{23,24}$,  
	\newauthor
	 Paola Santini$^{19}$, Mauro Stefanon$^{25}$, Tommy Wiklind$^{3}$.
	\\
	Affiliations are listed at the end of this paper\\
}
\date{Accepted: 13 December 2017}

\pubyear{2017}

\begin{document}
	\maketitle

	\begin{abstract}
		The rate of major galaxy-galaxy merging is theoretically predicted to steadily increase with redshift during the peak epoch of massive galaxy development ($1{\leq}z{\leq}3$). We use close-pair statistics to objectively study the incidence of massive galaxies (stellar $M_{1}{>}2{\times}10^{10}M_{\odot}$) hosting major companions ($1{\leq}M_{1}/M_{2}{\leq}4$; i.e., $<$4:1) at six epochs spanning $0{<}z{<}3$. We select companions from a nearly complete, mass-limited ($\geq5{\times}10^{9}M_{\odot}$) sample of 23,696 galaxies in the five CANDELS fields and the SDSS. Using $5-50$ kpc projected separation and close redshift proximity criteria, we find that the major companion fraction $f_{\mathrm{mc}}(z)$ based on stellar mass-ratio (MR) selection increases from 6\% ($z{\sim}0$) to 16\% ($z{\sim}0.8$), then turns over at $z{\sim}1$ and decreases to 7\% ($z{\sim}3$). Instead, if we use a major F160W flux ratio (FR) selection, we find that $f_{\mathrm{mc}}(z)$ increases steadily until $z=3$ owing to increasing contamination from minor (MR$>$4:1) companions at $z>1$. We show that these evolutionary trends are statistically robust to changes in companion proximity. We find disagreements between published results are resolved when selection criteria are closely matched. If we compute merger rates using constant fraction-to-rate conversion factors ($C_{\mathrm{merg,pair}}{=}0.6$ and $T_{\mathrm{obs,pair}}{=}0.65\mathrm{Gyr}$), we find that MR rates disagree with theoretical predictions at $z{>}1.5$. Instead, if we use an evolving $T_{\mathrm{obs,pair}}(z){\propto}(1+z)^{-2}$ from Snyder et al., our MR-based rates agree with theory at $0{<}z{<}3$.  Our analysis underscores the need for detailed calibration of $C_{\mathrm{merg,pair}}$ and $T_{\mathrm{obs,pair}}$ as a function of redshift, mass and companion selection criteria to better constrain the empirical major merger history.
	\end{abstract}
	\begin{keywords}
		Galaxies: evolution -- Galaxies: statistics -- Galaxies: high-redshift
	\end{keywords}
		
	\section{Introduction}
	In an hierarchical universe, collisions between similar-mass galaxies (major mergers) are expected to occur, and many theoretical studies predict such merging plays an important role in the formation and evolution of massive galaxies. A key measurement for quantifying the role of major merging in galaxy development is the merger rate and its evolution during cosmic history. A host of
	studies have measured major merger rates at redshifts $z\leq 1.5$, primarily based either on {close-pair statistics} \citep[e.g.,][]{Patton97,Lin04,kartaltepe_evolution_2007,bundy_greater_2009}, clustering statistics \citep[\eg][]{bell06b,robaina_merger-driven_2010}, and morphological disturbances and asymmetries \citep[\eg][]{lotz08,conselice_structures_2009}. These studies have all found higher incidences of major merging at earlier look-back times and a strong to moderate decrease to the present epoch, in broad agreement with many theoretical predictions \citep[\eg][]{Gottlober01,Bower06,hopkins_mergers_2010}.
	Despite these successes, large scatter (factor of 10) exists between even the most stringent individual constraints, owing to systematic uncertainties in different methodologies and merger timescales. These issues are compounded for empirical estimates at the epoch of peak galaxy development ($z\sim 2-3$; `cosmic high-noon'). Some early empirical estimates based on both methodologies found increasing major merger incidence  at $z>1.5$ \citep[\eg][]{bluck_surprisingly_2009}, but recent studies find a possible flattening or turnover in merger rates between $1<z<3$ \citep[\eg][]{Ryan08,man16,Mundy17}. These new empirical trends are in strong disagreement with recent theoretical models predicting that merger rates continue to rise from $z=1$ to $z=3$ and beyond \citep{hopkins_mergers_2010,lotz_major_2011,Rodriguez-Gomez15}. These discrepancies and the large variance between past measurements highlight the need for improved major merger constraints, especially during the critical high-noon epoch. 
	
	A host of selection-effect issues {has} plagued many previous attempts to constrain major merger statistics at high redshift, from low-number statistics and significant sample variance due to small-volume pencil-beam surveys, {and} rest-frame UV selections of both disturbed morphologies and {close pairs}. While the identification of {close pairs} is less prone to some systematics, the lack of statistically useful samples of spectroscopic redshifts or even moderately small-uncertainty photometric redshifts at $z>1$ until very recently have limited the usefulness of this method.
	Moreover, the wildly varying {close-companion} selection criteria among previous
	studies is a plausible explanation for tensions between empirical merger
	rates and theoretical predictions \citep{lotz_major_2011}.
	In this study, we will address many of these shortcomings and systematically
	explore the impact of major {close-companion} selection criteria by analyzing major companion fractions
	in a sample of 10,000 massive host galaxies (stellar mass $M_1\geq 2\times10^{10}M_{\odot}$) from the {\it five} Hubble Space Telescope ({\it HST}) legacy fields in CANDELS \citep{grogin_candels:_2011,Koekemoer} and the SDSS survey. This comprehensive sample provides
	statistically useful major companion counts, down to a mass limit of $M_2=5\times 10^{9}M_{\odot}$, from rest-frame optical images over a large volume out to $z=3$.
	
	{The} hierarchical major merging of similar mass halos via gravitational accretion is the underlying physical driver of galaxy-galaxy major merging. Cosmological simulations predict that the major halo-halo merger rate rises steeply with redshift as $R \propto (1+z)^{2-3}$ \citep[\eg][]{Fakhouri08,Genel09,Fakhouri10}, which {is} in agreement with simple analytical predictions based on Extended Press-Schechter (EPS) theory of $R\propto (1+z)^{2.5}$ \citep{Neistein08,Dekel13}. Cosmologically-motivated simulations of galaxy formation and evolution predict major galaxy-galaxy merger rates that follow $R \propto (1+z)^{1-2}$ over a wide redshift range \citep[\eg\,$z<6$;][]{Rodriguez-Gomez15}. While there is some debate on the increasing merger rate evolution among theoretical studies due to model-dependencies \citep[for review, see][]{hopkins_mergers_2010}, some works claim flattening of merger rates with increasing redshift \citep[\eg][]{Henriques15}, most agree with an increasing incidence (within a factor-of-two uncertainty). Not only are merger rates expected to be higher at early cosmic times, but major galaxy merging is predicted to play a crucial role in nearly all aspects of the formation and evolution of massive galaxies including  buildup of spheroidal bulges and massive elliptical galaxies \citep{Springel00,Kochfar03,Khockfar05,Naab06,Cox08}, triggering and enhancement of star formation (SF) including nuclear starbursts \citep{Sanders88a,Dimatteo07,DiMatteo08,Martig08}, and the fueling of active galactic nuclei (AGN) \citep{Hopkins06a,Younger09,Narayanan10,hopkins_mergers_2010} and subsequent SF quenching \citep[\eg][]{DiMatteo05,Hopkins08}.

Many empirical studies support the predictions that major merging may explain the documented build-up of massive and quenched (non-star-forming and red) galaxy number densities and their stellar content growth at $z<1$ \citep[\eg][]{bell_dry_2006,mcintosh_2008,van_der_wel09}, enhancement of SF activity \citep[\eg\,][]{jogee_history_2009,Patton11}, and elevation of AGN activity \citep{Treister12,weston17,Hewlett17}. Despite this agreement, some studies find a weak 
major merging-SF connection and suggest mergers may not be the dominant
contributor to in-situ galactic SF \citep{robaina09,Swinbank10,Targett11}.
Moreover, other studies find a lack of a merger-AGN connection
\citep{Grogin05,kocevski_candels:_2012,Villforth14,villforth17}. These conflicting observations lend support 
to theories that predict violent disk instabilities (VDI) due to {the} rapid hierarchical accretion of cold gas may be responsible for key processes like bulge formation and AGN triggering \citep{Bournaud11,Dekel14}. Indeed, a recent CANDELS study by \cite{Brennan15} found the observed evolution of massive quenched spheroids at $z<3$ is better matched to SAM predictions that include both mergers and {disk-instability} prescriptions. Therefore, the role of major merging in galaxy evolution remains a critical open question. Hence, measuring the frequency and rate at which major mergers occur at different cosmic times using large, uniformly selected {close-pair} samples is a key step towards answering the role played by them in massive galaxy development. 

Theoretical simulations predict that {galaxies involved in major close pairs} will interact gravitationally and coalesce over time into one larger galaxy, and thereby make them effective probes of ongoing or future merging. Many studies in the past have employed the close-pair method to estimate the frequency of major merging as a function of cosmic time. This typically involves searching for galaxies that host a nearby companion meeting a number of key criteria: (i) 2-dimensional projected distance, (ii) close redshift-space proximity, and satisfies a nearly-equal mass ratio $M_{1}/M_{2}$ between {the} host (1) and companion (2) galaxies. For each criterion, a wide range of choices {is} used in the literature. For projected separation $R_{\rm proj}$, a 
search annulus is often employed with minimum and maximum radii. Common choices
vary between $R_{\rm max}\sim 30-140$\,kpc \citep[\eg][]{Patton08,de_ravel09} and $R_{\rm min} \sim 0-14$ kpc \citep[\eg][]{bluck_surprisingly_2009,man16}.
Depending on available redshift information, the choice of physical proximity criterion ranges from stringent spectroscopic velocity differences \citep[commonly $\Delta v_{12}\leq 500$\,km\,s$^{-1}$; e.g.,][]{lin_redshift_2008} to a
variety of photometric redshift $z_{\rm phot}$ error overlaps 
\citep[e.g.,][]{bundy_greater_2009,man2012}  
To study similar-mass galaxy-galaxy mergers,
previous studies have adopted stellar-mass-ratio selections ranging from 
$2>M_{1}/M_{2} > 1$ \citep[or 2:1, \eg][]{de_propris_millennium_2007} to $5>M_{1}/M_{2} > 1$ \citep[5:1, \eg][]{Lofthouse17}, with 4:1 being by far the most common
mass ratio criterion. In the absence of stellar-mass estimates, flux ratio $F_{1}/F_{2}$ is often used as a proxy for $M_{1}/M_{2}$ \citep[\eg][]{bridge_role_2007}. 
The wide range of adopted {close-companion} selection criteria lead
to a large scatter in two decades of published pair-derived merger rates with redshift evolution spanning $R\propto (1+z)^{0.5-3}$ at $0<z<1.5$, and sometimes even indicating a flat or turnover in merger rates at $z>1.5$ \citep{williams_diminishing_2011,man2012,man16,Mundy17}. 
This large scatter in pair-derived merger rate constraints highlights the strong need for tighter constraints at cosmic high-noon, and motivates a careful analysis of selection effects.

Numerical simulations of the gravitation interactions between merging
galaxies can produce disturbed morphological features due to strong tidal forces \citep[\eg][]{BarnesHernquist94,Bournaud06,Peirani10}.
As such, morphological selections have also been used to empirically identify mergers. These selections are broadly divided into visual classifications
\citep[\eg][]{Darg10,kartaltepe15}, analysis of image$-$model residuals
\citep[\eg][]{mcintosh_2008,Tal09}, and automated measures of quantitative morphology such as Gini-M20 \citep{lotz04} and CAS \citep{conselice_03_cas}. 
Although morphology-based studies broadly find merger rates to be rising strongly with redshift as $(1+z)^{2-5}$ \citep[\eg][]{lopez-sanjuan_galaxy_2009,wen16}, sometimes finding as high as $25\%- 50\%$ of their sample as mergers \citep{conselice_structures_2008}, there are significant study-to-study discrepancies where some studies find no merger rate evolution \citep[\eg][]{cassata05,lotz08}. Morphology-based selections depend on identifying relatively fainter disturbances than the galaxy, which makes this method prone to systematics. The cosmological surface-brightness of galaxies falls off as $(1+z)^{-4}$, which can lead to {a} biased identification of faint merger-specific features as a function of redshift. In addition, most of these morphology-based merger rates are based on small-volume, pencil-beam surveys probing the rest-frame UV part of the spectrum, especially at $z>1$. This can lead to over-estimation of merger rates due to contamination from non-merging, high star-forming systems with significant substructure that can be confused as two merging galaxies.  Recent theoretical developments suggest that VDI can also cause disturbances in the host galaxy morphology and mimic merger-like features \citep[see][]{Dekel09,Cacciato12,Ceverino15}, which in principle may complicate the measurement of morphology-based merger rates. Thus, to robustly identify plausible merging systems out to high redshifts ($z\sim 3$) without having to rely on imaging-related systematics {strongly}, we resort to the close-pair method in this study. We acknowledge that the close-pair method has its limitations at high redshift where galaxies have large photometric redshift uncertainties, which may lead to incorrect merger statistics. In this study, we initially exclude the galaxies with unreliable redshifts from our analysis, but later add back a certain fraction of them by employing a statistical correction. 

In this paper, we analyze galaxy-galaxy close pairs in a large sample of 5698 {\it massive} galaxies ($M_{\rm stellar}\geq2\times10^{10}M_{\odot}$) from the state-of-the-art Cosmic Assembly Near-Infrared Deep Extragalactic Legacy Survey \citep[CANDELS-][]{grogin_candels:_2011,Koekemoer} of five highly-studied extragalactic fields at six epochs spanning $z\sim 0.5-3.0$ (with a width $\Delta z = 0.5$). To simultaneously anchor our findings to $z=0$, we take advantage of 4098 massive galaxies from the Sloan Digital Sky Survey (SDSS) \citep[][]{York00}, Data release 4 \citep[DR4;][]{Adelman-McCarthy06} at $z\sim 0.03-0.05$, which is matched in resolution to CANDELS and probes an average of $\sim 1.3\times 10^6$ Mpc$^3$ per redshift bin. With the available data, we also perform rigorous analyses to understand the impact of different {close-companion selection criteria} on the derived results.

We structure this paper as follows: In \cref{sample selection}, we provide {a} brief description of the CANDELS and SDSS data products (redshifts and stellar masses) and describe the selection of {massive galaxies hosting major companions} based on the stellar mass complete massive galaxy sample. In \cref{deriving_fmp}, we describe the calculation of {major companion fraction} and its redshift evolution including necessary statistical corrections. In \cref{impact_of_pair_selection}, we discuss the impact of {close-companion} selection choices on the derived {major companion fractions}. {In \cref{Discussion}, we calculate the major merger rates based on the {major companion fractions}. We synthesize detailed comparisons of the {companion fractions} and merger rates to other empirical studies and theoretical model predictions, and also discuss plausible reasons and implications of disagreement between the observed and theoretical merger rates}. We present our conclusions in \cref{Conclusions}. {Throughout this paper}, we adopt a cosmology of $H_0$ = $70 ~{\rm km}~ {\rm s}^{-1} {\rm Mpc}^{-1}$ ($h =0.7$), $\Omega_{\rm M} $ = $0.3$ and $\Omega_{\Lambda}$ = 0.7, and use the AB magnitude system \citep{oke83}.

\section{Data and Galaxy Sample }
\label{sample selection}
In this study, we analyze close galaxy-galaxy pairs selected from
a large sample of massive galaxies from the Cosmic Assembly Near-infrared Deep Extragalactic Legacy Survey \citep[CANDELS;][]{grogin_candels:_2011,Koekemoer},
spanning redshifts $0.5\leq z \leq 3$, and subdivided into five epochs
probing a volume of $\sim 1.3\times 10^6$ Mpc$^3$ each. We anchor our
findings to $z\sim0$ using a sample from the SDSS that is matched in
volume and resolution to the CANDELS sample.
To reliably track the major merging history since $z=3$
using {close-pair method}, we start with
a mass-limited sample of galaxies that will allow a complete selection of
massive $>2\times 10^{10}\,M_{\odot}$
galaxies with major companions meeting our chosen stellar mass ratio: $1\leq M_{1}/M_{2} \leq 4$ (1, 2 represent host and companion galaxies, respectively). In this section, we describe the relevant details of the data
necessary to achieve this sample selection.

\subsection{CANDELS : 5 LEGACY FIELDS}

\subsubsection{Photometric Source Catalogs}
\label{ss_sec_photometry}
The five CANDELS {\it HST} legacy fields -- 
UDS, GOODS-S, GOODS-N, COSMOS, and EGS -- have a wealth of multi-wavelength
data and cover a total area of $\sim 800 ~{\rm arcmin^{2}}$ ($\sim 0.22~{\rm deg}^2$).
The CANDELS survey observations and image processing are described in \cite{grogin_candels:_2011} and \cite{Koekemoer}, respectively.
We use the photometric source catalogs from
\citet[][UDS]{galametz_candels_2013}, \citet[][GOODS-S]{guo_candels_2013}, Barro et al., in prep (GOODS-N), \citet[][COSMOS]{Nayyeri17}, and \citet[][EGS]{Stefanon17}. 
Each catalog was generated with a consistent source detection algorithm
using SExtractor applied to the F160W ($H$-band) 2-orbit depth CANDELS mosaic image produced for each field. These authors used profile template fitting \citep[TFIT,][]{Liadler_tfit} to provide uniform photometry and spectral energy distributions (SEDs) for each galaxy at wavelengths spanning $0.4\mu {\rm m}$ to $1.6\mu{\rm m}$, supplemented by ground-based data \citep[for description, see][]{guo_candels_2013} and {\it spitzer}/IRAC photometry ($3.6\mu {\rm m}$ to $8.0\mu {\rm m}$) from the S-CANDELS survey \citep{Ashby15} . Each photometric object was assigned a flag (PhotFlag) to identify plausible issues using a robust automated routine described in \cite{galametz_candels_2013}. We use PhotFlag = 0 to remove objects with contaminated photometry due to nearby stars, image artifacts or proximity to the F160W coverage edges. This cut removes $\sim 3-5\%$ of raw photometric sources {depending on the field}. We also use the stellarity index from SExtractor (Class\_star $\geq0.95$) to eliminate bright star-like sources. We estimate that this additional cut removes active compact galaxies that {makeup} $\sim 1.3\%$ of our total desired mass-limited sample. We note that including these galaxies has no significant impact on our conclusions. 
We tabulate the total raw and good photometric source counts for the 
five CANDELS fields in
Table\,\ref{sample_info}.

\subsubsection{Redshifts \& Stellar Masses}
\label{redshifts}
We use the CANDELS team photometric redshift and stellar mass catalogs available for each field. For the CANDELS UDS and GOODS-S fields, the redshifts are published in \cite{dahlen_critical_2013}, and the masses are found in \cite{santini_stellar_2015}. For the remaining fields, we use the catalogs: GOODS-N (Barro et al., in prep), COSMOS \citep{Nayyeri17}, and EGS \citep{Stefanon17}.
As discussed extensively in \cite{dahlen_critical_2013}, photometric redshift probability distribution functions $P(z)$ were computed for each galaxy by fitting the SED data. 
{This exercise was repeated by six participants \citep[\#ID 4, 6, 9, 11, 12, and 13 in][]{dahlen_critical_2013} who performed SED fitting using different codes (EAZY, HyperZ) and template sets (BC03, PEGASE, EAZY). Additional detailed discussion on individual code functionality and their respective fitting priors can be found in \cite{dahlen_critical_2013}}. 
A team photometric redshift ($z_{\rm phot}$) was computed for each source equal to {the} median of the six $P(z)$ peak redshifts. 
When compared to a known spectroscopic sample, these photometric redshifts have an outlier removed RMS scatter $\sigma_{\rm z} \sim 0.029$ (see \citealp{dahlen_critical_2013} for definition). 
Additionally, spectroscopic redshifts ($z_{\rm spec}$) are also available for small subsets of galaxies in each field. The best available redshift $z_{\rm best}$ is cataloged as either the team $z_{\rm phot}$ or the good quality $z_{\rm spec}$ measurements when available, which are defined by the flag q\_zspec = 1 \citep{dahlen_critical_2013}. {Note that the compilation of redshifts included in our analysis sample does not include grism redshifts.} We limit our selection of massive galaxies to $0.5\leq z_{\rm best}\leq3.0$, and we employ a redshift bin size $\Delta z = 0.5$ to probe evolution between 5 and 11\,Gyr
ago using five roughly equal co-moving volumes ranging between $7\times 10^5 ~{\rm Mpc}^{3} -1.3 \times 10^6 ~{\rm Mpc}^{3}$. We exclude redshifts $z_{\rm best}<0.5$ since this volume is $\sim 10$ times smaller ($\sim 1.3\times10^{5} ~{\rm Mpc}^{3}$).

The stellar masses ($M_{\rm stellar}$) were estimated for each source by fitting the multi-band photometric data to SED templates with different stellar population model assumptions\footnote{Each model is defined by a set of stellar population templates, Initial Mass Function (IMF), Star Formation History (SFH), metallicity and extinction law assumptions; see \cite{Mobasher15}} fixed to the object's $z_{\rm best}$. The team stellar mass \citep[see][]{santini_stellar_2015,Mobasher15} for each source is chosen as the median of the estimates based on the same assumptions of IMF \citep{Chabrier03} and stellar population templates \citep{BC03}. Using the median mass estimate, we select a mass-limited ($M_{\rm stellar} \geq 5\times10^{9}M_{\odot}$) sample of 14,513 potential companion galaxies in a redshift range $0.5 \leq z_{\rm best} \leq 3$ (for breakdown, see Table\,\ref{sample_info}). As described in the next section, this provides a sample with
high completeness.

\subsubsection{$0.5\leq z \leq 3.0$ Sample Completeness}
\label{candels_sample_completeness}
\begin{figure}
	\centering
	\includegraphics[width=\columnwidth]{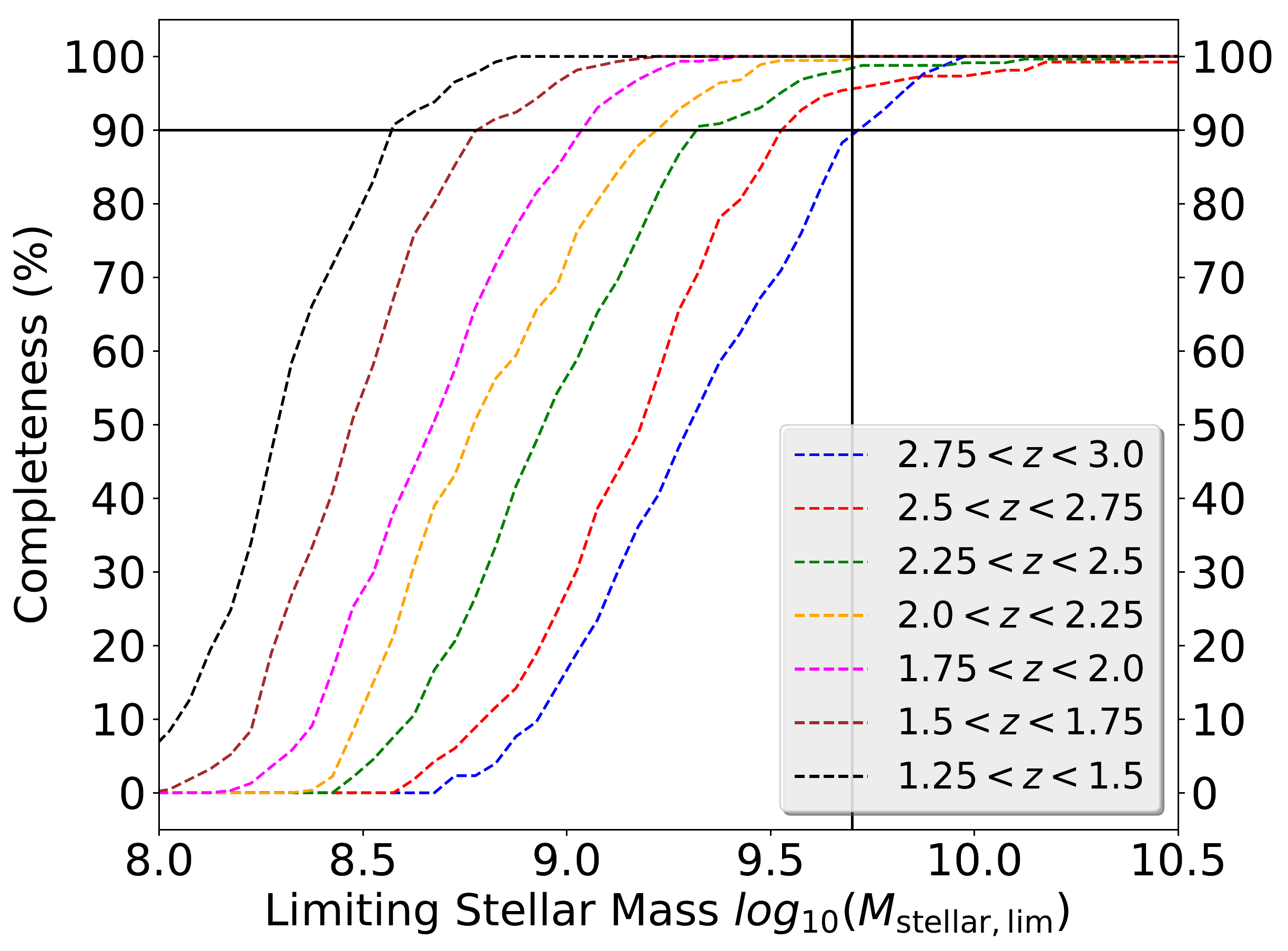} 
	\caption[]{Stellar-mass completeness of the $\logten(M_{\rm stellar}/M_{\odot})\geq9.7$ CANDELS galaxy sample. We show the normalized cumulative distributions of the limiting stellar masses ($M_{\rm stellar,lim}$) for the 20\% faintest galaxies of the desired mass-limited sample (dashed lines), color-coded according to their respective redshift slices ($\Delta z=0.25$) at $z\ga 1$ (see \cref{candels_sample_completeness} text for details). We show our desired mass limit in solid vertical black line and we mark the 90\% completeness in solid horizontal black line. At the highest redshift slice ($2.75<z<3$), we find that 90\% of the galaxies with $\logten(M_{\rm stellar}/M_{\odot})\geq9.7$ have their limiting stellar masses smaller than the desired major companion mass limit, implying that the CANDELS mass-limited sample of $\logten(M_{\rm stellar}/M_{\odot})\geq9.7$ galaxies is at least 90\% in the desired redshift range of this study $0.5\leq z\leq 3.0$.} 
	\label{fig:completeness} 
\end{figure}

{We demonstrate the completeness of massive CANDELS galaxies with redshifts $0.5\leq z \leq 3.0$ by adopting the method introduced in \cite{Pozzetti10} \citep[also see][]{Nayyeri17}. Briefly, \citeauthor{Pozzetti10} computes a stellar-mass limit as a function of redshift, above which nearly all the galaxies are observable and complete. They do so by estimating the limiting stellar-mass ($M_{\rm stellar,lim}$) distributions for the 20\% faintest sample population\footnote{By considering the 20\% faintest galaxy sample of the apparent magnitude distribution at each redshift bin, only those galaxies with representative mass-to-light ratios close to the $H_{\rm lim}$ are used towards estimating the $M_{\rm stellar,lim}$ \citep[see][for additional details]{Pozzetti10}.}, where $M_{\rm stellar,lim}$ of a galaxy is the mass it would have if the apparent magnitude ($H_{\rm mag}$) is equal to the limiting $H$-band magnitude ($H_{\rm lim}$). We estimate the $M_{\rm stellar,lim}$ by following \citeauthor{Nayyeri17} relation between the observed galaxy $M_{\rm stellar}$ and its $M_{\rm stellar,lim}$ as $\logten(M_{\rm stellar,lim}) = \logten(M_{\rm stellar})+0.4(H_{\rm mag}-H_{\rm lim})$ \citep[see][]{Nayyeri17} and use the published $H$-band $5\sigma$ limiting magnitudes \citep{grogin_candels:_2011,Koekemoer,galametz_candels_2013,Nayyeri17,Stefanon17}. 
	
In Figure\,\ref{fig:completeness}, we show the normalized cumulative distributions of $M_{\rm stellar,lim}$ for the 20\% faintest CANDELS $\logten(M_{\rm stellar}/M_{\odot})\geq9.7$ galaxy samples\footnote{We compute the distributions independently for the five CANDELS fields and present the mean of them at each redshift slice. We find that the behavior of individual field distributions is not significantly different from each other and with the mean distribution.} in narrow ($\Delta z = 0.25$) redshift slices at $z>1$. At all redshift bins up to $z = 2.25$, we find that all the galaxies in our mass-limited sample have $M_{\rm stellar}>M_{\rm stellar,lim}$, which implies $100\%$ completeness. At redshifts $2.25<z<2.75$ and $2.75<z<3$, we find that the desired sample selection is $>95\%$ complete and $90\%$ complete, respectively. Additionally, we test the impact of surface brightness on the measured stellar-mass completeness by analyzing the {\it effective} $H$-band surface brightness (SB$_{H}$) distributions of our desired mass-limited $\logten(M_{\rm stellar}/M_{\odot})\geq9.7$ galaxy sample at five redshift bins between $0.5<z<3$. We use a $H$-band surface brightness limit SB$_{H} = 26.45\,{\rm mag/arcsec^{2}}$ based on the model-galaxy recovery simulations by \cite{man16} and find that $100\%$ and $>95\%$ of our desired galaxies have SB$_{H}<26.45\,{\rm mag/arcsec^{2}}$ at redshifts $0.5<z<2$ and $2<z<3$, respectively. This implies that even the population that constitutes {\it lowest} 10\% of the SB$_{H}$ distribution (low surface brightness galaxies; hereby LSB galaxies) in our desired sample can be robustly detected up to $z=3$. As the LSB galaxies only make up a small fraction (less than 10\%) of our desired mass-limited sample, we expect that a smaller completeness among these LSB galaxies will not have an significant impact on the close-pair statistics presented in this study. These tests permit us to robustly search for major companions associated with $\logten (M_{\rm stellar}/M_{\rm \odot}) \geq 10.3$ galaxies unaffected by significant incompleteness. We include the breakdown of $N_{\rm m} = 5698$ massive galaxies per CANDELS field in  Table\,\ref{sample_info}.}

%\break
\subsection{SDSS}
\label{sec_nyuvagc_selection}
\subsubsection{Redshifts \& Stellar Masses}\label{sdss_redshifts_masses}

To anchor evolutionary trends to $z\sim 0$, we employ redshifts and stellar masses from Sample III
of the SDSS Group Catalog described in \citet{Yang_07}. Briefly, this
catalog contains $\sim 400,000$ galaxies spanning a redshift range $0.01<z<0.2$ from the $\sim4500$ square 
degree sky coverage of the {SDSS} Data Release 4 \citep[DR4,][]{Adelman-McCarthy06}. \citeauthor{Yang_07} computed $(g-r)$ color-based
$M_{\rm stellar}$ estimates using the \citet{bell_optical_2003} SED fitting based mass-to-light ratio calibrations
and K-corrections from the NYU-VAGC \citep{blanton_nyu-vagc:_2005}. For consistency, these masses were corrected by $-0.1$\,dex
to convert from a `diet' Salpeter IMF to a \cite{Chabrier03} IMF basis as in CANDELS. Besides {the} IMF, \citeauthor{bell_optical_2003} assumed similar exponentially declining star formation histories as {the} CANDELS team $M_{\rm stellar}$ participants, but used P\'EGASE stellar population models \citep{Fioc97} in contrast to \cite{BC03}, respectively \citep[for details, see][]{Mobasher15}. However, \cite{de_jong07} explored the impact of these model assumptions and found that both P\'EGASE and \cite{BC03} yield similar results in terms of \citeauthor{bell_optical_2003} color and mass-to-light ratio calibrations. In addition for a sample of galaxies with SDSS+GALEX photometry, \cite{moustakas13} found good agreement between SED {fitting-derived} stellar masses and independent SDSS photometry-based estimates. Hence, we conclude that the CANDELS and SDSS stellar mass estimates are not systematically different. 

We select Sample III galaxies within a redshift range $0.03\leq z \leq 0.05$ and sky area $1790$ sq.deg (${\rm RA = 100\deg-210\deg}$ \& ${\rm DEC} = 17\deg-69\deg$) to match the CANDELS sample in volume and resolution. Using these cuts, we find 9183 galaxies with $\logten (M_{\rm stellar}/M_{\odot}) \geq 9.7$. {We present the SDSS selection information in Table\,\ref{sample_info}}.  We are aware of more recent datasets than the SDSS-DR4; \eg the \,SDSS-DR7 \citep{abazajian09} has an improvement in photometric calibration from 2\% (DR4) to 1\% (DR7). However, owing to the contribution from $\sim 20\%$ random and $\sim 25\%$ model dependent systematic uncertainties for \citeauthor{bell_optical_2003} $M_{\rm stellar}$ estimates, we argue that these small photometric improvements have no significant impact on our results. Hence, we use the SDSS-DR4 because it is readily available and it meets our volume and resolution requirements.

\begin{figure}
	\centering
	\includegraphics[width=\columnwidth]{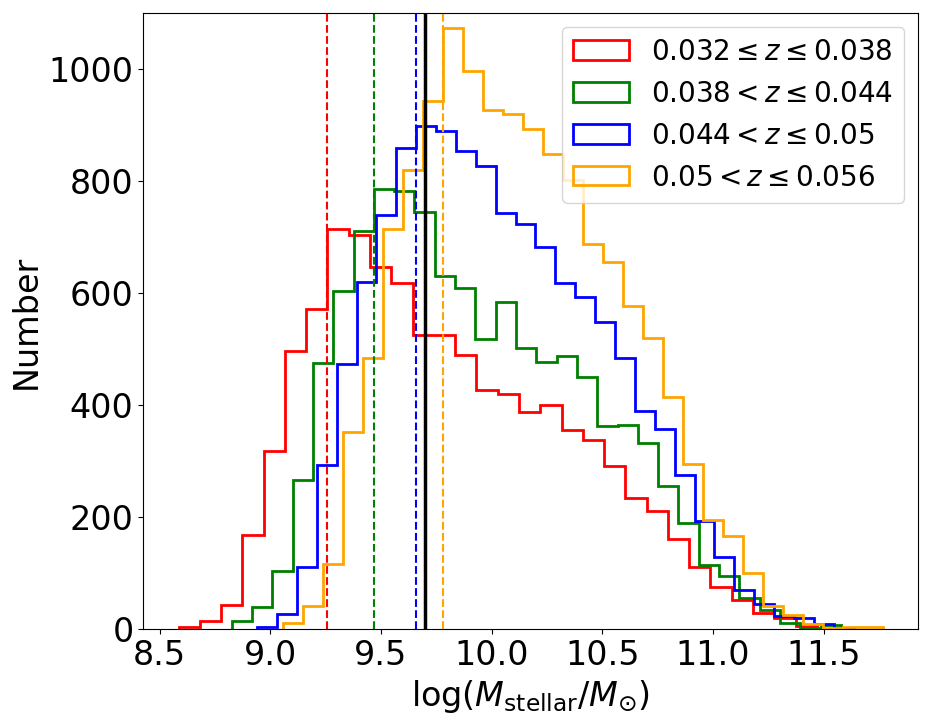} 
	\caption{ Stellar mass distributions for galaxies in the SDSS group catalog \citep{Yang_07} for four narrow redshift slices (see legend). The stellar mass at which the mass distribution turns over owing to the $r<17.77$ mag criteria for the SDSS spectroscopic target selection \citep{strauss02} is given by the vertical dashed lines. The solid black line indicates our desired stellar mass limit of the companion galaxies  ($ M_{\rm stellar} = 5\times10^{9}M_{\odot}$). We find that companion galaxies are highly complete at $z\leq 0.05$.  }  
	\label{fig:completeness_sdss}
\end{figure} 

\begin{table*}
	\centering
	\caption{Galaxy Sample Information. Columns: (1) name of the field/survey; (2) the total number of sources in photometry catalog before (after) applying good-source cuts described in Section\,\ref{ss_sec_photometry}; (3) the redshift range of interest in our study, used to select the mass-limited sample counts in (4,5), where the subsets with spectroscopic redshift information are given in parenthesis.} 
	\label{sample_info}
	\begin{tabular}{cc|cccc}
		\hline
		Name   &   Phot Sources   & Redshift Range  & $\logten (M_{\rm stellar}/M_{\rm \odot})\geq 9.7$ &   $\logten (M_{\rm stellar}/M_{\rm \odot})\geq 10.3$   &     \\
		(1) & (2) & (3) & (4) & (5) & \\
		\hline
		UDS  &  35932 (33998) &$0.5 \leq z \leq 3.0$ & 3019 (260)  &  1223 (141)  &     \\
		GOODS-S  &   34930 (34115) & $0.5 \leq z \leq 3.0$& 2491 (892) &  942 (403)  &     \\
		GOODS-N  &   35445 (34693) & $0.5 \leq z \leq 3.0$ & 2946 (494) &   1133 (209)  &       \\
		COSMOS  &   38671 (36753) &$0.5 \leq z \leq 3.0$ & 3232 (11) &  1307 (9)  &       \\
		EGS  &   41457 (37602) &$0.5 \leq z \leq 3.0$ & 2825 (199) &  1093 (72)  &      \\
		\hline 
		CANDELS (Total) & 186,435 (177,161) & $0.5 \leq z \leq 3.0$ &14,513 (1856) &  5698 (834) &   \\
		\hline 
		&&&&& \\
		SDSS-DR4 (1790 sq. deg)  &  141,564  & $0.03 \leq z \leq 0.05$ & 9183 (8524)  &  4098 (3859)   & \\
		\hline
	\end{tabular}
	
\end{table*}

\subsubsection{$z\sim 0$ Sample Completeness}
\label{sdss_sample_completeness}
The \citet{Yang_07} sample is magnitude-limited
($r<17.77$\,mag) {due} to the SDSS spectroscopic target selection; this provides a $\sim90$\% $z_{\rm spec}$ completeness \citep{strauss02}. Yang et al. included additional 
redshifts from supplementary surveys to improve the incompleteness due to spectroscopic fiber 
collisions \citep{blanton03a}. As such, our $z\sim0$ sample selection
has 92.8\% $z_{\rm spec}$ completeness (Table\,\ref{sample_info}). 
Nevertheless, several merger studies 
demonstrate that the {SDSS spectroscopic incompleteness grows with decreasing galaxy-galaxy separation} \citep{mcintosh_2008,weston17}.
We account for this issue and provide detailed corrections in \cref{fmp_for_sdss}. 
In addition, 
we demonstrate the stellar mass completeness of $\logten (M_{\rm stellar}/M_{\rm \odot}) \geq 9.7$ galaxies by employing the method by \cite{Cebri_sdss_completeness}. 
In Figure\,\ref{fig:completeness_sdss}, we show the stellar mass distributions for narrow redshift 
intervals ($\Delta z = 0.006$) at $z\leq 0.05$. We find that the $r<17.77$\,mag limit produces
a turnover in counts at different masses as a function of redshift. At $z\leq 0.05$, the mass at
which the distributions turn over (become incomplete) is well below our limit of 
$\logten (M_{\rm stellar}/M_{\odot}) = 9.7$. This indicates that our mass-limited sample is highly complete
for selecting possible major companions in a complete sample of $N_{\rm m} = 4098$ massive galaxies with
$\logten (M_{\rm stellar}/M_{\odot}) \geq 10.3$ and $0.03\leq z \leq 0.05$ (see Table\,\ref{sample_info}).

\subsection{Selection of Massive Galaxies Hosting Major Companions} 
\label{selection_of_massive_gals_in_maj_pairs}
\subsubsection{Projected Separation}
\label{major_projected_pairs}
With our well-defined mass-limited samples for CANDELS and SDSS in hand, {we start by identifying the massive ($M_{\rm stellar}\geq 2\times10^{10}M_{\odot}$) galaxies hosting a major projected companion satisfying} $1\leq M_{1}/M_{2}\leq 4$ and a projected physical separation of $5 ~{\rm kpc} \leq R_{\rm proj} \leq 50 ~{\rm kpc}$. The choice of $R_{\rm proj}\leq 50$ kpc is common in {close-pair} studies \citep{Patton08,lotz_major_2011,de_ravel_zcosmos_2011} which is supported by the numerical simulation {results showing} that major bound companions with this separation will merge within $\la 1$\,Gyr. Additionally, source blending from smaller separations ($\la 1.2~{\rm kpc}$) can cause incompleteness at $z\ga 0.04$ for SDSS and at $z\ga 2.5$ for CANDELS. Thus, we adopt a lower limit of $R_{\rm proj}=5$ kpc ($\sim 4\times$ the resolution), which also corresponds to the typical sizes of $\logten (M_{\rm stellar}/M_{\odot})\geq 9.7$ galaxies at $2.0\leq z \leq 2.5$. {In summary, we find $N_{\rm proj} = 318$ and $N_{\rm proj} = 2451$ unique (\ie\,duplicate resolved)  massive galaxies hosting major projected companions in SDSS ($0.03\leq z \leq 0.05$) and CANDELS (total of all five fields at $0.5\leq z \leq 3.0$), respectively}. We tabulate the breakdown of $N_{\rm proj}$ by redshift per each CANDELS field in Table\,\ref{summary_b09_selection}.

\subsubsection{Plausible Physical Proximity (SDSS)}
\label{plausible_physical_pairs}
We note that projected proximity does not guarantee true physical proximity as foreground and background galaxies can be projected interlopers. A common and effective method to define physical proximity is to isolate systems with a small velocity separation, which indicate that the host and companion galaxies are plausibly gravitationally bound. {For the SDSS sample, we employ the common criteria  $\Delta v_{12} = |v_{1}-v_{2}| \leq 500~{\rm km\,s^{-1}}$  \citep[\eg][]{kartaltepe_evolution_2007,Patton08,lin_redshift_2008}, where $v_{1}$ and $v_{2}$ are the velocities of the host and companion galaxies, respectively}. {Merger simulations find that {systems that satisfy} $\Delta v_{12} \leq 500~{\rm km\,s^{-1}}$ typically merge within $0.5-1$ Gyr \citep[\eg][]{conselice00}}. Other studies {show that close-pair systems with} $\Delta v_{12} > 500~{\rm km\,s^{-1}}$ are not likely to be gravitationally bound \citep[\eg][]{Patton00,de_propris_millennium_2007}. However, {owing to} spectroscopic redshift incompleteness (see \cref{sdss_sample_completeness}), we are only able to apply this velocity selection to a {subset of galaxies from \cref{major_projected_pairs}} which have spectroscopic redshifts. {In doing so, we find $N_{\rm phy} = 106$ massive galaxies hosting a major projected companion (in \cref{major_projected_pairs}) meeting $\Delta v_{12} \leq 500~{\rm km\,s^{-1}}$ criteria in the SDSS ($0.03\leq z \leq 0.05$) sample}. We describe the statistical correction for missing {major companions} due to spectroscopic incompleteness in \cref{fmp_for_sdss}. 

\subsubsection{Plausible Physical Proximity (CANDELS)}
\label{plausible_physical_pairs_can}
Most galaxies in the CANDELS catalogs do not have a spectroscopic redshift. Hence, we use a proximity method based on photometric redshifts and their uncertainties ($\sigma_{\rm z}$) to select plausible, physically close companions \citep[\eg][]{bundy_greater_2009,man_galaxy_2011,man16}.  
As described in \cref{redshifts}, each galaxy's $z_{\rm phot}$ value is the median of the peak values ($z_{\rm peak}$) of multiple photometric $P(z)$ distributions computed by the CANDELS team. However, the $P(z)$ data was not thoroughly analyzed to derive $z_{\rm phot}$ errors for all of the CANDELS fields. Thus, we compute $\sigma_{\rm z}$ values from a single participant $P(z)$ {dataset} that produces $z_{\rm peak}$ values that are consistent with the published team $z_{\rm best}$ values. This is necessary to achieve $z_{\rm phot}$ errors that are consistent with the $z_{\rm best}$ and stellar masses (calculated with $z_{\rm best}$) that we use in this study.
We find that the S. Wuyts\footnote{Method 13 as specified in \citealp{dahlen_critical_2013}.} photometric redshifts produced the best match to $z_{\rm best}$ (see Appendix\,I) after testing all participant $P(z)$ data. The Wuyts $P(z)$ distributions for each CANDELS galaxy were computed using the photometric redshift code EAZY \citep{brammer_eazy:_2008} and P\'EGASE \citep{Fioc97} stellar synthesis template models. We optimize\footnote{We shift the $P(z)$ distributions and raise them to a power such that when compared to the test set of spectroscopic redshifts ($z_{\rm spec}$), the 68\% confidence interval of the $P(z)$ should include $z_{\rm spec}$ 68\% of the time. A detailed description is given in Kodra \etal, in preparation.} the $P(z)$ for each galaxy and use this distribution to compute the uncertainty ($\sigma_{\rm z}$) defined as the 68\% confidence interval of the photometric redshift $z_{\rm phot}$ (see Kodra \etal, in preparation for details).

\begin{figure*}
	
	\centering
	\includegraphics[width=2\columnwidth]{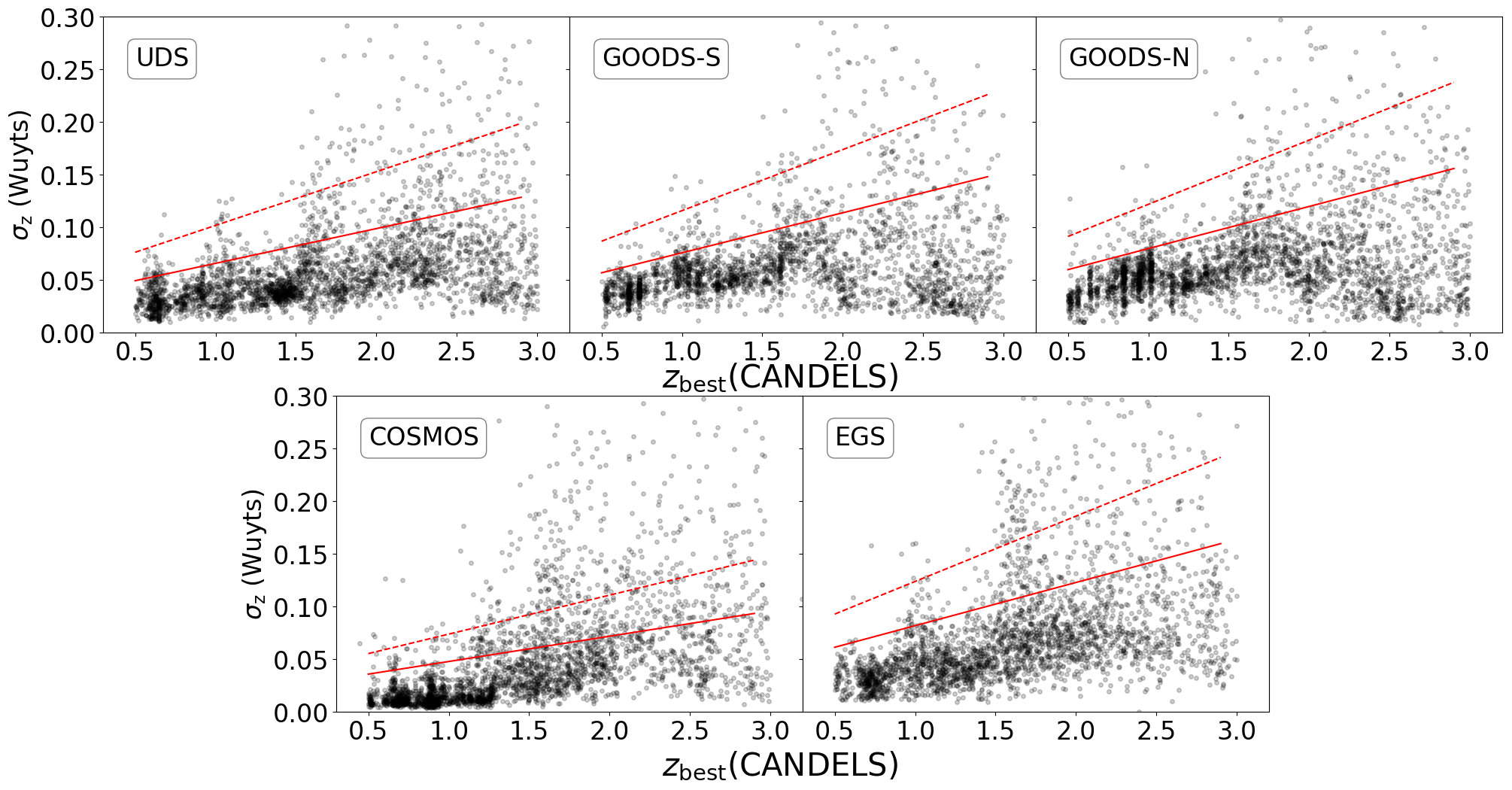}
	\caption{{Photometric redshift uncertainties ($\sigma_{\rm z}$) as a function of $z_{\rm best}$ for galaxies with $M_{\rm stellar}\geq5\times10^{9}M_{\odot}$ in each CANDELS field. The $\sigma_{\rm z}$ values are the $1\sigma$ photometric redshift errors from the optimized $P(z)$ distributions originally derived by S. Wuyts (see text for details). In each panel, we show the 80\% and 95\% outlier clipped limits of the redshift normalized uncertainty  $\sigma_{\rm z}/(1+z_{\rm best})$ distribution in solid and dashed lines, respectively. We {present} these limits in Table\,\ref{clipping_limits}.}}
	\label{fig_appendix_1}
\end{figure*}

In Figure\,\ref{fig_appendix_1}, we show the photometric redshift uncertainties ($\sigma_{\rm z}$) as a function of $z_{\rm best}$ for each galaxy in our sample ($M_{\rm stellar}\geq 5\times 10^{9}M_{\odot}$). We find that the $\sigma_{\rm z}$ distributions in the CANDELS fields are qualitatively similar to each other. We find the $\sigma_{\rm z}$ distributions have small scatter up to $z\sim 1.5$ with  their medians typically ranging between $0.02\leq \sigma_{\rm z,med}\leq 0.05$, and much larger scatter at $z\ga1.5$ with the medians ranging between $0.06\leq \sigma_{\rm z,med} \leq 0.08$. This large scatter is because the observed filters no longer span the 4000\AA\, break, which leads to larger uncertainties during template SED-fitting (for additional details, see Kodra \etal, in preparation). For each CANDELS field, {we show the 80\% and 95\% outlier limits} of the redshift normalized error [$\sigma_{\rm z}/(1+z_{\rm best})$] distribution and present their values in Table\,\ref{clipping_limits}. While the 95\% clipping limit rejects extreme outliers typically with $z_{\rm best}>1.5$, the 80\% limit does a reasonable job representing the upper envelope of the $\sigma_{\rm z}$ distribution at all redshifts. Therefore, to exclude {galaxies} with large $z_{\rm phot}$ errors, we elect to exclude those $\sigma_{\rm z}$ above the 80\% clipping limit. Hereafter, we define the large-error $z_{\rm phot}$ as {\it unreliable}.

For each galaxy in our CANDELS sample, we adopt the \cite{bundy_greater_2009} (hereafter, B09) redshift proximity criteria given by :
\begin{equation} \label{method_2_1}
\Delta z_{12}^{2} \leq \sigma_{z,1}^2 +\sigma_{z,2}^2 ~,
\end{equation}
where $\Delta z_{12} = (z_{\rm best,1}-z_{\rm best,2})$ is the redshift difference of the host and companion galaxies, and $\sigma_{z,1}$ and $\sigma_{z,2}$ are their photometric redshift errors, respectively. {It is important to note that projected pairs containing widely separated galaxies in redshift space that have large $z_{\rm phot}$ errors can satisfy Equation\,\ref{method_2_1}.} Hence, we apply the redshift proximity criteria only to those galaxies with reliable photometric redshifts. In summary, we select $N_{\rm phy} = 504$ {massive galaxies hosting major companions satisfying} $5~{\rm kpc} \leq R_{\rm proj} \leq 50$ kpc, $1 \leq M_{1}/M_{2} \leq 4$, and Equation\,\ref{method_2_1}. We present the breakdown of $N_{\rm phy}$ in each redshift bin per CANDELS field in Table\,\ref{summary_b09_selection}.

In \cref{fmp_for_candels}, we describe a statistical correction to add back a subset of galaxies excluded because of an unreliable $\sigma_{\rm z}$ that could be statistically satisfying the redshift proximity criteria. Additionally, owing to the possibility that some {companion galaxies} may satisfy the close redshift proximity criterion by random chance, we discuss the statistical correction for random chance pairing in \cref{correction_for_random_pairing}. We acknowledge the mismatch between the redshift proximity methods that we employ for the SDSS and CANDELS. In \cref{impact_of_pair_selection}, we test the impact of this mismatch and find that it does not significantly impact our results and conclusions. 

\section{Frequency of Major Merging}
\label{deriving_fmp}
To track the history of major merging, we start by analyzing the fraction of massive ($M_{\rm stellar}\geq 2\times 10^{10}M_{\odot}$) {galaxies hosting a major companion selected in \cref{selection_of_massive_gals_in_maj_pairs}. The {\it major companion} fraction\footnote{While the companion fraction is related to the pair fraction, it is important to be clear that it is not the same (see \cref{sec_comp_prev_results}).} is}
\begin{equation}\label{equation_fmp}
f_{\rm mc}(z) = \frac{N_{\rm mc}(z)}{N_{\rm m}(z)}~,
\end{equation}
{where at each redshift bin, $N_{\rm mc}$ is the number of massive galaxies hosting a major companion after statistically correcting the $N_{\rm phy}$ counts (see \cref{fmp_for_candels,fmp_for_sdss}), and $N_{\rm m}$ is the number of massive galaxies. The companion fraction $f_{\rm mc}$ is commonly used in the literature, and is the same as $f_{\rm merg}$ used by \cite{man16}}.  We use the samples described in \cref{selection_of_massive_gals_in_maj_pairs} to derive $N_{\rm mc}(z)$ separately for the five redshift bins in CANDELS ($0.5\leq z \leq 3.0$) and for the SDSS $z\sim 0$ anchor. We then discuss the application of a correction to account for galaxies satisfying the companion selection criteria by random chance. Finally, we characterize the companion fraction and use an analytical function to quantify the redshift evolution of $f_{\rm mc}$ during $0<z<3$.

\subsection{Deriving $N_{\rm mc}(z)$ for CANDELS}
\label{fmp_for_candels}

For redshifts $0.5\leq z \leq 3$, we compute $N_{\rm mc}$ corrected for incompleteness owing to unreliable photometric redshifts as
\begin{equation}\label{equation_Nmp}
N_{\rm mc} = N_{\rm phy}+C_{1}N_{\rm proj,unreliable}\, ,
\end{equation}
where $N_{\rm phy}$ is the {number of massive galaxies with reliable $z_{\rm phot}$ values that host a major companion} (\cref{plausible_physical_pairs_can}), $N_{\rm proj,unreliable}$ is the number of galaxies {hosting major projected companions} that are excluded because of unreliable $z_{\rm phot}$ values (\cref{major_projected_pairs}), and $C_{1}$ is the correction factor used to statistically add back a subset of excluded galaxies that are expected to satisfy the redshift proximity criteria we employ.
We estimate $C_1$ as
\begin{equation}
\label{equation_c1}
C_{1} = \frac{N_{\rm phy}}{N_{\rm proj}-N_{\rm proj,unreliable}}\, .
\end{equation}
To study $f_{\rm mc}(z)$ from the overall sample and also its field-to-field variations, we calculate $N_{\rm mc}$ for five $\Delta z=0.5$ bins separately for each CANDELS field and for the total sample using counts tabulated in Table\,\ref{summary_b09_selection}. For example, in the $0.5\leq z \leq 1$ bin, CANDELS contains 671 {massive galaxies with reliable $z_{\rm phot}$ values hosting a major projected companion}. This results in $C_{1} = 193/671 = 0.29$ and a corrected count of $N_{\rm mc}=230$; i.e., we add back 29\% of the previously excluded galaxies at these redshifts. We tabulate $C_{1}$ and $N_{\rm mc}$ values in Table\,\ref{summary_b09_selection}. We also note no significant difference in the computed $N_{\rm mc}$ values whether we use an 80\% or 95\% $\sigma_{z}$ clipping limit (\cref{selection_of_massive_gals_in_maj_pairs}) to remove unreliable photometric redshifts.

\subsection{Deriving $N_{\rm mc}(z)$ for SDSS}
\label{fmp_for_sdss}
To achieve an accurate low-redshift anchor for the fraction of {massive galaxies hosting a major companion} meeting our $\Delta v_{12}\leq 500\, {\rm km~s^{-1}}$ velocity separation criterion, we calculate $N_{\rm mc}$ corrected for the SDSS spectroscopic incompleteness that varies with projected separation using $\Delta R_{\rm proj} = 5\,{\rm kpc}$ bins as follows:
\begin{equation}\label{equation_Nmp_sdss}
N_{\rm mc} = N_{\rm phy} + \sum_{i=1}^{9} \left(C_2\,N_{\rm proj,nospec}\right)_i \, .
\end{equation}
For each of nine bins between $R_{\rm proj}=5-50$\,kpc, we compute a correction factor $C_{2,i}$ necessary to add back a statistical subset of the $N_{\rm proj,nospec}$ galaxies in the bin that lack spectroscopic redshifts but that we expect to satisfy the $\Delta v_{12}\leq 500{\rm km\,s^{-1}}$ criterion. 
{Following the same logic as in Equation\,\ref{equation_c1}, we estimate this correction at each $\Delta Rproj$ bin based on the counts of spectroscopic galaxies hosting a major projected companion and the fraction that satisfy $\Delta v_{12}\leq 500\,{\rm km\,s^{-1}}$. Owing to our well-defined sample volume ($0.03<z<0.05$), the total sample of spectroscopic hosts with plausible physical companions (\cref{plausible_physical_pairs}) is limited to $N_{\rm phy}=109$ over the nine separation bins.} 
To reduce random errors from small number statistics, we use a larger redshift range ($0.01\leq z \leq 0.05$) and SDSS footprint ($\sim 4000\deg^{2}$), to calculate the correction factor at each $\Delta R_{\rm proj}$ bin:
\begin{equation}\label{equation_c2}
C_{2,i} = \left(\frac{N_{\rm phy}^{\prime}}{N_{\rm proj}^{\prime}-N_{\rm proj,nospec}^{\prime}}\right)_{i}\, .
\end{equation}

\begin{figure}
		\centering
		\includegraphics[width=\columnwidth]{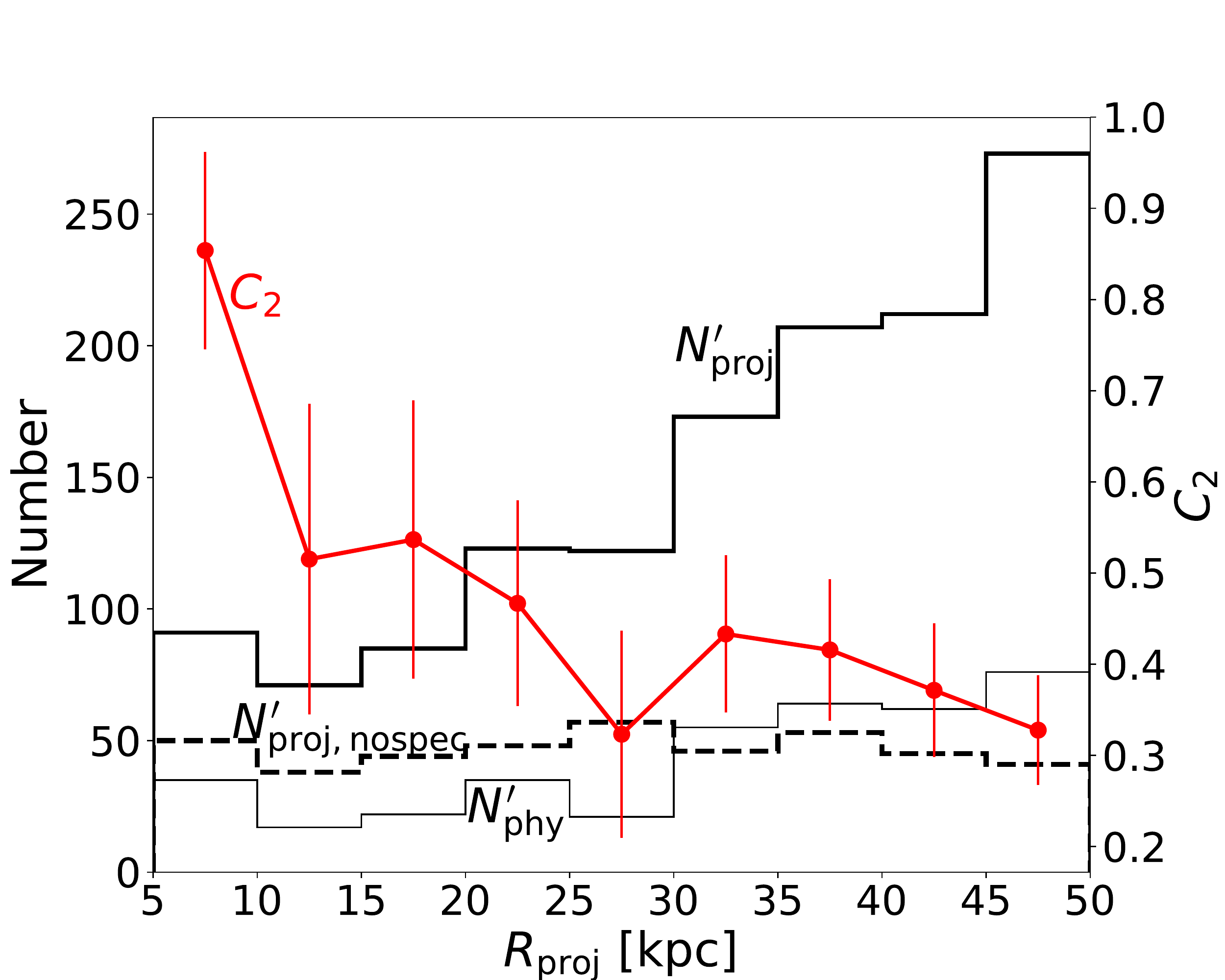} 
		\caption{The number of massive galaxies {hosting a major companion} in projected pairs as a function of projected separation {from the SDSS}: total ($N^{\prime}_{\rm proj}$; bold line), the subset of galaxies with no spectroscopic redshift information ($N^{\prime}_{\rm proj,nospec}$; dashed line), and the subset of galaxies satisfying $\Delta v_{12}\leq 500\,{\rm km\,s^{-1}}$ ($N^{\prime}_{\rm phy}$; thin line). The {\it prime} signifies that the quantities are derived using a larger SDSS sample spanning $0.01\leq z\leq 0.05$ and $\sim 4000\deg^{2}$ (see \cref{fmp_for_sdss}) for nine $R_{\rm proj}$ bins between $5-50$\,kpc. The correction factor $C_{2}$ per $R_{\rm proj}$ bin (Equation\,\ref{equation_c2}) is given by the {red} circles connected by a {red} solid line. The error bars represent 95\% binomial confidence of $C_{2}$.}
		\label{fig_sdss_spectroscopic_incompleteness}
	\end{figure}

In Figure\,\ref{fig_sdss_spectroscopic_incompleteness}, we plot $C_{2}$ and the factors in Equation\,\ref{equation_Nmp_sdss} as a function of $R_{\rm proj}$ using this larger sample (emphasized with a {\it prime}). For example, in the $20 \leq R_{\rm proj}\leq 25~{\rm kpc}$ separation bin, we find 119 {massive galaxies hosting a major projected companion}, of which 35 have small velocity separations and 44 lack  spectroscopic redshifts. This results in a 47\% correction ($C_{2} = 35/75 = 0.47$)
at this separation. We find the probability for a small-separation pair ($R_{\rm proj} = 5$\,kpc) to satisfy $\Delta v_{12}\leq 500~{\rm km~{s^{-1}}}$ (despite lacking the redshift information) is 85\% ($C_{2}=0.85$) and this rapidly decreases to $\sim30$\% ($C_{2}\sim0.3$) at $R_{\rm proj} = 30$\,kpc, and remains statistically constant between $R_{\rm proj} = 30-50$\,kpc. This correction is important since the spectroscopic incompleteness $N^{\prime}_{\rm proj,nospec}/N^{\prime}_{\rm proj}$ ranges from $>$0.6 ($\sim 5-10$\,kpc) to 0.2 ($\sim 45-50$\,kpc) over the separations we probe, which is in agreement with trends published in Figure\,2 from \cite{weston17}.

%We acknowledge a minor caveat that the calculated $C_{2}$ is an approximation of the true correction, as it assumes that the numerator and denominator properly sample their full parent distributions\footnote{To reduce the random errors when calculating $C_{2}$ as a function of $R_{\rm proj}$, we leverage the statistical power of the full redshift range and area of the SDSS Sample III ($0.01\leq z \leq 0.08$).}. 

%Also, for this part, 
	
\begin{table*}
	\centering
	\caption{Detailed breakdown of variables involved in estimating the $f_{\rm mc}$ and $f_{\rm mc,c}$ at five redshift bins between $0.5\leq z \leq 3$.  Columns: (1) name of the CANDELS field; (2) the CANDELS team $z_{\rm best}$ bin; (3) number count of massive ($\log M_{\rm stellar}/M_{\rm \odot} \geq 10.3$) galaxies; (4) number of massive galaxies {hosting a major projected companion} (\cref{major_projected_pairs}), those of which that have unreliable photometric redshift values are shown in parenthesis; (5) number of massive galaxies with reliable $z_{\rm phot}$ that {host a major projected companion} satisfying redshift proximity (Equation\,\ref{equation_methodII}) as described in \cref{plausible_physical_pairs_can}; (6) the correction factor computed using Equation\,\ref{equation_c1}; (7) the number of massive galaxies {hosting a major companion} after statistically correcting $N_{\rm phy}$ for incompleteness owing to unreliable $z_{\rm phot}$ values from Equation\,\ref{equation_Nmp}; (8) correction factor to account for random chance pairing as described in \cref{correction_for_random_pairing}; (9) the fraction of massive galaxies {hosting a major companion} ({\it major companion} fraction); (10) random chance pairing corrected $f_{\rm mc}$, as described in \cref{correction_for_random_pairing}.} 
	\label{summary_b09_selection}
	\begin{tabular}{cccccccccc}
		\hline
		Name & Redshift   &  $N_{\rm m}$ & $N_{\rm proj}\, (N_{\rm proj,unreliable})$ & $N_{\rm phy}$ & $C_{1}$ & $N_{\rm mc}$ & $C_{3}$   & $f_{\rm mc}$ (\%) & $f_{\rm mc,c}$ (\%)  \\
		(1) & (2)&   (3) & (4) & (5) &  (6) & (7) & (8) & (9)& (10)\\
		\hline
		& $0.5\leq z \leq1$ &256 & 132 (16) & 44 & 0.38& 50& & $20\pm5$&  \\
		& $1\leq z \leq1.5$ &304 & 130 (23) & 34 & 0.32& 41& & $14\pm4$&  \\
		UDS& $1.5\leq z \leq2$ &290 & 130 (30) & 25 & 0.25& 33& & $11\pm4$&  \\
		& $2\leq z \leq2.5$ &216 & 80 (31) & 14 & 0.29& 23& & $11\pm4$& \\
		& $2.5\leq z \leq3$ &157 & 82 (42) & 3 & 0.07& 6& & $4\pm3$& \\
		\hline
		& $0.5\leq z \leq1$ &216 & 107 (21) & 32 & 0.37& 40& & $18\pm5$& \\
		& $1\leq z \leq1.5$ &252 & 116 (14) & 34 & 0.33& 39& & $15\pm4$& \\
		GOODS-S& $1.5\leq z \leq2$ &213 & 87 (22) & 15 & 0.23& 20& & $9\pm4$& \\
		& $2\leq z \leq2.5$ &138 & 56 (14) & 3 & 0.07& 4& & $3\pm3$& \\
		& $2.5\leq z \leq3$ &123 & 57 (15) & 5 & 0.12& 7& & $6\pm4$& \\
		\hline
		& $0.5\leq z \leq1$ &333 & 195 (33) & 56 & 0.35& 67& & $20\pm4$& \\
		& $1\leq z \leq1.5$ &278 & 140 (27) & 40 & 0.35& 50& & $18\pm4$& \\
		GOODS-N & $1.5\leq z \leq2$ &209 & 99 (20) & 15 & 0.19& 19& & $9\pm4$& \\
		& $2\leq z \leq2.5$ &191 & 83 (17) & 6 & 0.09& 8& & $4\pm3$& \\
		& $2.5\leq z \leq3$ &122 & 61 (16) & 5 & 0.11& 7& & $6\pm4$& \\
		\hline
		& $0.5\leq z \leq1$ &448 & 244 (38) & 40 & 0.19& 47& & $11\pm3$& \\
		& $1\leq z \leq1.5$ &270 & 128 (37) & 1 & 0.01& 1& & $1\pm1$& \\
		COSMOS & $1.5\leq z \leq2$ &350 & 157 (75) & 15 & 0.18& 29& & $8\pm3$& \\
		& $2\leq z \leq2.5$ &153 & 86 (54) & 0 & 0.0& 0& & $0\pm0$& \\
		& $2.5\leq z \leq3$ &86 & 36 (22) & 6 & 0.43& 15& & $18\pm8$&  \\
		\hline
		& $0.5\leq z \leq1$ &224 & 120 (19) & 21 & 0.21& 25& & $11\pm4$& \\
		& $1\leq z \leq1.5$ &304 & 137 (19) & 36 & 0.31& 42& & $14\pm4$& \\
		EGS& $1.5\leq z \leq2$ &331 & 171 (50) & 39 & 0.32& 55& & $17\pm4$& \\
		& $2\leq z \leq2.5$ &167 & 94 (18) & 23 & 0.3& 28& & $17\pm6$& \\
		& $2.5\leq z \leq3$ &67 & 35 (13) & 2 & 0.09& 3& & $5\pm5$& \\
		\hline
		& $0.5\leq z \leq1$ &1477 & 798 (127) & 193 & 0.29& 230& 0.15 & $16\pm2$& $13\pm2$ \\
		& $1\leq z \leq1.5$ &1408 & 651 (120) & 145 & 0.27& 173& 0.17 & $12\pm2$& $10\pm2$ \\
		All fields& $1.5\leq z \leq2$ &1393 & 644 (197) & 109 & 0.24& 155& 0.18 & $11\pm2$& $9\pm2$ \\
		& $2\leq z \leq2.5$ &865 & 399 (134) & 46 & 0.17& 63& 0.25 & $7\pm2$& $5\pm2$ \\
		& $2.5\leq z \leq3$ &555 & 271 (108) & 21 & 0.13& 38& 0.22 & $7\pm2$& $5\pm2$\\
		\hline
	\end{tabular}
	
\end{table*}

\begin{table}
	\centering
	\caption{Photometric redshift uncertainty outlier limits that are used to determine reliable $z_{\rm phot}$ values for each CANDELS field. Columns: (1) name of the CANDELS field; (2,3) the 80\% and 95\% outlier clipped limits of the redshift normalized uncertainity $\sigma_{\rm z}/(1+z_{\rm best})$ distributions for galaxies in the mass-limited ($M_{\rm stellar}\geq 5\times10^{9}M_{\odot}$ sample for redshifts $0.5\leq z\leq 3.0$ as shown in Figure\,\ref{fig_appendix_1}.}
	\label{clipping_limits}
	\begin{tabular}{ccc} 
		\hline
		Name & 80\% limit   &   95\% limit\\
			(1) & (2)   &  (3) \\
			\hline
			UDS & 0.033 & 0.051 \\
			GOODS-S & 0.038 & 0.058\\
			GOODS-N & 0.04& 0.061\\
			COSMOS & 0.024& 0.037\\
			EGS & 0.041& 0.062\\
			\hline
	\end{tabular}
\end{table}

	\begin{figure*}
		\centering
		%	\subfloat[]{\includegraphics[width=3.in]{fmp_SDSS_CAN_3_proj_seps_Man14.png}} 
		%	\subfloat[]{\includegraphics[width=3.in]{M14_chance_pairing_corr.png}}\\
		\subfloat[]{\includegraphics[width=\columnwidth]{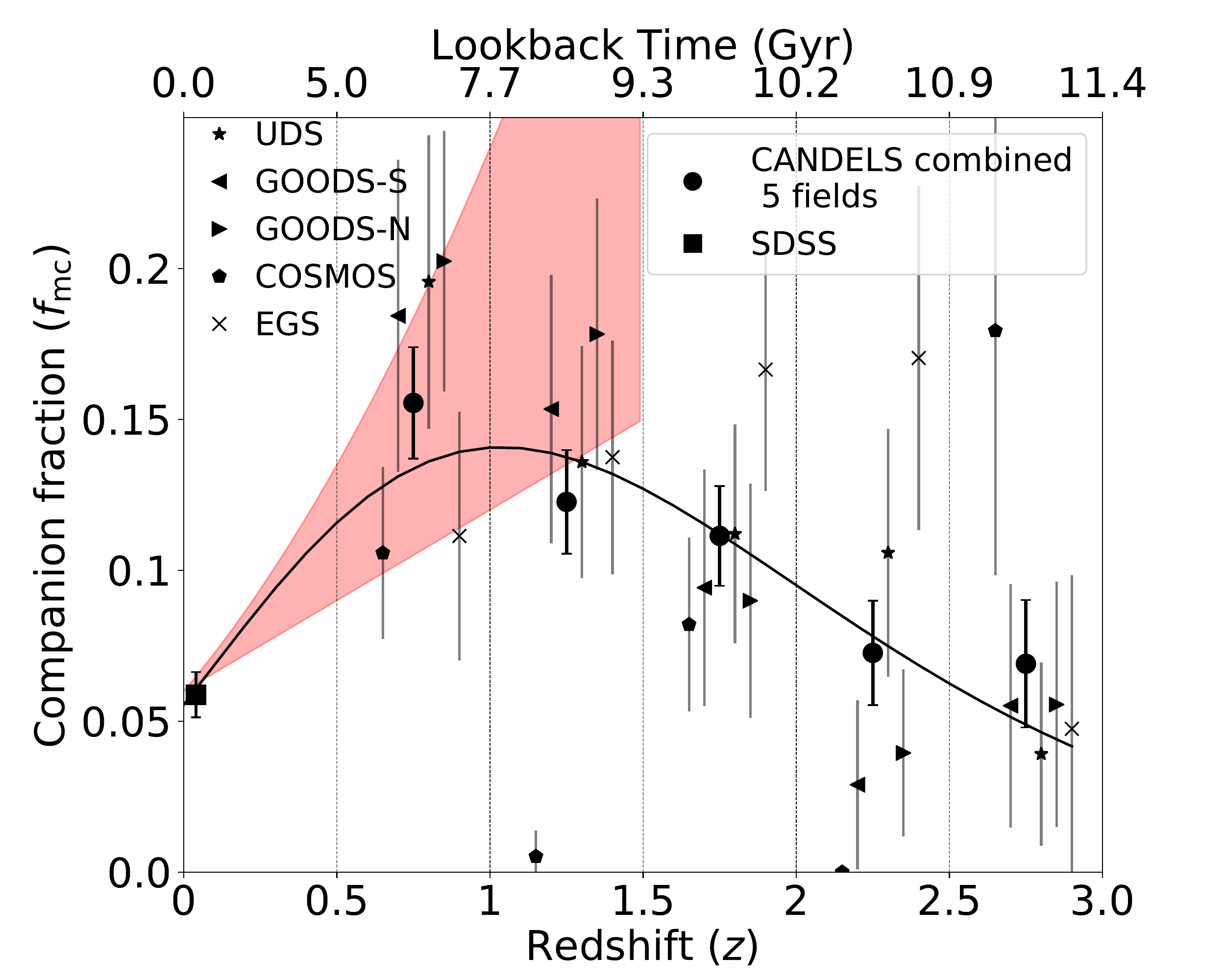}}
		\subfloat[]{\includegraphics[width=\columnwidth]{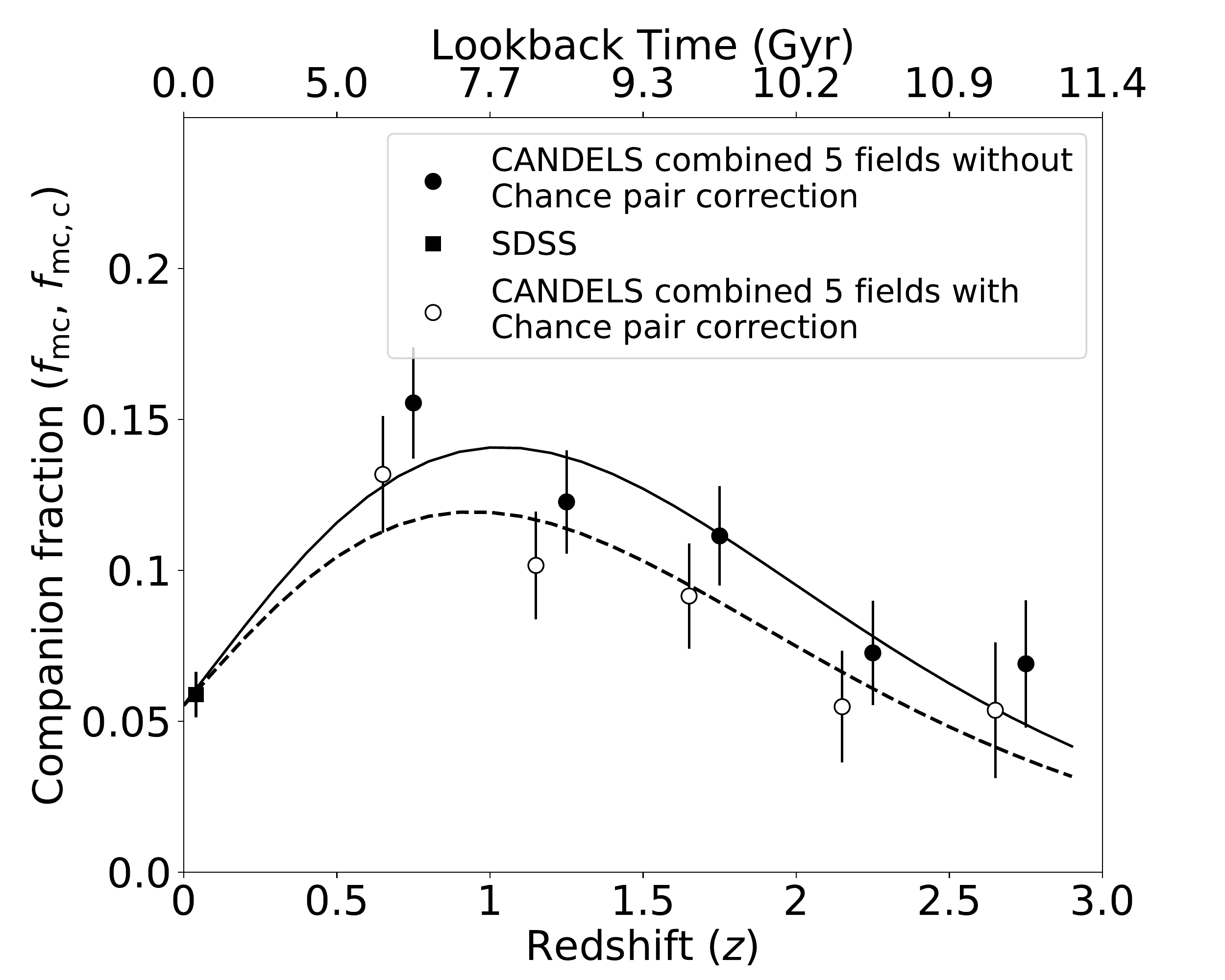}}
		%	\subfloat[]{\includegraphics[width=3.in]{fmp_SDSS_CAN_3_proj_seps_hybrid.png}}
		%	\subfloat[]{\includegraphics[width=3.in]{hybrid_chance_pairing_corr.png}}
		\protect\caption{ (a): The redshift evolution of the {\it major companion} fraction $f_{\rm mc}$ shown for the five CANDELS fields UDS (star), GOODS-S (left triangle), GOODS-N (right triangle), COSMOS (pentagon), EGS (cross). The combined CANDELS fractions in five redshift bins (circles) and the SDSS low-redshift anchor (square) include 95\% binomial confidence limit error bars. To place our finding in the context of common, close-pair-based evolutionary trends found in the literature, we plot the shaded region (red) encompassing a common range of power-law slopes $f_{\rm mc} = 0.06(1+z)^{1-2}$ at $0<z<1.5$. (b): The random chance corrected fractions ($f_{\rm mc,c}$) for the five CANDELS $\Delta z$ bins (open circles) are compared with the $f_{\rm mc}(z)$ from (a). For $f_{\rm mc,c}$, the binomial errors and scatter of $C_{3}$ (see \cref{correction_for_random_pairing}) are added in quadrature. Best-fit curves to the companion fraction ($f_{\rm mc}$) evolution data (see Equation\,\ref{equation_best_fit} and \cref{best_fit_formulation} for details) are shown in solid ($f_{\rm mc}$) and dashed ($f_{\rm mc,c}$) lines, respectively. In the case of SDSS, since the correction $C_{3}\sim 0.01$, we only plot $f_{\rm mc}$ for simplicity. From this figure, we conclude that the {major companion fraction} increases strongly from $z\sim 0$ to $z\sim 1$, and decreases steeply towards $z\sim 3$ (see text for details).} 
		\label{b09_with_and_without_correction}
	\end{figure*} 
	
\subsection{Redshift Evolution of Major Merging Frequency}
\label{modeling_fmp}
We use the corrected counts of {massive galaxies hosting a major companion to compute the companion fraction ($f_{\rm mc}$; see Equation\,\ref{equation_fmp}) in the SDSS and CANDELS}. We compare field-to-field variations of $f_{\rm mc}$ in CANDELS, quantify the redshift evolution of the $f_{\rm mc}$ from $z=3$ to $z=0$, and measure the impact of random chance pairing on the observed {major companion fraction} evolutionary trends.
	
\subsubsection{Field-to-Field Variations}
In Figure\,\ref{b09_with_and_without_correction}(a), we plot $f_{\rm mc}(z)$
for the combined CANDELS fields and compare this with the individual fractions
from each field at each redshift; these are also tabulated in 
Table\,\ref{summary_b09_selection}.
Despite noticeable variations between the fractions derived from each 
CANDELS field owing to small-number statistics, we find fair agreement between multiple fields at each redshift. We note that the combined CANDELS sample and three individual fields (UDS, GOODS-S, and GOODS-N) show consistent
trends with the highest merger fractions at $z\sim1$, which then steadily
decrease with increasing redshift. The EGS and COSMOS companion fractions 
exhibit different behavior with redshift, the former peaks at $z\sim 2$
while the latter has no trend with redshift owing to a lack of {galaxies hosting major companions} in
two different redshift bins. We compute the cosmic-variance ($\sigma_{\rm CV}$) on the combined CANDELS $f_{\rm mc}(z)$ values using the prescription by \citet{Moster11}. For $\logten (M_{\rm stellar}/M_{\odot}) \ga10$ galaxies at $0.5<z<3.0$ populating the five CANDELS fields each with an area of $160$\,arcmin$^2$ such that their cumulative area matches that of the total CANDELS coverage ($5\times160 = 800$\,arcmin$^2$), we find that the number counts of galaxies hosting major companions have $\sigma_{\rm CV}$ ranging from $11\%$ ($z=0.75$) to $18\%$ ($z=2.75$)\footnote{We take into account that $\sigma_{\rm CV}$ is smaller for multiple, widely separated fields when compared to the $\sigma_{\rm CV}$ of a single contiguous field. For additional details, see \cite{Moster11}}. While most of the individual CANDELS-field fractions are consistent with the $\sigma_{\rm CV}$ within their large uncertainties (owing to small sample size), few $f_{\rm mc}$ values (\eg\, at $z>1.5$ for the COSMOS and EGS fields; see Figure\,\ref{b09_with_and_without_correction}a) are significantly above the possible cosmic-variance limits.
	
\subsubsection{Analytical Fit to the Major Companion Fraction Evolution}\label{best_fit_formulation}
To characterize the redshift evolution of the companion fraction 
during $0<z<3$, we anchor the combined
CANDELS $f_{\rm mc}(z)$ measurements to the SDSS-derived data point at $z\sim0$.
As shown in Figure\,\ref{b09_with_and_without_correction}(a),
the low-redshift fraction is $\sim3\times$ lower than the maximum 
$f_{\rm mc}\sim0.16$ value at $0.5<z<1$, which then decreases to $f_{\rm mc}\sim 0.07$ at $2.5<z<3$. This suggests a turnover in the
incidence of merging sometime around $z\sim1$, {in agreement with some previous studies \citep{conselice03a,Conselice08}}.  Previous close-pair-based studies at $z\sim0$ find fractions $f_{\rm mc}\sim 2\% \pm 0.5\%$ \citep{Patton00,Patton08,Domingue09}, but they used criteria that are different from our fiducial selection. Similarly, many empirical, close-pair-based studies in the literature broadly agree that $f_{\rm mc}$ rises at $0<z<1.5$ but with a range of evolutionary forms $(1+z)^{1\sim2}$ owing to varying {companion} selection criteria \citep[for discussion, see][]{lotz_major_2011}. After normalizing for these variations, we note that our SDSS-based $f_{\rm mc}$ is in good agreement with previous close-pair-based estimates, and our rising trend (see shaded region, Figure\,\ref{b09_with_and_without_correction}a) between $0<z<1.5$ is in broad agreement with other empirical trends. In \cref{Discussion}, we will present detailed comparisons to other studies by re-computing $f_{\rm mc}$ based on different companion selection choices that match closely with others.

All studies that measured redshift evolution of merger frequency at $0<z<1.5$, irrespective of the methodology, have used the power-law analytical form $f_{\rm}(z)\propto (1+z)^{\rm m}$ to represent the best-fit of $f(z)$. This functional form cannot be used to represent the observed rising and then decreasing $f_{\rm mc}(z)$ for redshift ranges $0<z<3$. Therefore, following \cite{Conselice08} and initially motivated by \citet{carlberg90}, we use
a modified power-law exponential function given by
\begin{equation}\label{equation_best_fit}
f_{\rm mc}(z) = \alpha(1+z)^{\rm m}\exp^{\rm \beta(1+z)} \, .
\end{equation}
As demonstrated in Figure\,\ref{b09_with_and_without_correction}(a), this analytic function provides a good fit to the observed evolution. The best-fit curve to the fractions derived from the SDSS and CANDELS measurements from our fiducial {companion selection} criteria has parameters $\alpha = 0.5\pm0.2$, $m = 4.5\pm0.8$, and $\beta = -2.2\pm0.4$. We note that we will apply this fitting function for different {companion selection choices} throughout our comparative analysis described in \cref{impact_of_pair_selection,Discussion}.

\subsubsection{Correction for Random Chance Pairing}
\label{correction_for_random_pairing}
Finally, we note that a subset of massive galaxies hosting a major companion ($N_{\rm mc}$) can satisfy the {companion selection} criteria by random chance. To account for this
contamination, we apply a statistical correction and recompute the counts for the combined CANDELS sample per redshift bin as $N_{\rm mc,c} = N_{\rm mc}(1-C_{3})$ in each redshift bin. 
To compute $C_{3}$, we
generate 100 simulated Monte-Carlo (MC) randomized datasets\footnote{We generate these datasets by randomizing the positions of each galaxy in the $\logten(M_{\rm stellar}/M_{\odot})\geq 9.7$ mass-limited sample and repeating the selection process in \cref{selection_of_massive_gals_in_maj_pairs}}. We define $C_{3}$ in each redshift bin {as the ratio of massive galaxy number counts hosting major companions which satisfy our projected separation and photometric redshift proximity criteria in the MC datasets (\ie\, by random chance)} to the measured $N_{\rm mc}$ (\cref{fmp_for_sdss,fmp_for_candels}). For example, in redshift bin $1<z<1.5$, we find that 17\% of $N_{\rm mc}=173$ {galaxies statistically satisfy the companion selection criteria by random chance}. We tabulate $C_3$ values at each redshift for CANDELS in Table\,\ref{summary_b09_selection}. We repeat this process for the SDSS ($0.03< z <0.05$) and find a very small correction of $\sim 1\%$ ($C_{3} \sim 0.01$). This demonstrates the very low probability for two SDSS galaxies to satisfy both the close projected separation and stringent spectroscopic redshift proximity ($\Delta v_{12}\leq 500~{\rm km~s^{-1}}$) criteria.

In Figure\,\ref{b09_with_and_without_correction}(b), we compare the random chance corrected fractions $f_{\rm mc,c} = N_{\rm mc,c}/N_{\rm m}$ at each redshift bin from CANDELS, to the uncorrected $f_{\rm mc}$ values copied from the left panel. Owing to the insignificant 1\% correction at $z\sim 0$, we anchor both the corrected and uncorrected fits to the same SDSS data point. 
We find that $f_{\rm mc,c}(z)$ follows the same evolutionary trend as $f_{\rm mc}(z)$, in which the best-fit curve rises to a maximum fraction at $z\sim1$ and then steadily decreases to $z=3$. At all redshifts, $f_{\rm mc,c}$ is within the statistical errors of $f_{\rm mc}$.  The qualitatively {similar} redshift evolutionary trends of $f_{\rm mc}(z)$ and $f_{\rm mc,c}(z)$ is due to the nearly redshift independent amount of statistical correction for random pairing ($|\Delta f|/f_{\rm mc}\sim 20\%$) at $1<z<3$. This is because of the nearly invariant angular scale in this redshift range, which results in similar random chance pairing probabilities.
We note that some previous close-pair-based studies have applied this random chance correction \citep{kartaltepe_evolution_2007,bundy_greater_2009}, while others have not \citep[e.g.,][]{man2012}.
	
\section{Impact of Close-Companion Selection Criteria on Empirical Major Companion Fractions}
\label{impact_of_pair_selection}
So far, we have discussed the derivation and redshift evolution of
the major companion fraction $f_{\rm mc}$ based on
our {\it fiducial} selection criteria described in \cref{selection_of_massive_gals_in_maj_pairs}.
As illustrated in Table\,\ref{selection_crit_prev_stuides},
previous studies have employed a {variety of criteria to select companions}.
In this section, we study the impact of different {companion} selections
on $f_{\rm mc}(z)$ derived from our sample. We systematically vary each
criterion (projected separation, redshift proximity, and stellar mass ratio versus
flux ratio) individually, while holding the other criteria fixed to their
fiducial values. Then, we compare
each recomputed $f_{\rm mc}(z)$ to the fiducial result from Figure\,\ref{b09_with_and_without_correction}.
In \cref{correction_for_random_pairing}, we show that applying a statistical correction for {random chance pairing} will produce companion fractions that are $\sim 20\%$ lower at each redshift interval between $0.5\leq z \leq 3$.
Therefore, we focus the following comparative analysis on uncorrected fractions and note that our {qualitative} conclusions are robust to whether or not we apply this correction.

\subsection{Projected Separation}\label{sec_changing_rproj}
To quantify the impact that changing {\it only} the  criterion for companion projected separation will have on $f_{\rm mc}(z)$, we compare fractions based on the fiducial $R_{\rm proj}=5-50~{\rm kpc}$ selection to those derived from two common criteria: $R_{\rm proj}=14-43~{\rm kpc}$ \citep[e.g.,][]{lin_redshift_2008,man16} and $R_{\rm proj}=5-30~{\rm kpc}$ \citep[e.g.,][]{Patton08,man2012}. 
For each case, we hold all other {companion} selection criteria fixed such that we are strictly comparing fractions of
major (stellar mass-selected) companions in close redshift proximity that are found within projected annular areas of $2/3$ and $1/3$ the fiducial selection 
window ($2475\pi\,{\rm kpc}^{2}$), respectively. 

\begin{figure}
	\centering
	\includegraphics[width=\columnwidth]{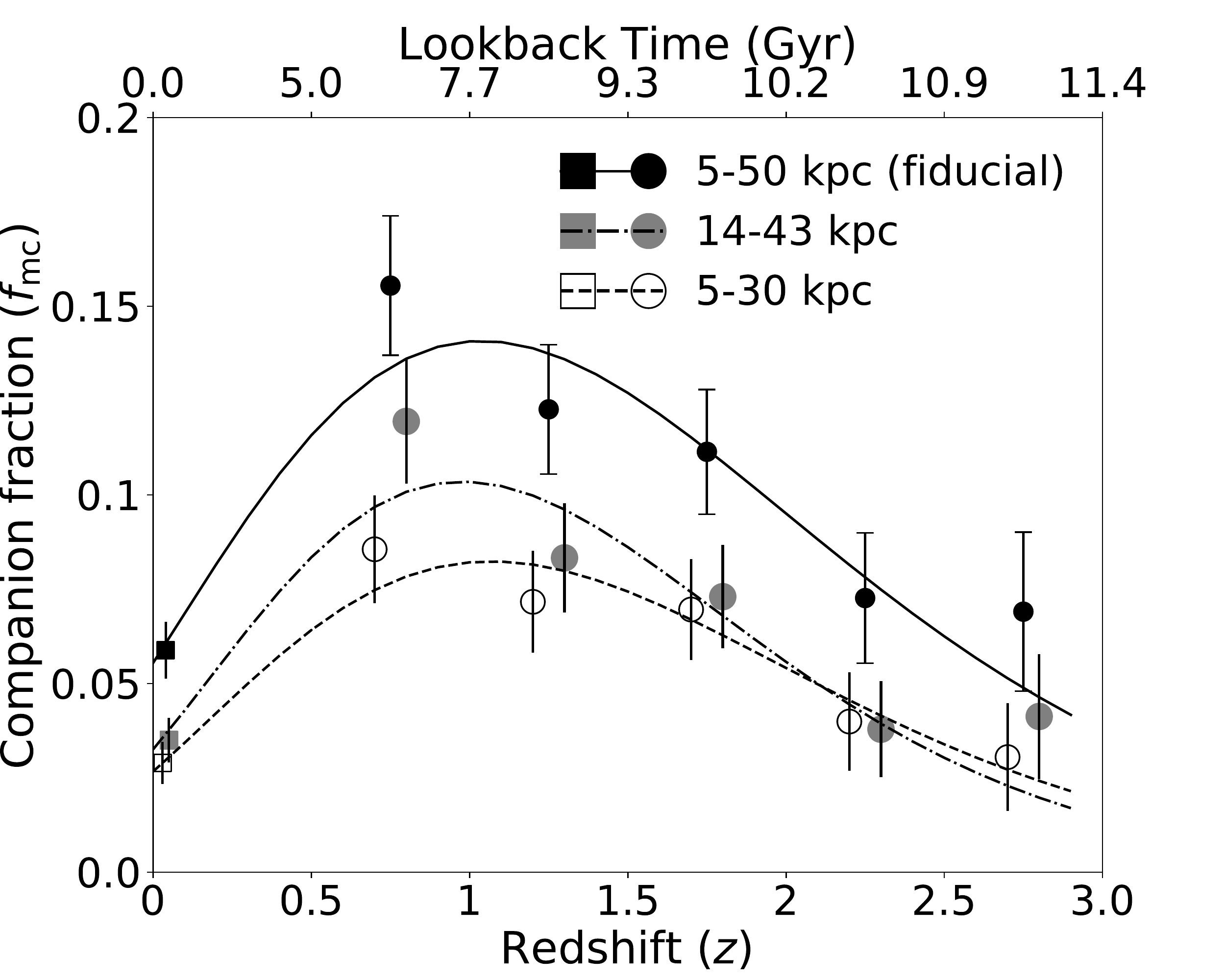}\\
	\caption{Comparison of the redshift evolution of the major companion fractions based on three projected separation criteria: $R_{\rm proj}=5-50$\,kpc (fiducial; black symbols, solid line), $R_{\rm proj}=14-43$\,kpc (grey symbols, dashed line), and $R_{\rm proj}=5-25$\,kpc (open symbols, dot-dashed line). Best-fit curves for each $f_{\rm mc}(z)$ spanning $0<z<3$ are a modified power-law exponential (Equation\,\ref{equation_best_fit}; see \cref{best_fit_formulation} for details). The low-redshift fractions are from the SDSS (squares) and the $z>0.5$ fractions are from CANDELS (circles).The error bar on each $f_{\rm mc}$ data point represents 95\% binomial confidence limit. The non-fiducial $f_{\rm mc}$ data points are offset by a small amount within each redshift bin for clarity. }	
	\label{changing_rproj}
\end{figure}

In Figure\,\ref{changing_rproj}, we plot the fiducial and non-fiducial $f_{\rm mc}(z)$ data and their best-fit curves.
The key result from this figure is the observed redshift evolution of major companion fractions shown in Figure\,\ref{b09_with_and_without_correction} are qualitatively robust to changes in the projected separation criterion. For each case, we find the fractions increase from low redshift to $z\sim1$, then turn over and decrease fairly steadily to $z=3$. Quantitatively, the non-fiducial fractions are well fit by the same $f_{\rm mc}(z)$ 
functional form as the fiducial results (see Table\,\ref{best_fits}). Owing to the size of the best-fit confidence intervals, the redshifts of the peak fractions are statistically equivalent. Larger samples of massive galaxy-galaxy pairs at redshifts
$0.5<z<1.5$ are needed to place stringent constraints on the peak or turnover redshift.

Besides the overall $f_{\rm mc}$ trends with redshift, we find that both
smaller projected separation criteria select smaller $f_{\rm mc}$ values than the fiducial
selection does, as we expect. Despite being a factor of two smaller in
projected area, the conservative $R_{\rm proj}=5-30~{\rm kpc}$ criterion
coincidentally 
produces fractions that are statistically matched to those from the larger
$R_{\rm proj}=14-43~{\rm kpc}$ selection criterion at most redshift intervals.
Only the $0.5<z<1$ bin has unequal companion fractions
between the two non-fiducial criteria. 
The coincidental finding of similar fractions using different projected
separation criteria is consistent with an increased probability of physical     
companions at smaller projected separations \citep{bell06b,robaina_merger-driven_2010}.

%Tab4
\begin{table*}
	\centering
	\caption{Modified power-law exponential function parameter values for best-fit models to major companion fraction evolution $f_{\rm mc}(z)$ based on different selection criteria; our fiducial criteria are given in bold. Columns: (1) major { companion} definition -- stellar mass ratio $1\leq M_{1}/M_{2}\leq 4$ (MR), or $H$-band flux ratio $1\leq F_{1}/F_{2}\leq 4$ (FR); (2) projected separation $R_{\rm proj}$ criterion; (3) redshift proximity method applied to CANDELS samples -- Equation\,\ref{equation_best_fit} adopted from B09, Equation\,\ref{equation_methodII} modified from \citet{man2012}, or `Hybrid' version of B09 (see text for details); and (4--6) best-fit parameter values ($\alpha$, $m$ and $\beta$) and their $1\sigma$ confidence limits for the model function given in  Equation\,\ref{equation_best_fit}.}
	\label{best_fits}
	\begin{tabular}{cccccc}
		\hline
		Major companion definition&  $R_{\rm proj}$& {Redshift Proximity}& $\alpha$& $m$ & $\beta$\\
		{(1)} & (2) & (3) & (4) & (5) & (6)\\
		\hline
		{\bf MR} & $\mathbf{5-50}${\bf \,kpc} & {\bf B09} & $\mathbf{0.5\pm0.2}$ & $\mathbf{4.6\pm0.9}$ & $\mathbf{-2.3\pm0.5}$   \\
		MR & $14-43$ kpc & B09 & $0.7\pm0.5$ & $6.2\pm1.4$ & $-3.1\pm0.8$  \\
		MR & $5-30$ kpc &  B09 & $0.4\pm0.1$ & $5.4\pm0.8$ & $-2.6\pm0.4$   \\
		\hline
		MR & $5-50$\,kpc & modM12 & $0.5\pm0.2$ & $4.6\pm0.9$ & $-2.2\pm0.5$  \\
		MR & $5-50$\,kpc & hybB09 & $0.4\pm0.1$ & $3.8\pm0.7$ & $-1.9\pm0.4$  \\
		\hline
		FR & $5-50$\,kpc & B09 &$0.1\pm0.1$ & $2.1\pm1.2$ & $-0.5\pm0.6$ \\
		\hline
	\end{tabular}
\end{table*}

% SAVED IN CASE FOR REFEREE   5-30 kpc
%& II & $0.3\pm0.1$ & $5.4\pm0.8$ & $-2.5\pm0.4$  \\
%& III & $0.2\pm0.1$ & $4.6\pm0.6$ & $-2.2\pm0.3$   \\

% SAVED IN CASE FOR REFEREE   14-43 kpc
%& II & $0.6\pm0.4$ & $6.0\pm1.3$ & $-2.9\pm0.7$   \\
%& III &$0.5\pm0.3$ & $5.3\pm1.2$ & $-2.6\pm0.7$   \\
%Tab5

\subsection{Redshift Proximity}
\label{changing_redshift_proximity}
Following a similar methodology as in \S\,\ref{sec_changing_rproj}, we quantify the impact that changing {\it only} the criterion for redshift proximity will have on $f_{\rm mc}(z)$. As described in \S\,\ref{plausible_physical_pairs}, we apply selections based on photometric redshifts and their uncertainties only to CANDELS data, and we anchor our evolutionary findings to stringent velocity-separation-based {companion fractions} at $z\sim0$ from the SDSS. As such, we first compare CANDELS fractions ($0.5<z<3$) based on our fiducial B09 selection given in Equation\,\ref{method_2_1} to fractions based on two other related methods.
Then we demonstrate that the low-redshift fraction from the
SDSS changes very little ($\Delta f/ f_{\rm fid}<5\%$) if we select major companions using simulated photometric redshift errors that are similar in quality to the CANDELS 
$\sigma_z$ values, rather than use our fiducial selection criterion
($\Delta v_{12}\leq 500$\,km\,s$^{-1}$).

\begin{figure}
	\centering
	\includegraphics[width=\columnwidth]{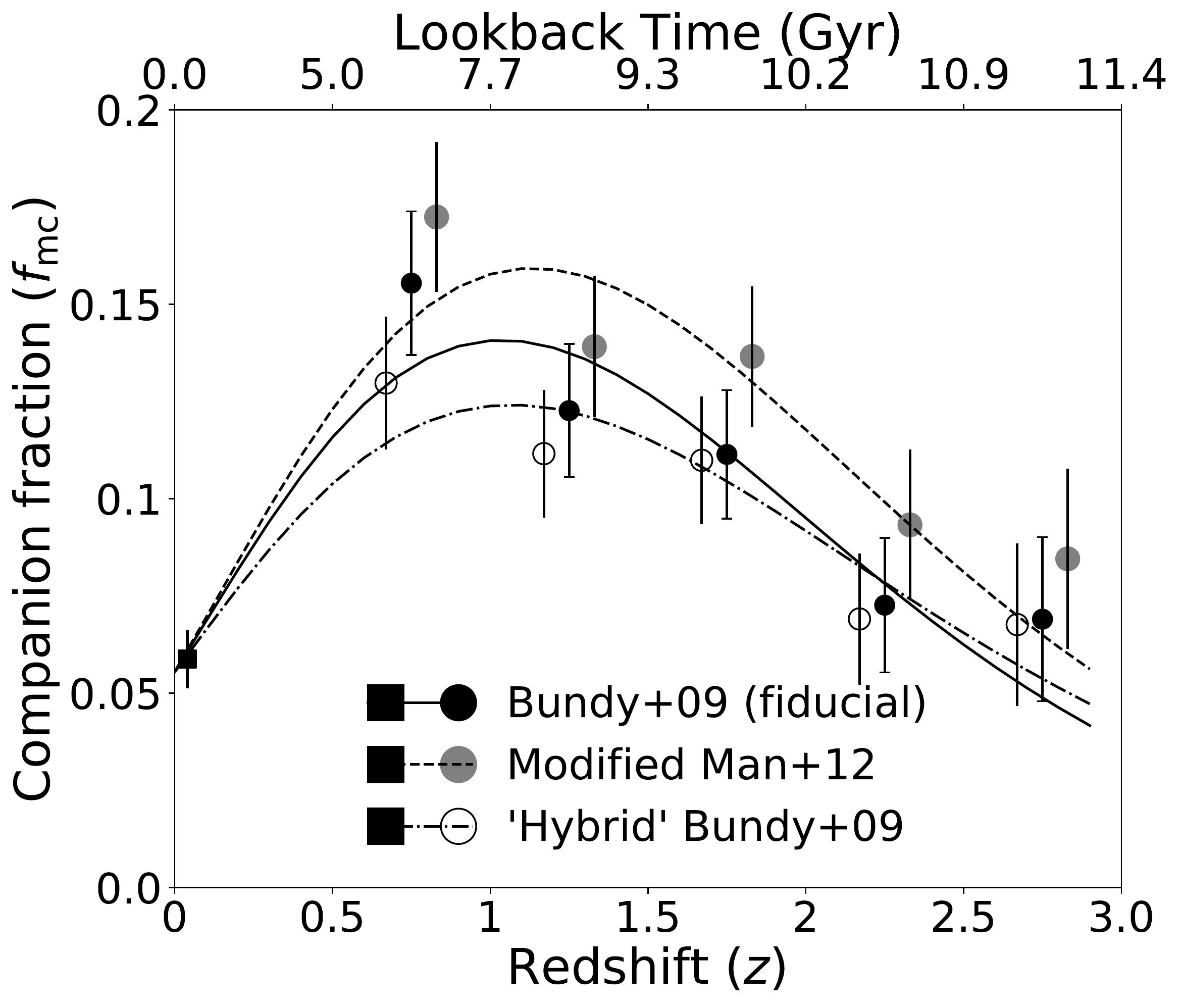} 
	\caption{Comparison of the redshift evolution of the major companion fractions $f_{\rm mc}(z)$ based on three redshift proximity criteria applied to $z>0.5$ galaxies from CANDELS: our fiducial (Equation\,\ref{method_2_1}; black circles, solid line) adopted from B09; Equation\,\ref{equation_methodII} (grey circles, dashed line) modified from \citet{man2012}; and a hybrid of the B09 criterion plus close velocity separation for the subset of CANDELS galaxies with spectroscopic redshifts (open circles, dot-dashed line). All three best-fit curves (Equation\,\ref{equation_best_fit}) to the $z>0.5$ data are anchored to the same $f_{\rm mc}(0)$ point (square symbol) based on SDSS companions with $\Delta v_{12}\leq 500~{\rm km~{s^{-1}}}$ velocity separation. 
		The error bars are defined as in Figure\,\ref{changing_rproj}.}
	\label{fig_changing_redshift_proximity_50kpc}
\end{figure}

In Figure\,\ref{fig_changing_redshift_proximity_50kpc}, 
we plot $f_{\rm mc}(z)$ data and best-fit curves for our 
fiducial selection and two additional methods. 
In each non-fiducial case,
we hold the projected separation and mass ratio 
selection criteria fixed to our fiducial choices.
The first method is a simple overlapping of the
host and companion galaxy $1\sigma$ photometric redshift errors 
\begin{equation}\label{equation_methodII}
|\Delta z_{12}|  \leq \sigma_{z,1} + \sigma_{z,2}\, .
\end{equation}
This criterion is modified from \citet{man2012} who used redshift overlap
at the $3\sigma$ level.
As with fiducial selection, we apply this criterion only to those galaxies
with reliable photometric redshifts (see \S\,\ref{plausible_physical_pairs}).
Second, we incorporate available spectroscopic redshifts for the host and/or 
companion galaxies from CANDELS into a 'hybrid'
of our fiducial B09 redshift proximity selection as follows:
\vspace{-0.2cm}
\begin{enumerate}
	\item[\bf a)]{if both galaxies have $z_{\rm spec}$ data, then the pair must satisfy $\Delta v_{12}\leq 500~{\rm km~s}^{-1}$;}
	\vspace{-0.1cm}
	\item[\bf b)]{or if only one galaxy has $z_{\rm spec}$ data, then $z_{\rm spec}$ must be within $z_{\rm phot}\pm\sigma_{\rm z}$ limit of the other galaxy;}
	\vspace{-0.1cm}
	\item[\bf c)] {otherwise use B09 criterion.}
\end{enumerate}
\vspace{-0.2cm}

As with changing the project separation for selecting close companions,
the key takeaway from Figure\,\ref{fig_changing_redshift_proximity_50kpc} is
the redshift evolution trends for $f_{\rm mc}(z)$ are robust to
slight modifications in the redshift proximity criterion. Both non-fiducial
companion fractions peak in the lowest-redshift CANDELS bin, {the} decline at $z>1$ and their redshift evolution anchored to the SDSS $z\sim0$
fraction are well fit by the same functional form with
statistically equivalent best-fit parameters (see Table\,\ref{best_fits}).
Besides the similar evolutionary trends, we notice that the modified Man et al. criterion results in $f_{\rm mc}$ values that are systematically $\sim 11-24\%$ higher than the fiducial fractions between $0.5<z<3$.
This is expected since simple photometric 
redshift error overlap (Equation\,\ref{method_2_1}) is less stringent than a quadrature overlap; e.g., two galaxies with $\sigma_z=0.06$ must have photometric redshifts 
within $\Delta z_{12} = 0.12$ compared to 0.085 (fiducial).
On the other hand, the hybrid B09 method yields $9-17$\% smaller major companion fractions than fiducial at $0.5<z<1.5$ owing to the availability of $z_{\rm spec}$ data for a fair fraction of CANDELS galaxies and the stringent $\Delta v_{12}$ criterion for this subset, but nearly identical fractions at $z>1.5$ for which the spectroscopic-redshift coverage becomes quite sparse.

We close our redshift proximity analysis by quantifying changes in the
$f_{\rm mc}(0)$ measurement from the SDSS data if we
employ either the B09 or modified Man et al. photometric redshift error overlap 
criteria. We opt to generate simulated $z_{\rm phot}$ and $\sigma_{\rm z}$ for
the SDSS sample such that these values mimic the quality of the CANDELS photometric data.
First, we construct the combined $\sigma_{\rm z}/(1+z_{\rm phot})$ distribution from all five CANDELS fields (see Fig.\,\ref{fig_appendix_1}), apply a $3\sigma$ outlier clipping,
and fit a normalized probability density function (PDF) to this distribution such that the area under the curve sums up to unity. Note that this PDF is the probability for each $\sigma_{\rm z}/(1+z_{\rm phot})$ value and is different from $P(z)$ discussed in \cref{sample selection}. 
For each galaxy, we set its SDSS redshift to $z_{\rm phot}$
and we assign it a $\sigma_{\rm z}$ value drawn randomly from
the PDF distribution. Then, we repeat our close companion selections
and find $f_{\rm mc}(0)=0.057$ (B09) and 0.061 (modified Man et al.).
The relative change in each fraction is roughly 3\% compared to our fiducial
selection of close velocity separation.

One can argue that actual $z_{\rm phot}$ data from the SDSS $ugriz$ 
photometry will produce larger redshift errors and higher companion fractions. 
Yet, the motivation of this exercise is to quantify the difference between
tight spectroscopic velocity separation versus photometric 
redshift proximity selection matched to the $z>0.5$ data, not to make our 
$z\sim 0$ companion fraction more uncertain.
We note that if we repeat this analysis by resampling each SDSS galaxy
redshift from a PDF defined by its simulated $\sigma_{\rm z}$, we find
{\it smaller} major companion fractions since this effectively makes
the redshift proximity worse (larger) and we demonstrated that companions have high probability of small velocity separations for the SDSS. Either way, all the SDSS $f_{\rm mc}$ anchors are significantly lower than the companion fractions at $z\sim1$, which means the turnover trend
in $f_{\rm mc}(z)$ is a robust result.

%Fig7

%Sec4.3
\subsection{Stellar Mass vs Flux Ratios}
\label{mass_vs_flux}
Our fiducial choice for selecting major companions is the stellar mass ratio criterion: $1\leq M_{1}/M_{2}\leq 4$. 
Some previous studies, in which galaxies lack stellar mass estimates, have used observed-band flux ratios of $1\leq F_{1}/F_{2}\leq 4$ as a proxy for selecting major companions \citep[\eg][]{bell06b,bluck_surprisingly_2009}. \cite{Bundy04} and B09 speculated that flux-ratio selection could preferentially lead to inflated {companion} counts especially at $z\ga2$ owing to increasing star formation activity. Two recent studies found conflicting trends in $f_{\rm mc}(z)$ estimates between $z\ga1$ and $z=3$ \citep{man2012,man16}. Here we quantify the impact that changing the major companion selection from 4:1 stellar mass ratio (MR) to 4:1 flux ratio (FR) has on companion fractions for massive galaxies in CANDELS and the SDSS while holding all other selection criteria fixed to fiducial values.

\begin{figure}
	\centering
	\includegraphics[width=\columnwidth]{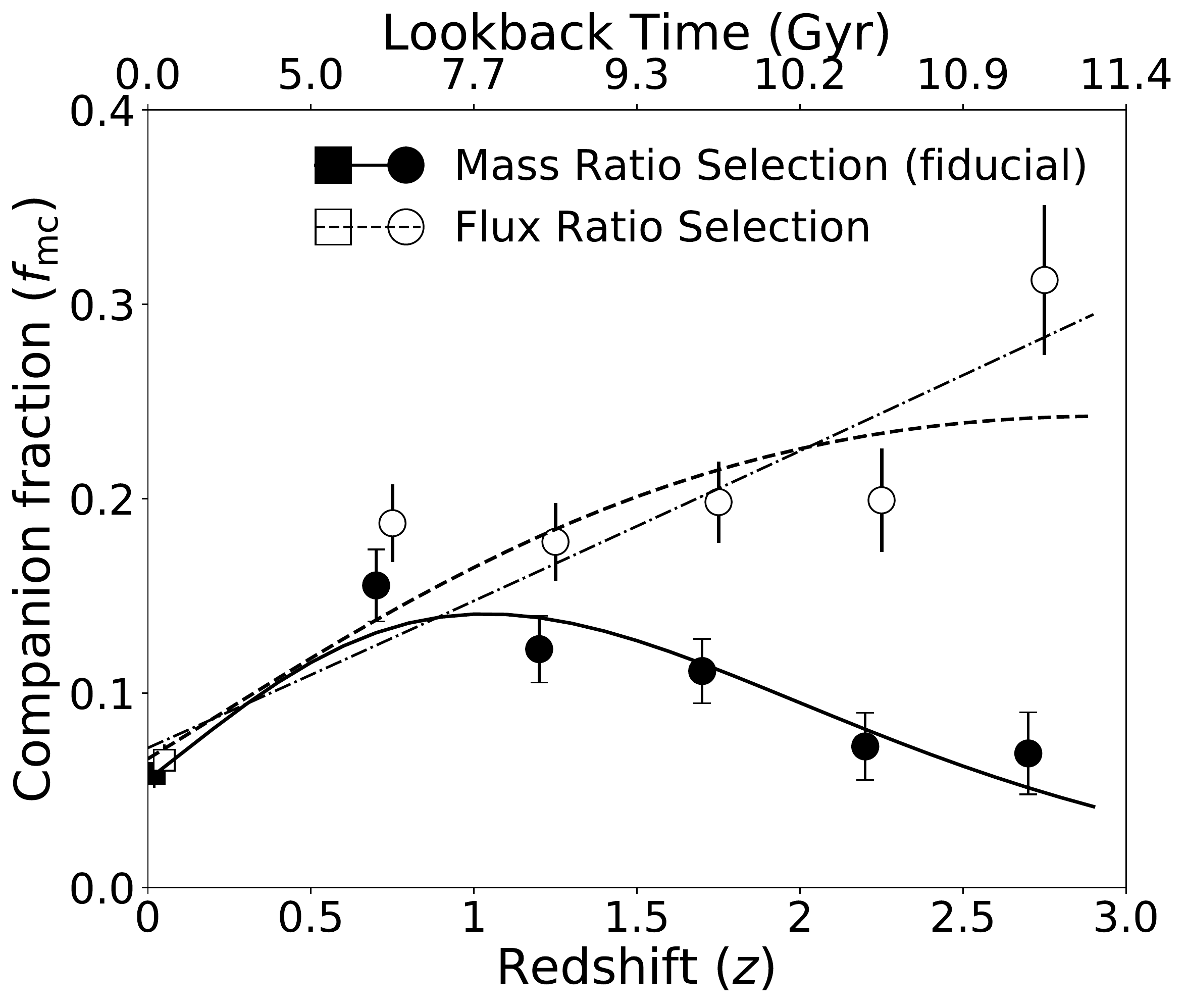} 
	\caption{Comparison of the redshift evolution of major companion fractions $f_{\rm mc}(z)$ based on our fiducial selection (stellar mass ratio $1\leq M_{1}/M_{2}\leq 4$; black symbols, solid line) and based on $H$-band flux ratios ($1\leq F_{1}/F_{2}\leq 4$; open points; slightly offset in $z$ direction for clarity). Best-fit curves to the flux-ratio fraction evolution are shown for modified power law (Eq.\,\ref{equation_best_fit}, dashed line) and a simple power law (dot-dashed line; $f_{\rm mc}(z) = 0.07(1+z)^{1}$). The data symbols distinguishing CANDELS and SDSS fractions, and the error bars are the same as in Fig.\,\ref{changing_rproj}. The flux ratio selection results in larger fractions at $z>1$ and an increasing power-law redshift dependence in contrast to the $f_{\rm mc}(z)$ derived from stellar-mass ratios.} \label{fmp_flux_ratio_LOS_B09_50kpc}
\end{figure}

In Figure\,\ref{fmp_flux_ratio_LOS_B09_50kpc}, we compare the fiducial MR-based $f_{\rm mc}(z)$ to fractions based on $H$-band (CANDELS) and $r$-band (SDSS) 4:1 flux ratios. 
We find the FR-based major companion fractions follow starkly different
evolutionary trends compared to MR-based fractions at $z\ga1$. 
At $0<z<1$, the FR and MR fractions both increase with redshift and agree
within their 95\% confidence limits. {However}, FR produces increasingly
larger fractions from $z\sim1$ to $z=3$ that diverge away from the fiducial
$f_{\rm mc}(z)$ trends and grow $1.5-4.5$ times {greater} than MR-based fractions
at these redshifts. 
We attempt to fit the FR-based $f_{\rm mc}(z)$ with
the same function form (Eq.\,\ref{equation_best_fit}) that we employ for MR
fraction redshift evolution, but the best-fit parameter values 
(see Table\,\ref{best_fits}) for the steadily increasing fractions between
$0<z<3$ are statistically consistent with a simple power law since the FR
fractions do not peak nor turn over at $z>1$. Therefore, we
refit the FR fraction evolution with a power law and find 
$f_{\rm mc}(z)= 0.07(1+z)^{1}$ for $0<z<3$.

\begin{figure*}
	\centering
	\includegraphics[height=2\columnwidth,width=2\columnwidth]{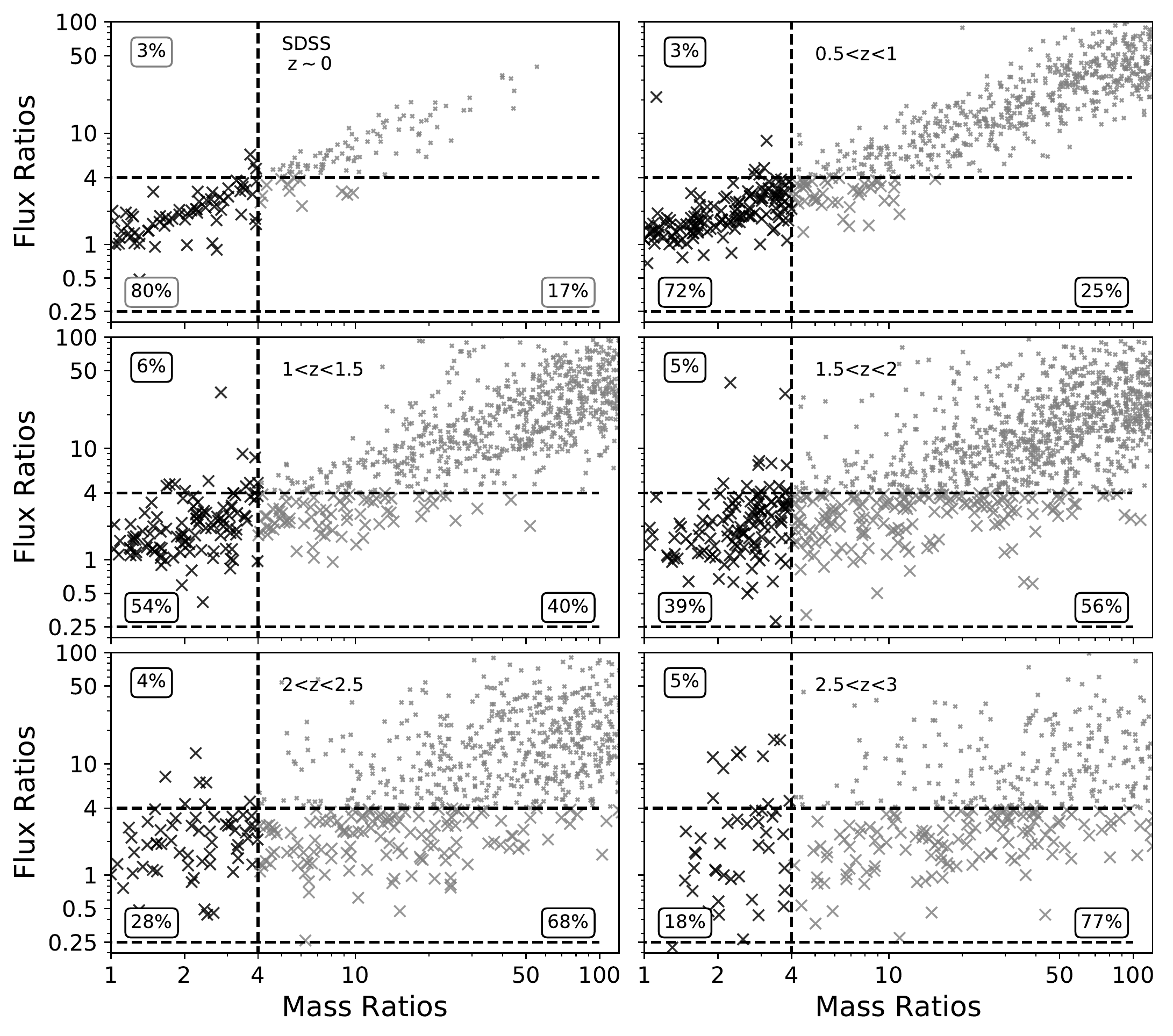}
	\caption{Stellar-mass-ratio vs. $H$-band flux-ratio for the close-pair systems satisfying our fiducial projected separation ($5-50$\,kpc) and redshift proximity choices for SDSS ($\Delta v_{12}\leq 500$\,km\,s$^{-1}$; $z\sim 0$; top left) and CANDELS (B09; $0.5\leq z \leq 3$; rest of the panels). We show the pairs that satisfy our fiducial mass-ratio criterion ($1\leq M_{1}/M_{2}{\leq}4$) in large-black points , those having a $M_{1}/M_{2}>4$ but satisfy the flux-ratio criterion ($1\leq |F_{1}/F_{2}|\leq 4$) in large-grey points, the remaining points in small-grey markers. To visualize the contamination from minor pairs ($M_{1}/M_{2}>4$), introduced by the flux-ratio criterion in each redshift panel, we show the percentage of pairs that satisfy either mass-ratio or flux-ratio selection that fall in each of the three quadrants. We notice that the contamination increases steadily from $z\sim 0$ to $z = 3$, causing inflated $f_{\rm mc}$ values observed in Figure\,\ref{fmp_flux_ratio_LOS_B09_50kpc} (see \cref{mass_vs_flux} for discussion). }	
	\label{flux_ratio_histograms}
\end{figure*}

To better understand the striking differences between the FR and MR
$f_{\rm mc}(z)$ trends, we analyze stellar mass ratios and flux ratios
of CANDELS and SDSS  {close-pair systems that satisfy our fiducial $5-50$\,kpc separation and redshift proximity selections} (see \S\,\ref{plausible_physical_pairs}) 
in Figure\,\ref{flux_ratio_histograms}.
At $z<1$, we find good agreement between MR and FR values for a majority
of major and {\it minor} ($M_{1}/M_{2}>4$) close-pair systems.
The agreement suggests that $r$-band 
($H$-band) flux ratios
are a good approximation for stellar-mass ratios at $z\sim0$ ($0.5<z<1$).
In detail, 80\% (72\%) of SDSS (CANDELS) pairs are considered major pairs according to {\it both} FR and MR criteria. We find 17\% (25\%)
of FR-based major pairs have a $>4:1$ (minor) stellar mass ratio, and only
3\% of MR-selected major pairs have minor flux ratios at $z<1$.

In the bottom four panels of Figure\,\ref{flux_ratio_histograms}, we notice
that the FR-based major companion selection suffers a
steadily increasing contamination by minor companions according to CANDELS
stellar mass ratios. This contamination rises from 40\% at $1<z<1.5$ to over
three-quarters in the highest redshift interval. In contrast, our fiducial
selection of major companions maintains a very small ($\sim5\%$) constant contamination of $F_1/F_2>4$ pairs between $1<z<3$. This analysis clearly demonstrates that FR selection results in inflated major companion counts at $z\ga 1$, confirming the result that significantly different mass-to-light ratio properties of the companion galaxies at $z>1$ may be the main cause {of} the contamination by minor {companions} \citep{man16}.

%Fig8

%The redshift evolution of the fraction of massive galaxies in major, plausibly, physical pairs $f_{\rm mc}(z)$ satisfying our fiducial selection in filled legend, in comparison to the $f_{\rm mc}$ using a different major pair selection criteria ($1\leq F_{1}/F_{2}\leq 4$) in open legends, and holding all other criteria constant. The CANDELS points at $0.5<z<3$ are shown in circles, and the SDSS anchor is shown in square. Their corresponding best-fits are represented in solid and dashed lines, respectively. The error bars correspond to 95\% binomial confidence limits of $f_{\rm mc}$. The flux ratio selection results in larger $f_{\rm mc}$ and an increasing power-law dependence in contrast to the fiducial $f_{\rm mc}(z)$ using stellar-mass ratio criteria (see text for details).

%Fig9

\begin{table*}
	\centering
	\caption{Compilation of {companion selection criteria employed by previous empirical close-pair-based studies}. Columns: (1) names of the studies -- presented in the order of their mention in Figure\,\ref{fig_comp_prev_studies}; (2) the choice of projected separation $R_{\rm proj}$ annulus; (3) the criterion to choose major {companions}, where $\mu_{12} = M_{1}/M_{2}$ is the stellar mass ratio, flux-ratio is given by $F_{1}/F_{2}$, and the luminosity ratio is given by $L_{1}/L_{2}$; (4) the choice of redshift proximity criterion to select companions in plausible close pairs, where $\Delta z_{12}$ is the photometric redshift difference $z_{1}-z_{2}$, $\Delta v_{12}$ is the spectroscopic velocity difference, LOS is the statistical correction for line-of-sight projections, and $CDF(z_{1},z_{2})$ is the cumulative probability of the galaxies involved in a close pair; (5) the applied selection criterion to select the primary galaxy sample, which is either a stellar mass-limited or flux-limited selection; (6) the redshift range in which the study presented their findings. We divide the table into two parts, where studies based on mass ratios are above the solid-dashed line and vice-versa. \protect\cite{newman_can_2012} employed a redshift proximity where $\Delta z_{12}/1+z_{1}<0.1$ for $z<1$ and $\Delta z_{12}/1+z_{1}>0.2$ when $z_{1}>1$. The dagger represents that the studies derived a `pair' fraction rather than the companion fraction ($f_{\rm mc}$), see \cref{sec_comp_prev_results} for details. }
	\label{selection_crit_prev_stuides}
	\begin{tabular}{cccccc}
		\hline
		Study & $R_{\rm proj}$ selection & Major Companion& Redshift Proximity & Primary galaxy selection & Redshift  \\ 
		(1) & (2) & (3) & (4) & (5) & (6)\\
		\hline
		Ryan$+$08 & $5-30$\,kpc & $1\leq \mu_{12} \leq 4$ & $2\sigma$ $z_{\rm phot}$ error overlap & $\logten(M_{\rm stellar}/M_{\odot})\geq 10$ & $0.5<z<2$ \\
		Bundy$+$09$^\dagger$ & $5-30$\,kpc & $1\leq  \mu_{12} \leq 4$ & $\Delta z_{12}^2 \leq \sigma_{\rm z,1}^2 + \sigma_{\rm z,2}^2$ & $\logten(M_{\rm stellar}/M_{\odot})\geq 10$ & $0.4 \leq z \leq 1.4$ \\ 
		Mundy$+$17$^\dagger$ & $5-30$\,kpc & $1\leq \mu_{12} \leq 4$ & $CDF(z_{1},z_{2})$ & $\logten(M_{\rm stellar}/M_{\odot})\geq 10$ & $0<z<2$ \\ 	
		Williams$+$11 & $14-43$\,kpc & $1\leq \mu_{12} \leq 4$ & $|\Delta z_{12}|/1+z_{1}<0.2$ & $\logten(M_{\rm stellar}/M_{\odot})\geq 10.5$ & $0.4\leq z \leq2$ \\
		Newman$+$12 & $14-43$\,kpc & $1\leq \mu_{12}\leq 4$ & see caption & $\logten(M_{\rm stellar}/M_{\odot})\geq 10.7$& $0.4\leq z \leq2$  \\
		Man$+$16 & $14-43$\,kpc & $1\leq \mu_{12}\leq 4$ & same as Newman$+$12 & $\logten (M_{\rm stellar}/M_{\rm \odot}) \geq 10.8$ & $0\leq z \leq3$ \\
		\hline
		Kartaltepe$+$07 & $5-30$\,kpc & $1\leq L_{1}/L_{2} \leq 4$ & $\Delta v_{12}\leq 500$\,km\,s$^{-1}$ & $M_{\rm V}<-19.8-1.0z$ & $0.1 \leq z \leq 1.2$\\
		Bundy$+$09$^\dagger$ & $5-30$\,kpc &  $1\leq F_{1}/F_{2}\leq 4$ & LOS & $\logten (M_{\rm stellar}/M_{\odot}) \geq 10$ & $0.4 \leq z \leq 1.4$ \\
		Man$+$12 & $5-30$\,kpc & $1\leq F_{1}/F_{2}\leq 4$ & $3\sigma$ $z_{\rm phot}$ error overlap & $\logten (M_{\rm stellar}/M_{\rm \odot}) \geq 11$ & $0\leq z \leq3$ \\
		Bluck$+$09,$+$12 & $5-30$\,kpc & $1\leq F_{1}/F_{2}\leq 4$ & line-of-sight correction & $\logten (M_{\rm stellar}/M_{\rm \odot}) \geq 11$ & $1.7\leq z \leq3$ \\
		Lin$+$08 & $14-43$\,kpc & $1\leq L_{1}/L_{2} \leq 4$ & $\Delta v_{12}\leq 500$\,km\,s$^{-1}$ & $-21<M_{\rm B}+1.3z<-19$ & $z\leq 1.2$ \\
		L\'opez-Sanjuan$+$11$^\dagger$ & $14-43$\,kpc & $1\leq L_{1}/L_{2} \leq 4$ & $\Delta v_{12}\leq 500$\,km\,s$^{-1}$ & $M_{\rm V}<-20-1.1z$ & $z=0.5$, $z=0.8$ \\
		Man$+$16 & $14-43$\,kpc & $1\leq F_{1}/F_{2}\leq 4$ & same as Newman+12 & $\logten (M_{\rm stellar}/M_{\rm \odot}) \geq 10.8 $ & $0 \leq z \leq 3$ \\
		\hline
	\end{tabular}
\end{table*}
		
\section{Discussion}
\label{Discussion}
Our extensive analysis of the full CANDELS sample and a well-defined
selection of SDSS galaxies in the preceding sections provides a new baseline for the evolving frequency of massive galaxies
with close major companions (and of major mergers by extrapolation) spanning epochs from the start of cosmic high noon 
to the present-day. In this section, we present a comprehensive comparison
of our measurements to those
from previous studies by taking advantage of our analysis of varying the {companion} selection criteria from \cref{impact_of_pair_selection}. We derive the empirical major merger rates based on a {\it constant} observability timescale and find {significant} variation in the rates depending on our stellar-mass ratio or $H$-band flux-ratio choice. We discuss plausible reasons for these variations by re-computing merger rates based on theoretically-motivated, redshift-dependent timescale prescriptions. Finally, we describe the need for detailed calibrations of the complex and presumably redshift-dependent conversion factors required to translate from fractions to rates, which is key to improve major merger history constraints.

\subsection{Comparison to previous empirical close-pair-based studies} \label{sec_comp_prev_results}
We compare our CANDELS$+$SDSS major companion fractions from  to the results from previous stellar-mass and flux selected, empirical close-pair-based studies tabulated in Table~\ref{selection_crit_prev_stuides}. {We note that some studies quote fraction of pairs ($f_{\rm pair}$), which is the fraction of close-pair systems among a desired massive galaxy sample of interest. This is different from the companion fraction $f_{\rm mc}$.} To correctly compare to previous studies which used a $f_{\rm pair}$ definition, we derive a simple conversion between the two definitions by separately computing the ratio of $f_{\rm mc}/f_{\rm pair} = 1.5$, which is constant for the CANDELS and SDSS samples at $0<z<1.5$. Overall, our $f_{\rm mc}(z)$ results are in good agreement with previous studies when we properly convert the published pair fractions into companion fractions.
We separate our comparisons into stellar mass-ratio and flux-ratio based studies in Figure\,\ref{fig_comp_prev_studies}. For each set of comparisons, we further split them into matching $5-30$\,kpc and $14-43$\,kpc projected separation selections. We discuss each comparison with respect to any differences in the details of the redshift proximity selection.

\subsubsection{Stellar Mass-Ratio Selected Studies}
We use the $5-30\,$kpc and $14-43\,$kpc fractions from \cref{sec_changing_rproj} (Figure\,\ref{changing_rproj}) and plot them in Figure\,\ref{fig_comp_prev_studies} {alongside} previous studies that employed a similar projected separation criterion and 4:1 stellar mass-ratio selection ($1\leq M_{1}/M_{2}\leq 4$).  We start our discussion by comparing to the studies that employed a $R_{\rm proj} = 5-30$\,kpc criterion. \cite{Ryan08} studied major close pairs among galaxies more massive than $10^{10}M_{\odot}$ over the redshift range $0.5<z<2.5$, and found an un-evolving fraction ($\sim 10\%$). We agree with these estimates within their quoted uncertainties; however, we acknowledge that the consistency at $2<z<2.5$ is because of their large error bar ($>50\%$). We also note that \citeauthor{Ryan08} used a $2\sigma_{z}$ overlap as their redshift proximity, which is less restrictive than our modified \cite{man2012} redshift proximity criteria in Equation\,\ref{equation_methodII}. Based on our analysis in \cref{changing_redshift_proximity}, this implies that the \citeauthor{Ryan08} fractions may be over-estimated. Next, we compare to B09 who studied $5-30\,$kpc pairs among $\logten(M_{\rm stellar}/M_{\odot})\geq10$ galaxies between $0.4<z<1.4$ using a redshift proximity that we have adopted as our fiducial choice. It is important to note that B09 elected a 4:1 $K_{s}$-band flux ratio as their fiducial major companion selection criterion. However, in their analysis, they found that 80\% of their flux-ration-selected companions also met $<4:1$ stellar-mass ratio major {companion} selection. To appropriately compare to B09 pair fractions, we include a 0.8 multiplicative factor to covert them to mass-ratio-based $f_{\rm mc}$. We find that their estimates are on average smaller than our fractions, marginally agreeing at $0.7<z<1.4$, and disagreeing at $0.4<z<0.7$. 

{We now compare to the most recent study by \cite{Mundy17}, who analyzed close pairs ($R_{\rm proj} = 5-30\,{\rm kpc}$) among a sample of $\logten(M_{\rm stellar}/M_{\odot})\geq10$ galaxies at $0<z<2$ from the UDS, VIDEO, GAMA, and COSMOS datasets.} In Figure\,\ref{fig_comp_prev_studies}, we show all their pair fractions, except for an upper-limit of $17.5\%$ from COSMOS at $1.5<z<2$ and convert them to $f_{\rm mc}$. We find excellent agreement between  GAMA-survey-based, spectroscopically derived fraction at $0<z<0.2$ and our SDSS-based fraction at $z\sim 0$. For the same GAMA sample, they also computed the fractions using photometric redshifts and found an average of $\sim 5\%$ at $0<z<0.2$. We separately computed the SDSS $f_{\rm mc}$ using B09 redshift proximity and matching their $R_{\rm proj}=5-30$\,kpc selection, and find that our fraction $f_{\rm mc}(0)\sim 2.8\%$ is $\sim 1.8$ times smaller and marginally disagrees within their quoted uncertainties. At $0.2<z<1$, we find that \citeauthor{Mundy17} fractions are in good agreement with our $f_{\rm mc}(z)$, except for their COSMOS-survey-based value at $0.2<z<0.5$, which is marginally smaller than our $f_{\rm mc}$. At $1<z<1.5$, their fraction ($\sim 15\%$) is in disagreement with our findings, while their upper-limits are consistent with our results.  Overall, we demonstrate good agreement with \cite{Mundy17}.
	
Now, we discuss our comparison to the studies that employed a $R_{\rm proj} = 14-43$\,kpc separation selection. We start with \cite{williams_diminishing_2011}, who studied major ($<$4:1 stellar-mass ratio) close pairs among $\logten (M_{\rm stellar}/M_{\odot})\ga 10.5$ galaxies at $0.4<z<2$, and found a diminishing redshift evolution of the fractions. We find disagreement with the \citeauthor{williams_diminishing_2011} fractions at $0.4<z<1.2$, where our $f_{\rm mc}$ are higher by 25\%, however, we show good agreement within their diminished fractions at $1.2<z<2$. We note that our sample is almost two times larger than their galaxy sample, and a likely reason for this discrepancy is that their uncertainties may have been under-estimated. Next, we compare to \cite{newman_can_2012}, who quantified major companion fraction among $\logten (M_{\rm stellar}/M_{\odot})\ga 10.7$ galaxies, and found that $f_{\rm mc}$ is flat (at $\sim 10\%$) at $0.4<z<2$. We find good agreement with their findings within the quoted uncertainties and in that redshift range. 

{Finally, we compare to a recent study by \cite{man16} who study a sample of $\logten (M_{\rm stellar}/M_{\odot})\ga 10.8$ galaxies to constrain major companion fractions at $0<z<3$ using two separate datasets, namely 3D-HST and UltraVISTA. Firstly, we find that our $f_{\rm mc}(z)$ values are in excellent agreement with the \citeauthor{man16} 3D-HST estimates over their redshift range, which is not surprising as 3D-HST covers $75\%$ of the CANDELS fields. We find that the \citeauthor{man16} UltraVISTA fractions follow an increasing trend that is qualitatively similar to our $f_{\rm mc}(z)$ at $0<z<1$, however, we notice that the former fractions are quantitatively smaller than the latter. At $z>1$, we find that the UltraVISTA fractions are in good agreement with our results, except at $z\sim 2.25$, where our $f_{\rm mc}$ is two times smaller. Overall, we note that our companion fractions are in good agreement with \cite{man16} estimates. Moreover, \cite{man16} applied a parabolic fitting function to their UltraVISTA major companion fractions and report that their redshift evolution may be peaking around $z\sim 1-1.5$, in contrast to our fractions that peak around $0.5<z<1$ and diminish up to $z=3$. This may be a consequence of more-massive galaxies experiencing growth earlier in cosmic-time than the less-massive population, which is often referred to as the galaxy-downsizing phenomenon.} In summary, we conclude that our stellar-mass ratio based major companion fractions $f_{\rm mc}(z)$ are in good agreement with previous empirical close-pair-based estimates, once they are closely-matched in selection criteria. 
	
\subsubsection{Flux Ratio Selected Studies}\label{flux_ratio_selected_studies}
\begin{figure*}
	\centering
	\subfloat[]{\includegraphics[width=\columnwidth]{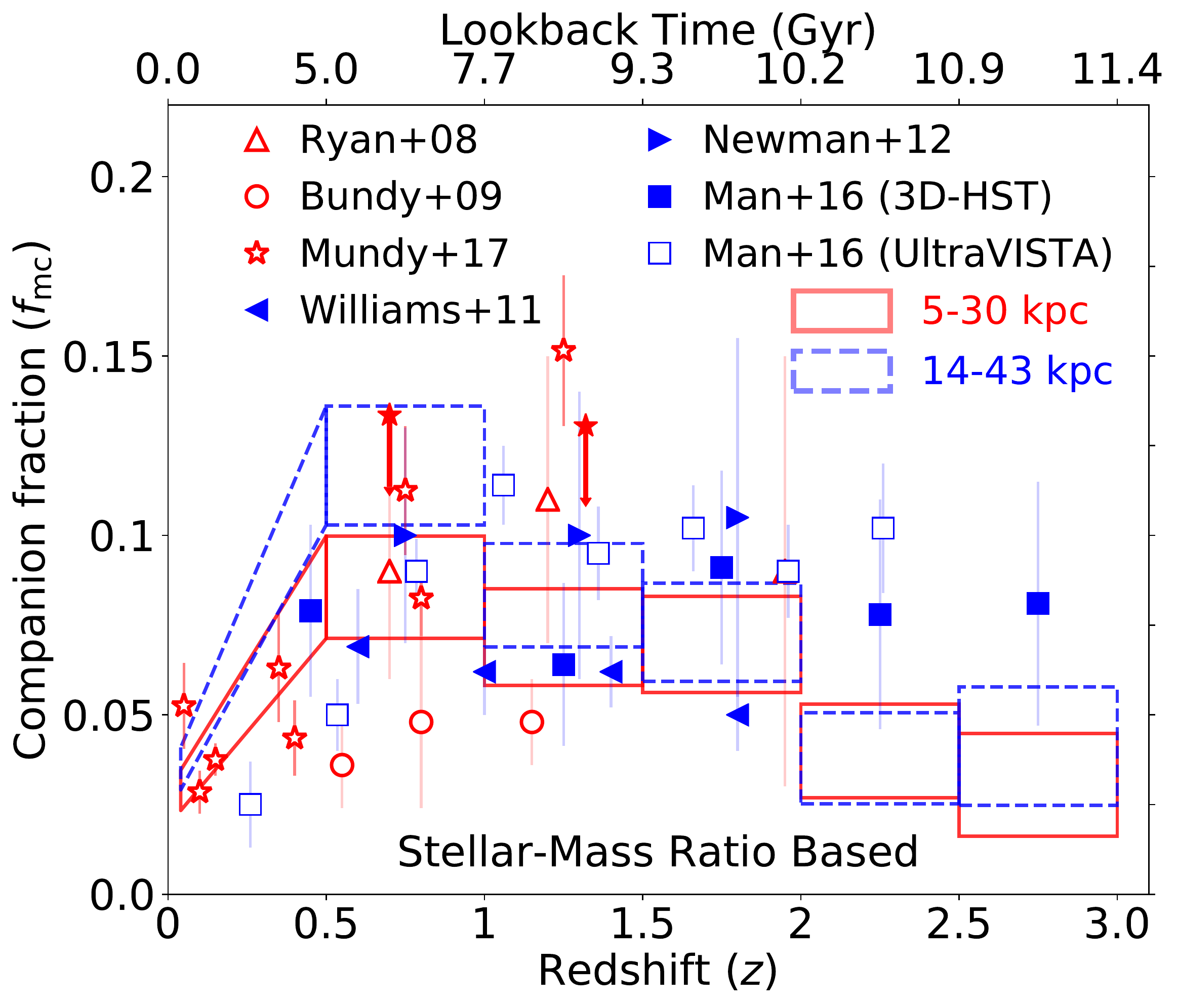}}
	\subfloat[]{\includegraphics[width=\columnwidth]{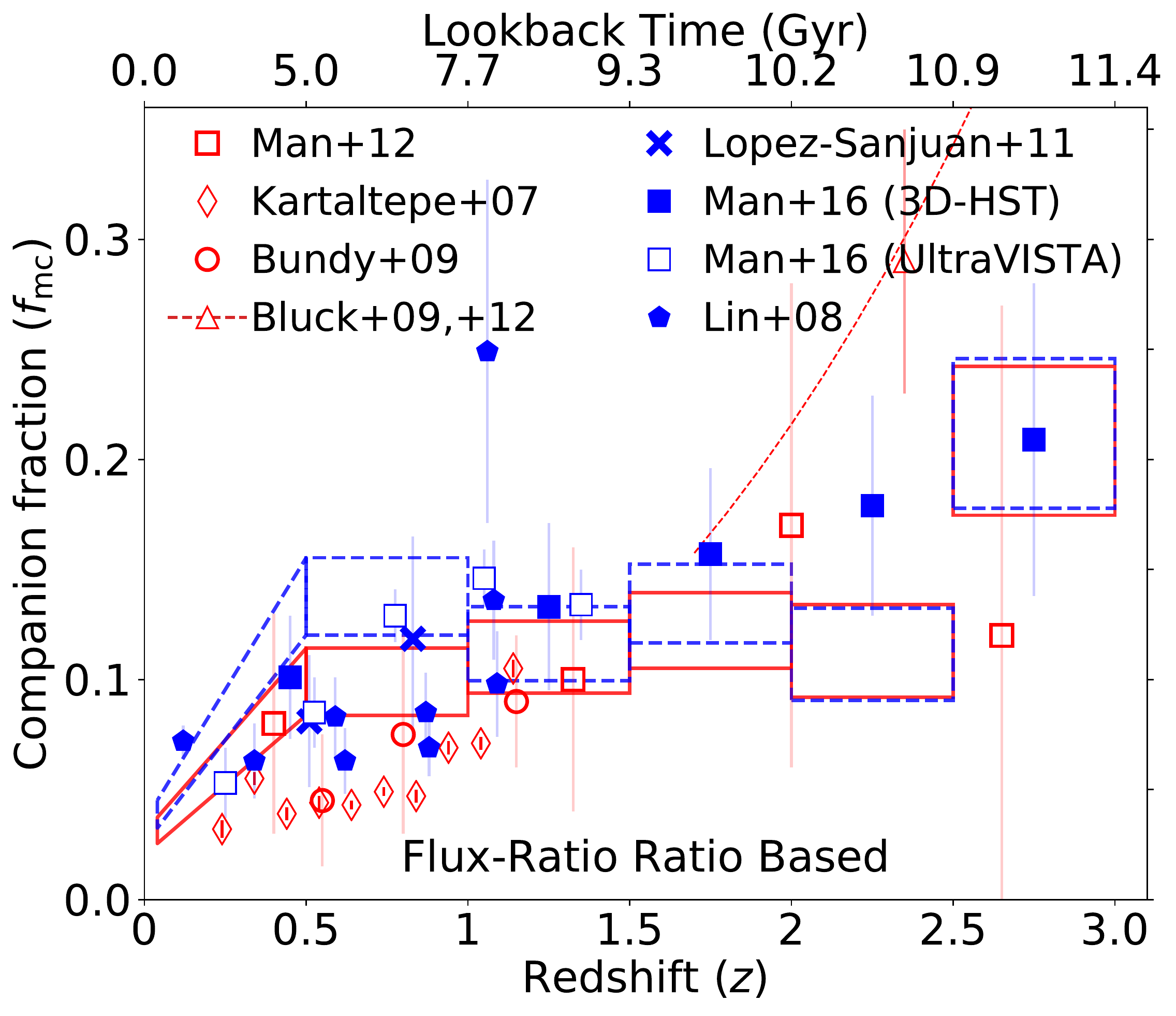}}
	\\
	\caption{Comparison of major companion fractions from CANDELS$+$SDSS to those from previous studies that employed 4:1 stellar-mass ratio (a) and 4:1 flux-ratio (b) selections. In both panels, we outline the $f_{\rm mc}$ measurements data points as rectangles where their height represents the 95\% binomial confidence limits per redshift bin (width of the rectangles). The data-points of previous empirical studies are given in the panel keys (see Table\,\ref{selection_crit_prev_stuides}). We compare fractions based on different projected separation criteria as follows: $5-30$\,kpc (solid-red line; open-red markers), and $14-43$\,kpc (blue-dashed line, filled-blue markers). We find our major companion fraction estimates are in good agreement with previous empirical constraints when the companion selection criteria are closely matched. We off-set multiple fields from \citet{Mundy17} by a small amount, and show the upper-limits in filled markers with bold arrow for clarity.}
	
	\label{fig_comp_prev_studies}
\end{figure*}

We separately derive the major companion fractions for $R_{\rm proj} = 5-30$ and $14-43$\, kpc selections using $H$-band ($r$-band) flux ratio selection for CANDLES (SDSS) and plot them in Figure\,\ref{fig_comp_prev_studies} along side some published studies that used a similar projected separation and $<4:1$ flux (luminosity) ratio selection (see Table\,\ref{selection_crit_prev_stuides}). First, we discuss the comparison to those studies that employ $R_{\rm proj} = 5-30$\,kpc. We start with \cite{kartaltepe_evolution_2007}, who studied close pairs using a $V$-band luminosity-limited galaxy sample at $0.2\leq z \leq1.2$, and found that the companion fraction evolves as $f_{\rm mc} \propto (1+z)^{3.1}$. We find that \citeauthor{kartaltepe_evolution_2007} estimates are $\sim 2$ times smaller than our results at $0.5<z<1$, however, they start to converge at $z<0.5$ and $z>1$, and agree with our $f_{\rm mc}\sim 10\%$ at $z=1.2$. We note that \citeauthor{kartaltepe_evolution_2007} uses a smaller separation annulus of $R_{\rm proj}=5-20$\,kpc, and in addition they also apply a random chance pairing correction. Based on our analysis from \cref{correction_for_random_pairing,sec_changing_rproj}, we note \citeauthor{kartaltepe_evolution_2007} fractions may be systematically smaller, and most plausibly are causing the observed discrepancy at $0.5<z<1.0$.

We now compare to B09, and remind our reader that they used a 4:1 flux-ratio as their fiducial major companion criterion.  We find good agreement with the \citeauthor{bundy_greater_2009} estimates at $0.7<z<1.4$, but find marginal disagreement at $0.4<z<0.7$. Next, we compare to \cite{man2012}, who analyzed close pairs among $\logten(M_{\rm stellar}/M_{\odot})\geq11$ galaxies using 4:1 $H$-band flux-ratio criterion at $0<z<3$, and found fairly high fractions up to $z=3$. We find good agreement with their fractions over the full redshift range, however, we acknowledge that a proper comparison with our highest-redshift bin results is not possible owing to their large uncertainty. 

{Finally, we compare to \cite{bluck_surprisingly_2009,Bluck12}, who analyzed the incidence of close companions which satisfy a $4:1$ $H$-band flux ratio selection among a GOODS NICMOS survey sample of $\logten(M_{\rm stellar}/M_{\odot}>11)$ galaxies at $1.7<z<3$. \citeauthor{bluck_surprisingly_2009} reports that the fraction of galaxies hosting a nearby major companion evolves steeply with redshift as $f\propto(1+z)^{3}$ (dashed line in Figure\,\ref{fig_comp_prev_studies}b), reaching up to $f = 0.19\pm0.07$ ($1.7<z<2.3$) and $f = 0.40\pm0.1$ ($2.3<z<3$), with a total fraction of $f = 0.29\pm0.06$. While we find consistent agreement at $1.7<z<2.3$, we note that their fractions are a factor of two higher at $2.5<z<3$. Similarly, we also find that their total fraction, which spans over $1.7<z<3$, is also larger by a factor of two when compared to our total $f_{\rm mc}$ at $2<z<3$. Although we qualitatively agree with the strongly increasing redshift evolution of \citeauthor{bluck_surprisingly_2009}, we note that our $f_{\rm mc}(z)$ trend is shallower with a simple power-law slope of $m\sim1$ (see Figure\,\ref{fmp_flux_ratio_LOS_B09_50kpc}). We argue that the steeply increasing fraction evolution by \citeauthor{bluck_surprisingly_2009} may be a consequence of the increased contamination from minor companions by their flux-ratio selection (see \S\,\ref{mass_vs_flux}). Also, significant field-to-field variance due to their small sample size and the use of a relatively less restrictive line-of-sight contamination correction may be contributing towards the observed (factor of two) disagreement to our fractions. }
	
We now focus our comparisons to studies that employ a $R_{\rm proj} = 14-43$\,kpc selection (see Table\,\ref{selection_crit_prev_stuides}). We start our comparison with \cite{lin_redshift_2008}, who studied close pairs among luminosity-limited sample of galaxies at $0.12<z<1.1$. We find that our fractions are only marginally consistent with their results at $z<0.5$ and disagree with them at $0.5<z<1$ as our fractions are $\sim 2$ times larger than their values. We note that \citeauthor{lin_redshift_2008} used a stringent spectroscopic redshift proximity criterion of $\Delta v_{12}\leq 500$\,km\,s$^{-1}$, which may be the reason for their smaller fractions and the observed discrepancy. Nevertheless, we notice that \citeauthor{lin_redshift_2008} fractions start to converge with our estimates at $z>1$, which may be due to the rising minor-companion contamination by flux-ratio selection leading to larger fractions. We now compare to \cite{sanjuan11}, who study a spectroscopically confirmed sample of close pairs among $B$-band luminosity-limited galaxy sample to quantify pair fractions at two redshifts $z=0.5$ and $z=0.8$. We find that our $f_{\rm mc}$ estimate at $z=0.5$ is marginally higher than their fraction, but is in good agreement at $z=0.8$. Also, we qualitatively agree with their increasing redshift evolutionary trend of the fractions at $0.5<z<0.8$. 

{Finally, we compare to \cite{man16}, who also derived 4:1 $H$-band flux-ratio based companion fractions for $\logten (M_{\rm stellar}/M_{\odot})\ga 10.8$ galaxies spanning $0.1<z<3$ among the 3D-HST and UltraVISTA datasets. We find that our $f_{\rm mc}(z)$ values trace very closely with the \citeauthor{man16} 3D-HST estimates and are in excellent agreement within the uncertainties. We also find that \citeauthor{man16} UltraVISTA estimates qualitatively agree with the increasing redshift evolution of our flux-ratio selected $f_{\rm mc}(z)$. However, we notice that their estimates at $z<0.5$ are marginally smaller than our $f_{\rm mc}$ extrapolation between $z\sim 0$ and $0.5<z<1$. In summary, we conclude that our $H$-band flux-ratio based $f_{\rm mc}(z)$ are in good agreement with previous empirical studies once we match closely the choices of companion selection criteria.}
% Fig10

\vspace{1cm}

\subsection{Major Merger Rates}
\label{Merger Rates}
As outlined in the Introduction, empirical merger rates provide a fundamentally crucial measure of the importance of major merging in the evolution of massive galaxies. Our new major companion fractions from CANDELS$+$SDSS allow us to place clearly-defined new constraints on the evolution of major merger rates
between $z=3$ and $z=0$. Here, we elect to focus on a {straightforward} comparison of merger rates derived from our major companion fractions with the largest systematic differences: stellar mass-ratio versus flux-ratio selection. Then we compare our merger rates to several recent empirical studies and cosmologically-motivated predictions that employ similar companion selection criteria as
we do.

The major merger rate based on close-companion (pair) statistics is defined as the number of major galaxy-galaxy merger events per unit time per selected galaxy \citep[see][]{man16}: 
\begin{equation}
\label{merger_rate_equation}
R_{\rm merg,pair} = \frac{C_{\rm merg,pair}\times f_{\rm mc}}{T_{\rm obs,pair}}~[{\rm Gyr^{-1}}]~,
\end{equation}
where $T_{\rm obs,pair}$ is the average observability timescale during which a galaxy-galaxy pair satisfies a given selection {criterion}, and the multiplicative factor $C_{\rm merg,pair}$ attempts to account for the fact that not all such companions will merge within the $T_{\rm obs,pair}$ interval  \citep[\eg][]{lotz_major_2011}. 
%Thus, $C_{\rm merg,pair}$ can span values from $\sim0$ to 1.
It is clear that the  $C_{\rm merg,pair}$ and $T_{\rm obs,pair}$ factors are 
{\it crucial} assumptions involved in converting observed companion fractions into merger rates. As such, these factors are a major source of systematic uncertainty in merger rate calculations. A few analyses of theoretical simulations have provided fairly broad guidelines for these key assumptions.  \cite{hopkins_mergers_2010} find that different merger timescale assumptions can induce up to a factor of 2 uncertainty in the merger rates. 
Others have attempted to constrain the fraction-to-rate conversion factors
and quote limits of $C_{\rm merg,pair} = 0.4$ to $1$ \citep{Kitzbichler_White_08,jiang14}.
Despite the need for improved constraints on these key merger rate
factors, attempts to calibrate $C_{\rm merg,pair}$ and $T_{\rm obs,pair}$
in detail with regard to changing companion selection criteria and as a function of fundamental galaxy properties such as redshift and stellar mass, are lacking.

In practice, previous empirical studies have used merger rate factors that
span a wide range.
For example, some studies have adopted $C_{\rm merg,pair} = 1$ for simplicity \citep[\eg][]{man16}, while others have employed an intermediate value of 0.6 \citep[\eg\, B09,][]{lotz_major_2011} based on limits from simulations. Similarly, some empirical studies adopt their $T_{\rm obs,pair}$ values based on analytical fitting function provided by \cite{Kitzbichler_White_08}, while others adopt values provided by \cite{lotz_effect_2010} that range from 0.3 to 2\,Gyr depending on the close-companion selection criteria. For the major merger rate calculations below, we make a simple assumption that these factors are constant over all redshifts and masses we probe. We adopt $C_{\rm merg,pair} = 0.6$ and $T_{\rm obs,pair} = 0.65$\,Gyr, which is suitable for our fiducial $R_{\rm proj} = 5-50$\,kpc criterion \citep[][]{lotz_effect_2010}. Exploring the detailed impact of these assumptions is beyond the scope of this paper. However, in \cref{evo_timescales}, we test two theoretically motivated redshift-dependent timescale prescriptions to re-derive the merger rates and discuss their implications.

\begin{figure*}
	\centering
	\includegraphics[width=7in]{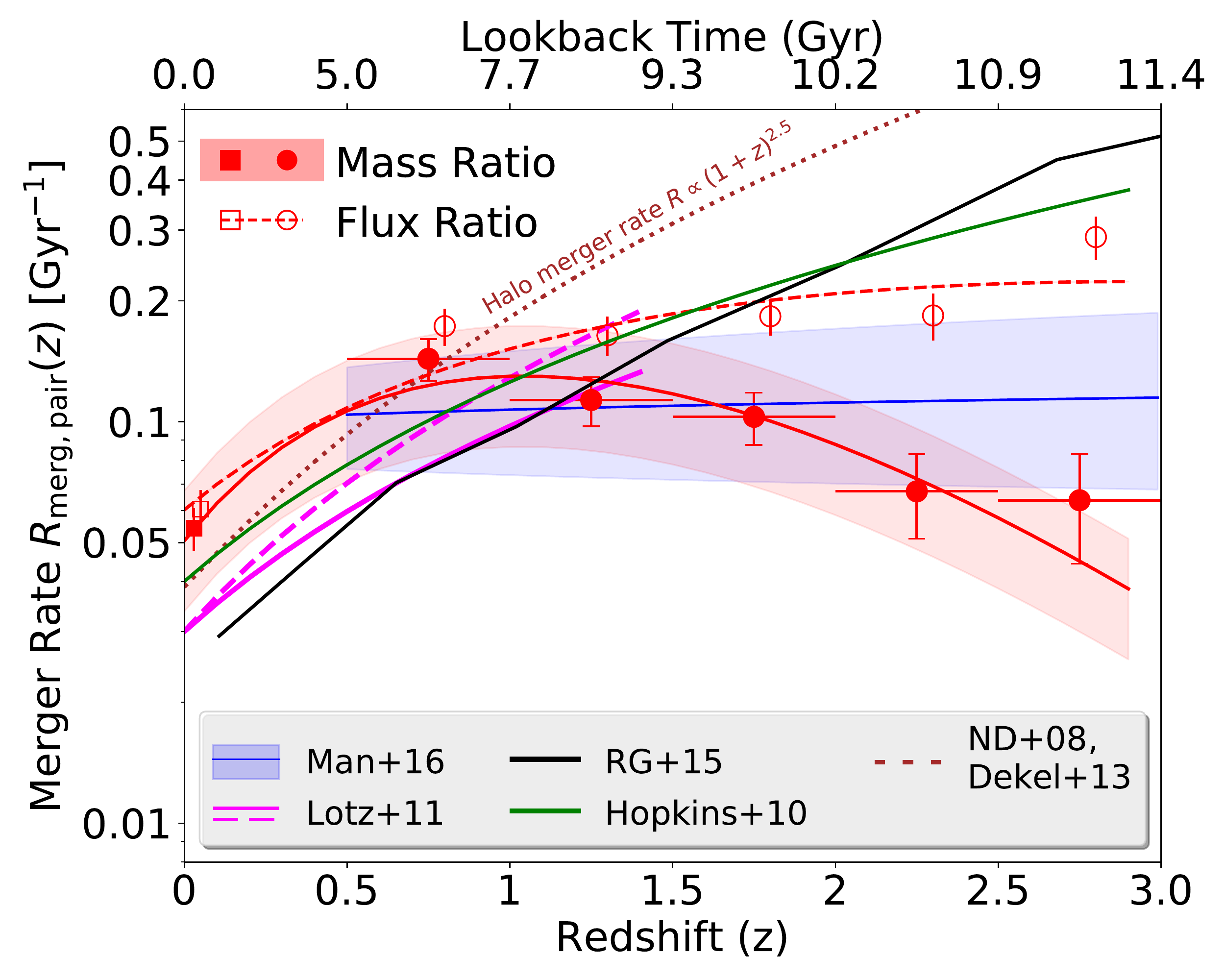}\\
	\caption{Comparison of CANDELS$+$SDSS galaxy-galaxy major merger rates $R_{\rm merg,pair}(z)$ (number of mergers per galaxy per Gyr) for massive ($M_{\rm stellar}\geq 2\times 10^{10}M_{\odot}$) galaxies 
	at $0<z<3$, to rates from previous empirical studies and theoretical model predictions. We show the $R_{\rm merg,pair}(z)$ computed using major companion fractions ($f_{\rm mc}$) based on fiducial projected separation ($5-50$\,kpc) and redshift proximity (CANDELS: Equation\,\ref{method_2_1}; SDSS: $\Delta v_{12}\leq 500$\,km\,s$^{-1}$) split into stellar mass-ratio (filled points; solid line) and flux ratio (open points; dashed line) for CANDELS (circles) and SDSS (square). We employ simplistic assumptions for fraction-to-rate conversion factors $T_{\rm obs,pair} = 0.65$\,Gyr and $C_{\rm merg,pair} = 0.6$, and  show the variation of $R_{\rm merg,pair}(z)$ for $C_{\rm merg,pair} = 0.4$ to $0.8$ in red-shading. The error-bars on the data points indicate their 95\% binomial confidence limits, solely based on the observed number counts. The solid (dashed) magenta line is the empirical merger rate evolution from \protect\cite{lotz_major_2011} for stellar-mass (luminosity) limited selections, the empirical merger rate from \protect\cite{man16} is shown in solid blue line along with its blue shaded uncertainty. {We compare to theoretical galaxy-galaxy major merger rate predictions of $M_{\rm stellar}\geq 2\times 10^{10}M_{\odot}$ galaxies by \protect\cite{hopkins_mergers_2010} (solid green lines), \protect\cite{rodriguez-gomez_merger_2015} (solid black), and to the analytical halo major merger rate prediction of $M_{\rm halo} \sim 10^{12}\,M_{\odot}$ dark-matter halos by \protect\cite{Neistein08,Fakhouri10,Dekel13} ($R\propto(1+z)^{2.5}$; dashed brown line).} We strongly advise against trusting flux-ratio-based galaxy merger rates despite their agreement with the theoretical models, owing to the notable contamination from non-major mergers (see \cref{mass_vs_flux}). Our fiducial mass-ratio based empirical major merger rates show broad agreement up to $z\sim 1.5$, but demonstrate a strong tension at $z>2$ when compared to previous empirical and theoretical constraints. }	
	\label{merger_rates}
\end{figure*}

In Figure\,\ref{merger_rates}, we show the major merger rate evolution during $0<z<3$, which we derive from CANDELS+SDSS major companion fractions for stellar mass-ratio (MR) and flux-ratio (FR) selections with matched fiducial projected separation and B09 redshift proximity criteria.  Holding aside the factor-of-two systematic uncertainty contribution from $T_{\rm obs,pair}$ to merger rate calculations, we find both the MR and FR-based merger rates evolutionary trends mimic their respective $f_{\rm mc}(z)$ trends discussed in \cref{modeling_fmp,mass_vs_flux}, which is expected given our simplistic fraction-to-merger-rate conversion factors. We notice that both the MR and FR selections yield consistent major merger rates at $z<1$, where they increase strongly from $\sim 0.05$ ($z\sim 0$) to $\sim 0.15$\,Gyr$^{-1}$ ($0.5<z<1$). At $z>1$, the different selections have divergent merger rates. The MR-based merger rates decline with redshift to $z=3$, indicating a turnover at $z\sim 1$. In contrast, the inflated companion fractions owing to contamination by minor {companions}, the flux-ratio selection yields $1.5-4.5$ times larger rates between $1\la z \la 3$ with an increasing power-law trend $R_{\rm merg,pair}\propto(1+z)^1$.

Next, we compare our empirical MR-based merger rates to previous empirical estimates  (Figure\,\ref{merger_rates}). For $z\la 1.5$, we compare to \cite{lotz_major_2011} who compile close-pair fractions from several {observational} stellar-mass and luminosity-selected studies (magenta line) and found strongly increasing major merger rate evolution with redshift $R\propto(1+z)^{1.7-2.1}$, which qualitatively agrees with both our MR and FR rate evolutions over this redshift range. The \citeauthor{lotz_major_2011} merger rate evolutions have systematically smaller normalization than ours because their compilation includes close-pair-based fractions that may be systematically smaller than our $f_{\rm mc}$ {due to the use of different selection criteria}. For example \cite{kartaltepe_evolution_2007,lin_redshift_2008} use a stringent proximity selections (see \cref{flux_ratio_selected_studies}, and Table\,\ref{selection_crit_prev_stuides}) and \cite{Patton08} uses a stringent major companion selection criterion \footnote{\cite{Patton08} selects major companions using a 2:1 luminosity ratio as opposed to our 4:1 criterion from a luminosity-limited galaxy sample at $z\sim0$.}. For merger rates at $0.5<z<3$, we compare our fiducial MR-based merger rate to a recent study of close pairs by \cite{man16}. We find that our individual $R_{\rm merg,pair}$ data points agree well with the \citeauthor{man16} results (blue) within the measurement uncertainties at these redshifts. Yet, our standalone rising-diminishing merger rate evolution disagrees with the \citeauthor{man16} flat trend. If we were to extend their rates to $z\sim 0$, we would disagree with them; however, we note that \citeauthor{man16} do not attempt to anchor to a low-redshift (for their 3D-HST sample), as such, claim a flat trend with redshift in agreement with \cite{Mundy17}.  {Recently, \cite{Mundy17} in conjunction with Duncan et al., in prep find that the  \cite{Henriques15} semi-analytic model mock light-cones accurately reproduce a flat pair-fraction trend that is consistent with the recent findings by \cite{man16,snyder17}, and the results of this work.  Employing an un-evolving close-pair observability timescale assumption, \citeauthor{Mundy17} reports a flat major merger rate evolution. Our fiducial MR-selected merger rates shown in Figure\,\ref{merger_rates} agree with the \citeauthor{Mundy17} result, which is not surprising owing to the assumption of constant observability timescale.}

%and  with our FR rates over the full redshift range $0<z<3$

Here, we discuss our comparison to theoretical major merger predictions. {Since these are based on stellar-mass ratios, we refrain from comparing them to our FR-based rates.} \cite{hopkins_mergers_2010} provides {a} comprehensive analysis of major galaxy-galaxy merging using $\Lambda$CDM motivated simulation and Semi-Analytic Modeling to derive major merger rate predictions, and quantified systematic error contribution from various theoretical model-dependent assumptions. {We find that the \citeauthor{hopkins_mergers_2010} major merger rate evolution of $\logten(M_{\rm stellar}/M_{\odot})\geq10.3$ galaxies ($R = 0.04(1+z)^{1.35}$; green line in Figure\,\ref{merger_rates}), within a factor of two uncertainty, agrees qualitatively and quantitatively with our MR-based rates up to $z<1.5$. At $z>1.5$, our fiducial MR-selected, diminishing merger rate trend disagrees with the \citeauthor{hopkins_mergers_2010} predictions}. We also compare our merger rates to a recent theoretical prediction by \cite{Rodriguez-Gomez15} from the Illustris numerical hydrodynamic simulation \citep{vogelsberger14}. {We note that their $\logten(M_{\rm stellar}/M_{\odot})\geq10.3$ galaxy merger rates (black line in Figure\,\ref{merger_rates}) follow an even stronger increasing redshift dependence ($R\propto(1+z)^{2.4-2.8}$) than the \citeauthor{hopkins_mergers_2010} predictions, which also agrees qualitatively with our $f_{\rm mc}(z)$ trend at $z< 1.5$. However, at $z<1.5$, we note that the \citeauthor{rodriguez-gomez_merger_2015} predictions are smaller than our empirical estimates.} Similar to the conclusion from the \citeauthor{hopkins_mergers_2010} comparison, our fiducial MR-based merger rates disagree with predictions from \citeauthor{rodriguez-gomez_merger_2015} at $z>1.5$. {We present further discussion on the plausible reason for the observed discrepancy of flat vs. rising merger rate evolution in \cref{evo_timescales}.}

%While the FR-selected merger rate qualitatively agrees with the halo merger rates, the former follows a shallower trend than the later
Being the primary driver for galaxy-galaxy merging, halo-halo merger rates serve as a ceiling for both empirical measurements and theoretical galaxy merging predictions, and their qualitative trend is expected to mimic the galaxy merger rates closely. We find that our empirical MR-selected merger rate trend broadly agrees with the analytical merger rates of $M_{\rm halo}\sim10^{12}\,M_{\odot}$ dark-matter halos ($R\propto(1+z)^{2.5}$; brown line in Figure\,\ref{merger_rates}) predicted by \cite{Neistein08,Fakhouri10,Dekel13} \citep[also see][]{Maulbetsch07} at $z\la 1$ but starts to deviate significantly at $z>1$. {Since the translation from a halo-halo merger rate to a galaxy-galaxy merger rate depends on the redshift-dependent mapping of galaxies onto their respective dark-matter halos (halo-occupation statistics), the galaxy merger rates are expected to follow a shallower redshift evolution than the ones set by the halo merger rates \citep[for discussion, see][]{hopkins_mergers_2010}. Although different assumptions of halo-occupation models may contribute up to a factor of two uncertainty in the theoretically predicted galaxy merger rates \citep[\eg][]{Hopkins2010b}, they all hint towards a rising incidence with increasing redshift. As such, a deviation of more than factor of five (\eg\, see redshift bin $2<z<3$ in Figure\,\ref{merger_rates}) between the constant timescale-based fiducial MR-selected empirical rates (diminishing) and the analytical halo merger rate evolution (rising; $R\propto (1+z)^{2.5}$) is puzzling.} Based on this analysis, we conclude that a straight-forward MR-based estimation of major merger rates using a constant fraction-to-merger-rate observability timescale ($T_{\rm obs,pair} = 0.65$\,Gyr) are in disagreement with the theoretical predictions at $z>1.5$. We discuss the implications of these conclusions in \cref{reasons_tension}.

%On the other hand, the $H$-band FR-based rates agree with theoretical predictions up to $z=3$ but, {\bf they} may be less trustworthy owing to the contamination from stellar-mass based minor-pairs ($M_{1}/M_{2}>4$) as demonstrated in \cref{mass_vs_flux}

%Fig11

\subsection{Implications of the Disagreement between Simplistic Empirical Merger Rates and Model Predictions}  \label{reasons_tension}
So far, we have demonstrated that major merger rates based on a rather straight-forward MR-selection start to disagree with theoretical predictions of major galaxy-galaxy merging at $z\ga1.5$. This indicates there may be one or many redshift-dependent effects plaguing the key variables ($C_{\rm merg,pair}$, $f_{\rm mc}$, and $T_{\rm obs,pair}$) in Equation\,\ref{merger_rate_equation}, causing the observed merger rates to disagree with simulations. Here, we first investigate for plausible systematic under-estimation of $f_{\rm mc}$ values at $z\ga1.5$ as a reason for the observed discrepancy by testing if we are missing close companions by not choosing the full $z_{\rm phot}$ PDF when applying the B09 redshift proximity. Then, we discuss another possibility that the simplistic, constant value assumptions of $T_{\rm obs,pair}$ may be incorrect, and test two theoretically-motivated, redshift-dependent $T_{\rm obs,pair}$ prescriptions to re-derive merger rates. We also elaborate on the remnant formation times, which is a key difference between the merger rates measured in theoretical simulations and empirically calculated close-pair-based rates, and comment on how it may be {affecting} the data-theory comparison.
	
\subsubsection{Are We Missing Companions by Not Accounting for the Full $z_{\rm phot}$ PDFs?} \label{plausible_fmp_systematics}
One plausible reason for the $z>2$ data-theory discrepancy may be that the adopted B09 redshift proximity criterion (Eq.\,\ref{method_2_1}) is missing plausible physical close-pair systems because we are using only the $1\sigma$ photometric uncertainties and not the full $z_{\rm phot}$ PDF.  To test if this is the case we carry out a simple test, where we start by simplifying Equation\,\ref{method_2_1} as $\xi =  \Delta z_{12}/\sqrt{\sigma_{\rm z,1}^2+\sigma_{\rm z,2}^2}$; therefore, our fiducial
criterion equates to $\xi \leq 1$. The distribution of $\xi$ values for physical close-pair systems is Gaussian, assuming $\sigma_{\rm z}$ values are Gaussian. Therefore, using $\xi \leq 1$ is equivalent to selecting 68\% of the full $\xi$ distribution. Ideally, if one assumes the contribution from non-physical pairs (hereafter called as the background) per $\xi$ bin is negligible, as much as $32\%$ of the close pairs (thereby companions) can be missed when using $\xi \leq 1$. For this exercise, we allow for a non-zero background to properly estimate the missing pairs at five redshift bins between $0.5<z<3.0$.

\begin{figure*}
	\centering
	\includegraphics[width=2\columnwidth]{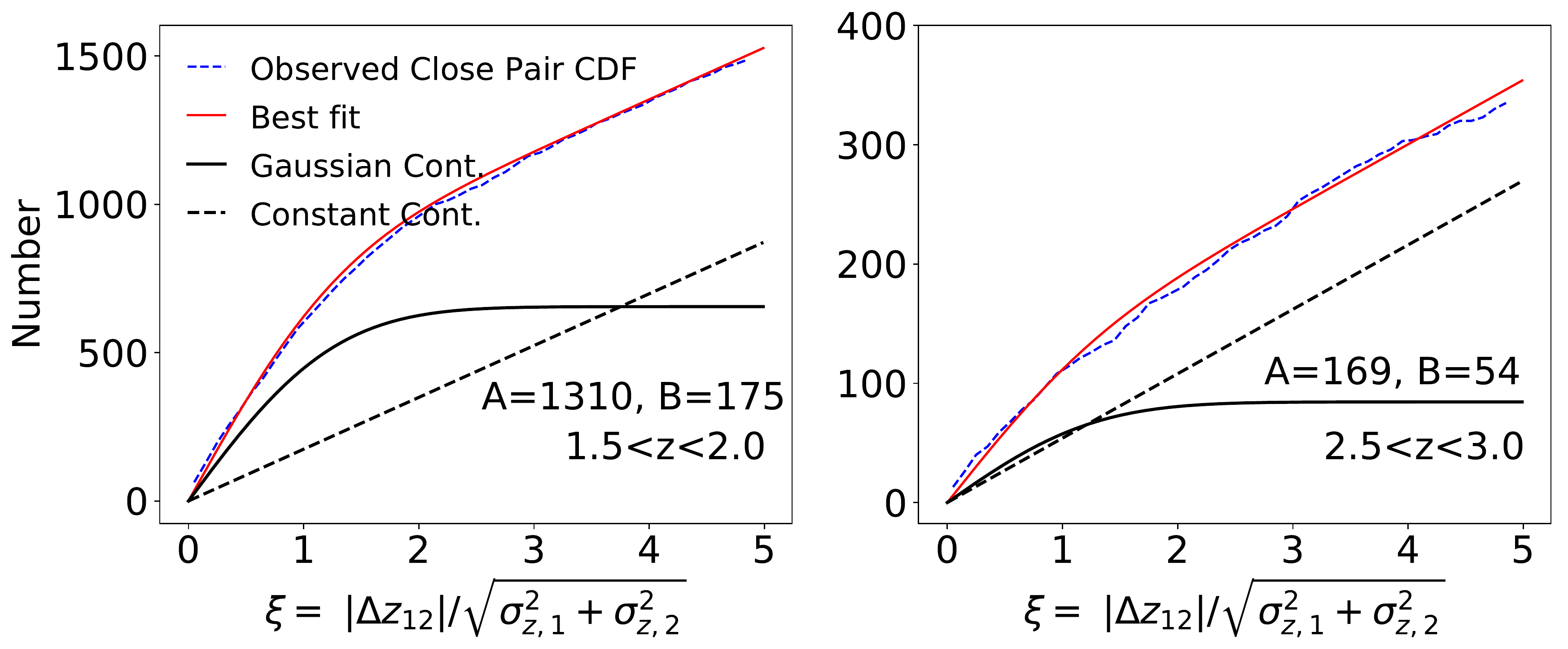}\\
	\caption{The cumulative distribution functions (CDFs) of $\xi = \Delta z_{12}/\sqrt{\sigma_{z,1}^2+\sigma_{z,2}^2}$ (from Eq.\,\ref{method_2_1}) for projected pairs satisfying our fiducial $5\,{\rm kpc}\leq R_{\rm proj}\leq {50\,{\rm kpc}}$ selection (blue dashed, narrow line) as described in \cref{plausible_fmp_systematics} for two redshift bins: $1.5<z<2$ (left) and $2.5<z<3$ (right). In each redshift-bin panel, following the equation $N(\xi\leq c) = A/\sqrt{2\pi}\int_{0}^{c}\exp^{-x^2/2} dx + Bc$, the best-fit curve to the CDF is shown in narrow solid-red line. The break-down of the best-fit curve into Gaussian and background contributions  are shown in bold solid-black and bold-dashed lines, respectively, and we print the best-fit values ($A$ and $B$) in each panel.}	
	\label{fig_appendix_6a} 
\end{figure*}

\begin{figure}
	\centering
	\includegraphics[width=\columnwidth]{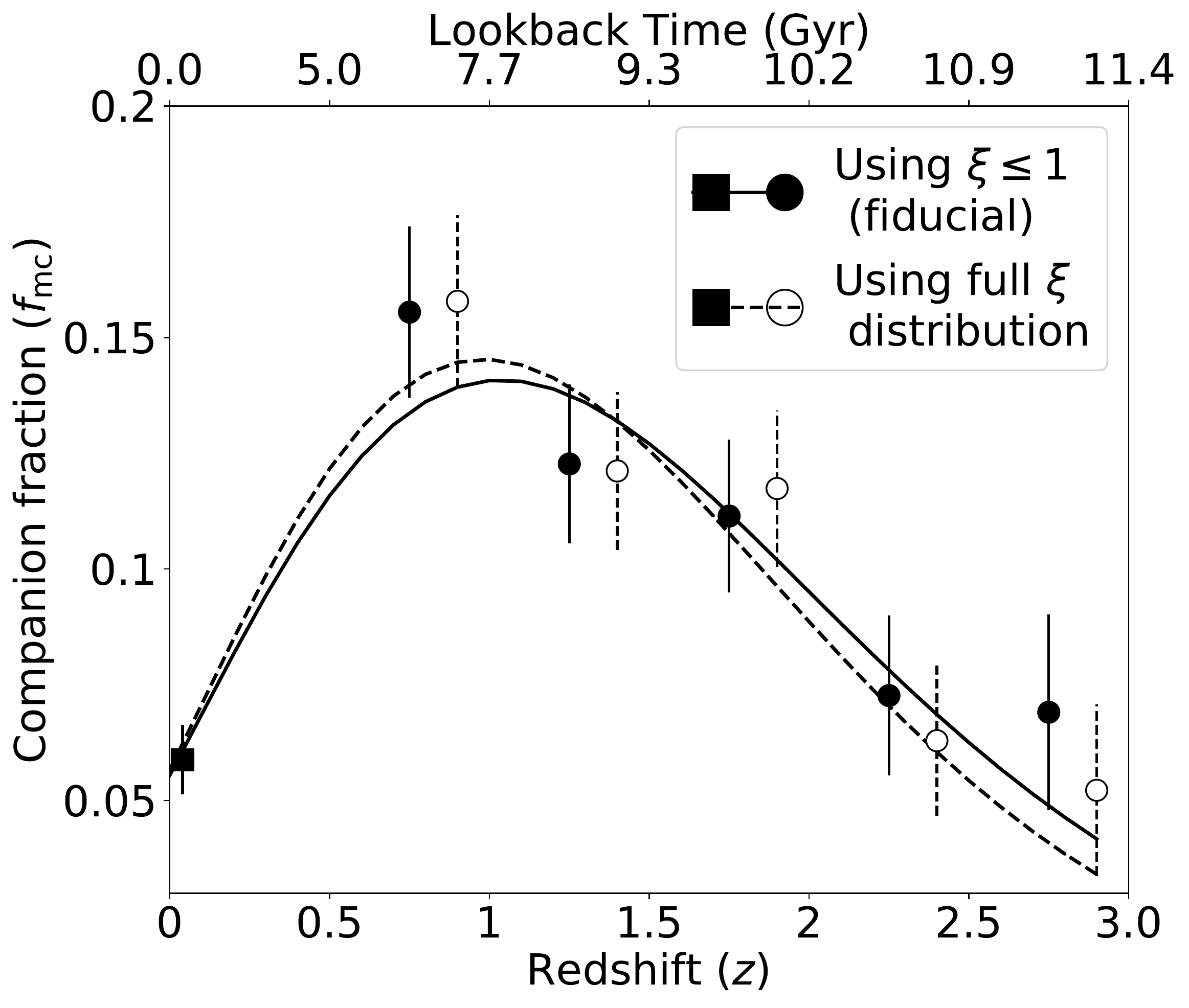}
	\caption{Comparing the redshift evolution of the fiducial major companion fraction $f_{\rm mc}(z)$ (filled circles; copied from Figure\,\ref{b09_with_and_without_correction}) to the fractions after correcting for full $\xi = |\Delta z_{12}|/\sqrt{\sigma_{z,1}^2+\sigma_{z,2}^2} $ distribution usage (open circles; see \cref{plausible_fmp_systematics} for details) . For this exercise, we anchor both CANDELS $f_{\rm mc}(z)$ to the same fiducial selection based SDSS data-point at $z\sim 0$. The error-bars correspond to 95\% binomial confidence limits on the $f_{\rm mc}$.}	
	\label{fig_appendix_6b}
\end{figure}

We begin with the cumulative distribution function (CDF) of $\xi$ for all the close-pair systems selected using our fiducial $R_{\rm proj} = 5-50\,{\rm kpc}$ projected separation criterion. To avoid errors from small number statistics, we do not limit close-pair systems {with a} stellar-mass-ratio selection. {However, we require that} both the host and companion galaxies to be brighter than $H = 25$\,mag to exclude large $\xi$ values from {affecting} the statistics. We now model CDF with a Gaussian integral function in conjunction with a constant background contribution, following the fitting formula: $N(\xi\leq c) = A/\sqrt{2\pi}\int_{0}^{c}\exp^{-x^2/2} dx + Bc$. Here, $N(\xi\leq c)$ is the number of close-pair systems that satisfy a cutoff significance ($c$) as $\xi\leq c$, and the variables $A$ and $B$ are the contributions from plausible physical close pairs and the background, respectively. For demonstrative purposes, we present the fits to the CDF and their best-fit parameter values of $A$ and $B$ in Figure\,\ref{fig_appendix_6a} for two redshift bins between $1.5<z<3$. At each redshift bin, we define the ratio $(A/2)/N(\xi\leq 1)$ as the multiplicative correction factor to our fiducial $f_{\rm mc}$ that corrects for the missing close-pair systems due to Method I. When the background is negligible ($B\sim 0$), the correction factor is $\sim 1.44$. However, if $B$ is significant, then the correction factor may become less than unity. In Figure\,\ref{fig_appendix_6b}, we plot the corrected $f_{\rm mc}$ alongside the fiducial fractions. We note that fractional change in the major companion fraction after applying the correction factor is less than 5\% at $0.5<z<2$. However, at $z>2$ we find the relative contribution from the background starts to dominate the Gaussian contribution and therefore causes the fiducial $f_{\rm mc}$ to decrease by more than 10\%. Despite these changes, we note that the overall corrected $f_{\rm mc}(z)$ values do not significantly deviate from the fiducial $f_{\rm mc}(z)$ values.  Therefore, we conclude that we are not missing close pairs (thereby companions), by using $68\%$ photometric-redshift errors for our fiducial redshift proximity criteria.

\begin{figure}
	\centering
	\includegraphics[width=\columnwidth]{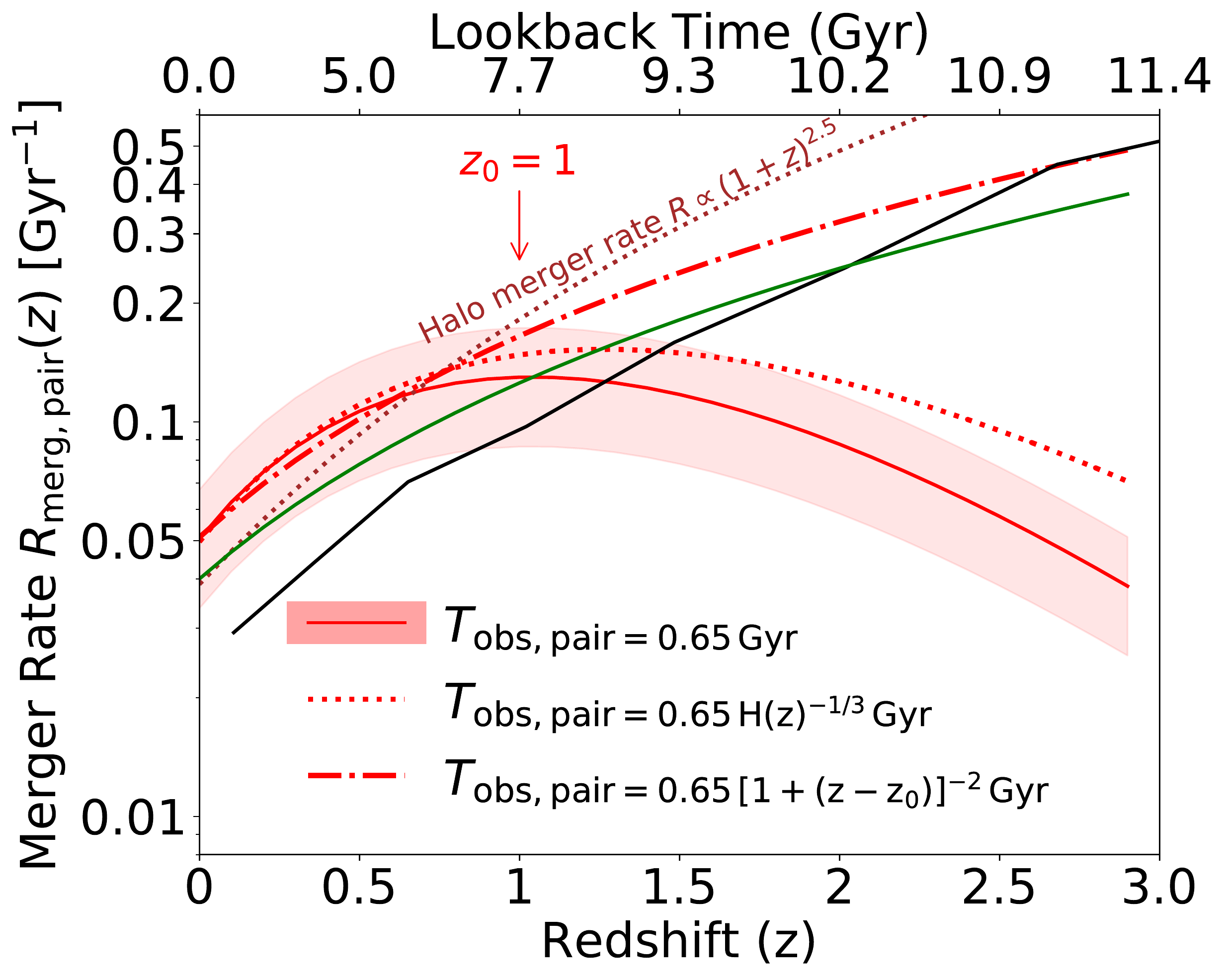}\\
	\caption{Comparing the redshift evolution of major merger rate of $M_{\rm stellar}\geq 2\times10^{10}M_{\odot}$ galaxies based on our fiducial close-pair timescale assumption $T_{\rm obs,pair} = 0.65$\,Gyr copied from Figure\,\ref{merger_rates} (solid red line, shading) to rates from different timescale choices. We show the rates based on \protect\cite{jiang14} scaling relation $T_{\rm obs,pair}\propto H(z)^{-1/3}$ in red dotted line, and \protect\cite{snyder17} relation $T_{\rm obs,pair}\propto(1+z)^{-2}$ in red dot-dashed line (starting at $z_{0}=1$). We also plot the theoretical merger rate predictions shown in Figure\,\ref{merger_rates}.}	
	\label{fig_evolving_timescales}
\end{figure}

\subsubsection{Evolving Observability Timescale and Remnant Formation Times} \label{evo_timescales}
It is evident by now that the close-pair observability timescale ($T_{\rm obs,pair}$) is a key parameter that dictates the measurement of merger rates. Yet, a wide range of values can be found in the literature (see \cref{Merger Rates}), often assumed to be constant over a wide range of redshifts. Therefore, with the growing evidence \citep{man16,Mundy17} for flat or possibly diminishing close-pair-based fractions at $1<z<3$, it is expected that a constant timescale assumption-based merger rates to disagree with the continually rising merger-rate trends from the theoretical predictions. A possible implication of this disagreement is that the timescale may be evolving with redshift, such that an ongoing merger can be observed as a close pair for a shorter period at earlier cosmic times.  This may not be surprising as simulation-based merging-timescale calibrations \citep[e.g.,][]{Kitzbichler_White_08,jiang14}, motivated by dynamical friction time for a satellite DM-halo to merge with the parent, are complex functions that depend on key merger properties such as the mass ratio, galaxy mass, $R_{\rm proj}$, and $H(z)^{-1/3}$ ($H(z) = H_{0}E(z)$; $H_{0}$ is the present-day Hubble constant; $E(z)$ is the evolution of $H_{0}$ as a function of redshift). Moreover, the dynamical timescale based on simple cosmological arguments is also redshift dependent, and can be approximated as $T\propto (1+z)^{-1.5}$ \citep[see][]{snyder17}. However, as per the discussion in \cite{snyder17}, any processes that can act faster than the dynamical timescale and may impact the variables of close companion selection criteria, which in turn can change the $T_{\rm obs,pair}$. In this context, the galaxies are undergoing {a} rapid transformation {concerning} their stellar-mass assembly, owing to rapid ongoing star-formation \citep{madau14} at $z>1$. \citeauthor{snyder17} found that this rapid mass assembly can shorten the close-pair observability timescale towards earlier redshifts, which evolves as $T\propto (1+z)^{-2}$ at $z>1$. Moreover, they measured the close-pair fractions within the Illustris simulation \citep{vogelsberger14} and inferred that the timescale should evolve as $T\propto (1+z)^{-2}$ to match the intrinsic merger rates trend from \cite{rodriguez-gomez_merger_2015}. 

In Figure\,\ref{fig_evolving_timescales}, we re-compute the merger rates for the fiducial mass-ratio based companion fractions by adopting the two evolving timescale assumptions $T_{\rm obs,pair} \propto H(z)^{-1/3}$ and $T_{\rm obs,pair}\propto (1+z)^{-2}$ from \cite{jiang14,snyder17}, respectively. Since \citeauthor{snyder17} asserts that the close-pair observability is mainly dynamical time dominated at $z<1$, which results in an un-evolving timescale value, we choose to evolve the $T_{\rm obs,pair}$ only at $z>1$ when re-computing the merger rates. We find that the major merger rates based on \citeauthor{jiang14} evolving timescale still does not sufficiently agree with the theoretical predictions. On the other hand, when we use simulation-tuned \citeauthor{snyder17} prescription of $T_{\rm obs,pair}\propto (1+z)^{-2}$, we find that the merger rates agree closely to model predictions of \cite{hopkins_mergers_2010,rodriguez-gomez_merger_2015}, and closely mimic the \cite{Neistein08,Dekel13} halo-halo merger rates. {The fact that our evolving timescale based empirical merger rates agree with the theoretical predictions from \citeauthor{rodriguez-gomez_merger_2015} implies a very good agreement between our $f_{\rm mc}(z)$ and \citeauthor{snyder17} fractions. Additionally, we argue that the disagreement between \cite{Henriques15,Mundy17} flat merger rate evolution and the \cite{hopkins_mergers_2010,rodriguez-gomez_merger_2015} theoretical predictions (see \S\,5.2) may be due to the un-evolving close-pair observability timescale assumption.} These results imply that an evolving timescale assumption may be more appropriate when measuring merger rates at $z>1$, and it {may be} necessary to explain the discrepancy between a constant timescale based rates and theoretical {model} predictions. 

Additionally, we also stress the difference in the measured quantities between the simulations and empirical close-pair-based studies. In theoretical simulations, a merger is counted towards the intrinsic merger rate when the merging process is concluded, \ie\, when the remnant of the progenitor galaxies is formed, and therefore it is an instantaneous measure. On the other hand, the close-pair method probes future merging systems, within some time-frame after observing them. \citeauthor{snyder17} found that the remnant formation time also evolves as $(1+z)^{-2}$, which is defined as the time prior to the formation of the remnant, when the merger is observed as a close pair. We follow Section\,4.1 in \citeauthor{snyder17} and compute instantaneous remnant formation rates at $z>1$, which {is} conceptually similar to the instantaneous merger rates quoted by the simulations. We find that remnant formation rate follows an increasing redshift trend, which is in qualitative agreement with finding by \cite{snyder17}. We acknowledge that a comprehensive analysis of the empirical and theoretical measurement methods is necessary to make quantitative data-theory comparisons {confidently}. However, it is beyond the scope of the current study to fully address them. We reserve further discussions on this analysis to an accompanying paper (Paper\#2 in the series; Mantha et al., in prep).

\subsection{Future Work}
So far, we have discussed plausible evolutionary observability timescale prescriptions that may be necessary when deriving merger rates at $z>1$. While this may be true,  $C_{\rm merg,pair}$ is another key parameter in Equation\,\ref{merger_rate_equation} which may also depend on the companion selection criteria. In fact, recent studies \citep[\eg][]{lin10,de_ravel_zcosmos_2011} find that $C_{\rm merg,pair}$ varies as a function of $R_{\rm proj}$, redshift, and local over-density (often represented by $\delta$). Although, a redshift-dependent timescale prescription sufficiently brings empirical and theoretical merger rates to agreement, it is possible that $T_{\rm obs,pair}$ may be a simultaneous function of several key close-pair variables. As such, these conversion factors may be  better represented as $C^{\prime}_{\rm merg,pair} = C_{\rm merg,pair}(z,R_{\rm proj},\delta)$, $T^{\prime}_{\rm obs,pair} = T_{\rm obs,pair}(z,R_{\rm proj},M_{1}/M_{2},M_{\rm stellar},\delta)$, where $C^{\prime}_{\rm merg,pair}$ and $T^{\prime}_{\rm obs,pair}$ are the evolving prescriptions of $C_{\rm pair}$ and $T_{\rm obs,pair}$, respectively. To fully quantify the interplay among the companion selection criteria and the fraction-to-merger-rate conversions, it is important to derive detailed calibrations for $C^{\prime}_{\rm merg,pair}$ and $T^{\prime}_{\rm obs,pair}$ using large-scale cosmological simulations. {We will discuss this in a follow-up Paper\#2 (Mantha et al., in prep). }
	
Furthermore, it has been speculated that stellar-mass ratio may be a poor representer of ongoing 
``significant'' mergers, owing to the dominant cold-gas contribution to the total baryonic mass (gas$+$stellar mass) of the galaxy at $z\sim 2-4$ \citep{lotz_major_2011,man16}. If this is true, a galaxy merger with $M_{1}/M_{2}>4$ may in fact be significant if the total baryonic (stellar+cold gas) mass ratio of the galaxies is considered (\ie\,$M_{1,{\rm bar}}/M_{2,{\rm bar}}\leq 4$). This may cause for many of the mergers that can be significant contributors to several aspects of galaxy evolution to be missed because of the selection bias. Empirical confirmation of this speculated selection bias deserves a dedicated analysis of its own, and we will discuss this in an accompanying paper (\#3 of this series).

Semi-Analytic Models (SAMs) are showing a great promise as the testing grounds for many galaxy evolution related questions \citep[\eg][]{Brennan15,Pandya17}. Three studies, \cite[][Somerville et al., in prep]{Somerville08b,lu14,croton16} have independently developed SAMs based on Bolshoi-Plank N-body simulation \citep{klypin16} and produced mock datasets that mimic observational data in the five CANDELS fields. Additionally, multiple realizations of each CANDELS mock field are also made available. We will take advantage of the intrinsic merger history information and apply close pair analysis on these mock datasets to derive the necessary calibrations. With the help of multiple realizations per field, we will be able to quantify the impact of sample variance towards close-pair statistics and merger rates. Also, by introducing realistic, CANDELS-like random and systematic errors onto simulated  $z_{\rm phot}$ and $M_{\rm stellar}$ quantities, we will {carry out} a comprehensive investigation of their impact on the measured merger rate evolution. These calibrations and analyses will help fill the gaps in our understanding of the merger rate measurements and {will provide} most up-to-date major merger rate constraints with future state-of-the-art telescopes.

%Fig12
%Fig13	
	
\section{Conclusions}
\label{Conclusions}
In this paper we analyze a large sample of nearly 9800 {\it massive} ($M_{\rm stellar}\geq 2\times10^{10}M_{\odot}$) galaxies spanning $0<z<3$ using the five {\it Hubble Space Telescope} CANDELS fields (totaling $\sim 0.22\deg^{2}$) and volume-matched region of the SDSS surveys to quantify the redshift evolution of major companion fraction $f_{\rm mc}(z)$. We adopt a fiducial selection criteria of projected separation ($5\leq R_{\rm proj}\leq 50$\,kpc), redshift proximity $\Delta z_{12}\leq \sqrt{\sigma^{2}_{z,1}+\sigma^{2}_{z,2}}$ (CANDELS) and $\Delta v_{12}\leq 500$\,km\,s$^{-1}$ (SDSS), and stellar mass-ratio major companion selection criterion $1\leq M_{1}/M_{2}\leq 4$ (MR). 

Our key result is the MR-based major companion fraction increases from 6\% ($z\sim 0$) to $\sim 16\%$ ($0.5<z<1$) and decreases to $\sim 7\%$ ($2.5<z<3$), indicating a turnover at $z\sim 1$ with power-law exponential (Equation\,\ref{equation_best_fit}) best-fit values $\alpha = 0.5\pm0.2$, $m = 4.6\pm0.9$, and $\beta = -2.3\pm0.5$. We perform comprehensive tests demonstrating that these evolutionary trends are robust to changes in companion selection criteria except for $H$-band flux ratio (FR) selection, where FR-based $f_{\rm mc}(z)$ agree with MR-based fractions up to $z\sim 1$, but disagree at $z>1$ as they increase steadily with redshift as $f_{\rm mc}(z) \propto(1+z)^{1}$ up to $z=3$, confirming previous speculations \citep{Bundy04,bundy_greater_2009}. This disagreement is due to increasing contamination of FR selection by minor companions (MR$>4$) from 40\% at $1<z<1.5$ to over
three-quarters at $2.5<z<3$, confirming the result by \cite{man16} that significantly different mass-to-light ratio properties of the companion galaxies at $z>1$ may be the main cause for the contamination by minor companions. 

We compute major merger rates for MR and FR-based fractions by using a constant fraction-to-merger-rate conversion $C_{\rm merg,pair} = 0.6$ and a non-evolving close-pair observability timescale $T_{\rm obs,pair} = 0.65\,$Gyr \citep{lotz_effect_2010}. {We find that the MR-based rates follow an increasing trend with redshift up to $z\sim 1.5$, after which they decline up to $z\sim 3$. On the other hand, the FR-based rates follow a rising trend with redshift over our full redshift range $0<z<3$.} We also re-compute the MR-based rates using two additional theoretically-motivated, redshift dependent timescales: $T_{\rm obs,pair}\propto H(z)^{-1/3}$ \citep{jiang14} and $T_{\rm obs,pair}\propto(1+z)^{-2}$ \citep{snyder17}. 

{If recent cosmologically-motivated merger simulations are representative of nature,
our results indicate the strong need for improved understanding of how MR and FR estimates trace the galaxy mass (gas+stars) and halo mass ratios of merging systems as a function of redshift. This analysis underscores the strong need for detailed calibrations of these complex, presumably evolving prescriptions of $C_{\rm merg,pair}$ and $T_{\rm obs,pair}$ to constrain the empirical major merger rates confidently. In addition, $C_{\rm merg,pair}$ and $T_{\rm obs,pair}$ may be evolving not only as a function of redshift, but also stellar-mass, companion selection criteria ($R_{\rm proj}$, $M_{1}/M_{2}$), and environment (local over-density), which we plan to pursue through a comprehensive analysis of close-pair statistics using mock datasets from the Semi-Analytic Models in Paper\#2.}

Finally, we recompute $f_{\rm mc}(z)$ for different companion selections and we find good agreement with previous empirical estimates when we match their companion selection criteria. This confirms the report by \cite{lotz_major_2011} that part of the reason for the observed study-to-study differences is due to close-pair selection mismatch. Additionally, our comprehensive analysis of the impact of different selection criteria and statistical corrections on $f_{\rm mc}(z)$ measurements yields the following results:

\begin{itemize}
	\item The redshift evolution of major companion fractions {is} qualitatively robust to changes in projected separation criterion, with smaller $R_{\rm proj}$ resulting in lower fractions as expected. {However,} the $5-30$\,kpc selection ($1/2$ the area of $5-50$\,kpc selection) yields similar results to $14-43$\,kpc criterion owing to the increased probability of physical companions at smaller projected separations.
	\item Moreover, the redshift evolution of major companion fractions {is} qualitatively robust to changes in redshift proximity criterion. While the modified \citeauthor{man2012} (\cref{changing_redshift_proximity}) proximity-based fractions are systematically $\sim 11-24\%$ higher than the fiducial values, the hybrid B09-based proximity (\cref{changing_redshift_proximity}) yields $9-17\%$ smaller major companion fractions at $0.5<z<1.5$ and nearly identical values at $z>1.5$ when compared to fiducial B09 (Equation\,\ref{method_2_1}) based fractions, owing to the sparsity of spectroscopic-redshift coverage in CANDELS at these redshifts.
	\item We recompute the SDSS $f_{\rm mc}(0)$ evolutionary anchor using the modified \citeauthor{man2012} and B09 redshift proximity criteria applied to simulated $z_{\rm phot}$ errors similar to those for $z>0.5$ CANDELS galaxies. We find the recomputed $z\sim0$ companion fractions are statistically equivalent ($\Delta f/f_{\rm fid}\sim3\%$) to the fiducial $f_{\rm mc}(0)$ values, which demonstrates a high probability of small velocity separations ($\Delta v_{12}\leq 500$\,km\,s$^{-1}$) for companions in $5-50$\,kpc projected pairs in the SDSS. 
	\item When major companion fractions are corrected for random chance pairing, we find negligible corrections ($1\%$) for SDSS owing to our stringent $\Delta v_{12}\leq 500$\,km\,s$^{-1}$ proximity selection, and approximately equal corrections of $\sim 20\%$ at $0.5<z<3$ for CANDELS. Importantly, our key results regarding the evolutionary trends of major companion fractions are qualitatively the same with or without this statistical correction. 
	\item  We also demonstrate that the B09 redshift proximity method does not exclude close-pair systems owing to our use of a simple 68\% confidence limit to represent the full $z_{\rm phot}$ PDF.
	\item {Finally, we demonstrate that MR-selected merger rates using un-evolving prescriptions of $C_{\rm merg,pair}$ and $T_{\rm obs,pair}$ agree with the theoretical predictions at $z<1.5$ but strongly disagree at $z\gtrsim1.5$. If we re-compute our MR-based rates using an evolving, dynamical-time motivate timescale by \citeauthor{jiang14}, we find the merger rates are still lower than and disagree with the theoretical merger rate predictions at $z>1.5$. However, if we use a redshift-dependent timescale $T_{\rm obs,pair}(z)\propto(1+z)^{-2}$ motivated by \cite{snyder17}, our MR-based rates agree with the theory at redshifts $0<z<3$.}
	
\end{itemize}
	
\section*{Acknowledgements}
This work is dedicated in memory of K. N. V. S. Kishore Babu, a dear friend who inspired KBM to pursue Astrophysics. We are grateful to Peter Behroozi, Mark Brodwin, Philip Hopkins, Kartheik Iyer, Allison Kirkpatrick, Viraj Pandya, Vicente Rodriguez-Gomez, Brett Salmon, Raymond Simons for helpful comments and discussions during different manifestations of this work. Special thanks to Daniel Shanaberger for carefully reading the manuscript, thanks to Gillen Brown, Cody Ciaschi, Bandon Decker, Benjamin Floyd, Ripon Saha, and Madalyn Weston for their valuable inputs during the UMKC Galaxy Evolution Group (GEG) discussions. DHM and KBM acknowledge support from the Missouri
Consortium of NASA's National Space Grant College and Fellowship Program, and funding from the University of Missouri Research Board. RSS acknowledges support from the Downsbrough family, and from the Simons Foundation. SL acknowledges support from the National Research Foundation of Korea (NRF) grant, No. 2017R1A3A3001362, funded by the Korea government (MSIP). This work is based on observations taken by the CANDELS Multi-Cycle Treasury Program with the NASA/ESA HST, which is operated by the Association of Universities for Research in Astronomy, Inc., under NASA contract NAS5-26555. Support for Program number HST-GO-12060 was provided by NASA through a grant from the Space Telescope Science Institute, which is operated by the Association of Universities for Research in Astronomy, Incorporated, under NASA contract NAS5-26555. This publication makes use of the Sloan Digital Sky Survey (SDSS). Funding for the creation and distribution of the SDSS Archive has been provided by the Alfred P. Sloan Foundation, the Participating Institutions, the National Science Foundation, the US Department of Energy, the National Aeronautics and Space Administration, the Japanese Monbukagakusho, the Max Planck Society and the Higher Education Funding Council for England. This publication also made use of NASA's Astrophysics Data System
Bibliographic Services, TOPCAT \citep[Tools for OPerations on Catalogues And Tables,][]{Taylor05}, the core python package for the astronomy community \citep[{\it Astropy 1.2.1;}][]{Robitaille13}.
	
\appendix	
\section{Checking Consistency of CANDELS team Redshifts to Single Participant Estimates}
\begin{figure*}
	\centering
	\includegraphics[width=2\columnwidth]{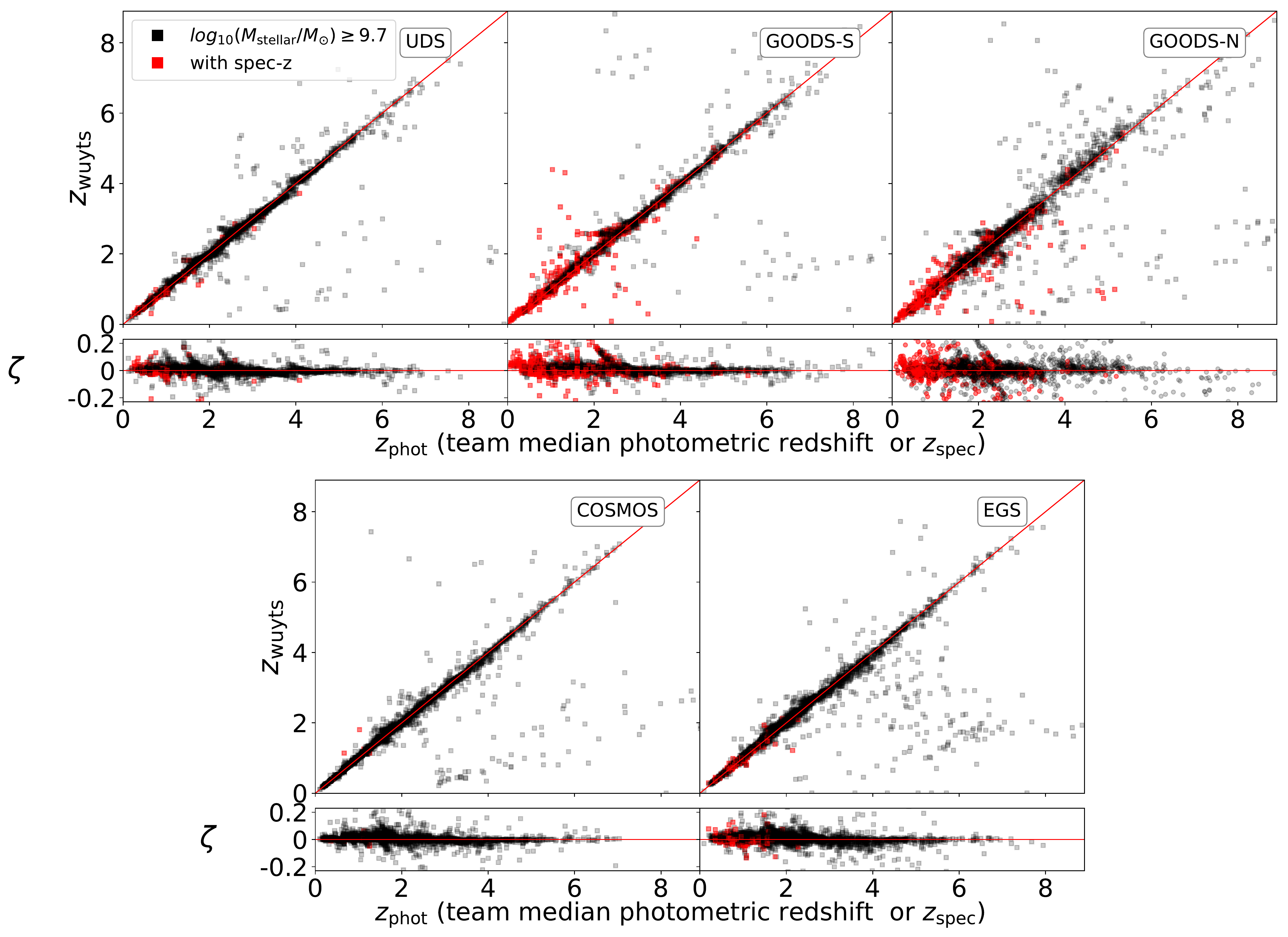}\\
	\caption{Comparison between the single participant (S. Wuyts) $P(z)$-based redshifts ($z_{\rm wuyts}$) to the CANDELS team $z_{\rm phot}$ values for $\logten M_{\rm stellar}/M_{\odot}\geq9.7$ galaxies (black squares) and those with spectroscopic redshifts (red squares) for five CANDELS fields. In each panel, the top portion of the figure compares both redshift estimates with one-to-one correspondence (red) line , and the bottom portion visualizes the redshift normalized scatter defined as $\zeta = (z_{\rm wuyts}-z_{\rm phot})/(1+z_{\rm phot})$ centered on zero. We find that the S. Wuyts redshift estimates best match to the CANDELS team values with a median outlier fraction of $1.5\%$ and $3\%$ for $z_{\rm phot}$ and $z_{\rm spec}$ samples, respectively (see text in Appendix B for details). }	
	\label{fig_appendix_1b}
\end{figure*}

In \cref{plausible_physical_pairs}, we have discussed adopting the $\sigma_{z}$ values based on single participant $P(z)$ as the photometric-redshift errors of $z_{\rm phot}$ to be able to apply the redshift proximity criterion {successfully}. To choose an appropriate participant for this analysis, we test the consistency of redshift estimates from six participants to the CANDELS team $z_{\rm phot}$ values for the sample of $\logten(M_{\rm stellar}/M_{\odot}\geq9.7)$ galaxies. We find the redshifts based on S. Wuyts $P(z)$s agree best with the CANDELS team $z_{\rm phot}$ and available $z_{\rm spec}$ values, where the median (of the CANDELS five fields) outlier fraction\footnote{It is defined as the fraction of $\logten(M_{\rm stellar}/M_{\odot}\geq9.7)$ galaxies that are outliers. We consider a galaxy to be an outlier if $(z_{\rm participant}-z_{\rm phot})/(1+z_{\rm phot})>0.1$} of 1.5$\%$ and 3$\%$, respectively. This agreement is key {to} our analysis, and thus we choose the S. Wuyts $\sigma_{z}$ values centered on their redshifts to estimate the $\sigma_{z}$ for the CANDELS team $z_{\rm phot}$ values. For simplicity, in Figure\,\ref{fig_appendix_1b}, we only show the S. Wuyts redshift estimates vs. team $z_{\rm phot}$. 
	
\section*{AFFILIATIONS}
$^1${Department of Physics and Astronomy, University of Missouri-Kansas City, Kansas City, MO 64110, USA}\\
$^2${Department of Physics and Astronomy, Rutgers, The State University of New Jersey, NJ 08854-8019, USA}\\
$^3${Space Telescope Science Institute, 3700 San Martin Drive, Baltimore, MD 21218.}\\
$^4${Department of Physics and Astronomy and PITT PACC, University of Pittsburgh, Pittsburgh, PA 15260, USA}\\
$^5${University of Nottingham, School of Physics \& Astronomy, Nottingham, NG7 2RD UK}\\
$^{6}${University of California, Santa Cruz, USA}\\
$^{7}${Oxford Astrophysics, Denys Wilkinson Building, Oxford OX1 3RH, UK}\\
$^{8}${Astrophysics Science Division, Goddard Space Flight Center, MD 20771, USA}\\
$^{9}${Department of Physics, University of Bath, Bath, UK}\\
$^{10}${The University of Michigan, 300E West Hall, Ann Arbor, MI 48109-1107}\\
$^{11}${Racah Institute of Physics, The Hebrew University, Jerusalem 91904 Israel}\\
$^{12}${School of Physics and Astronomy, Rochester Institute of Technology, Rochester, NY 14623, US}\\
$^{13}${Department of Physics and Astronomy, Colby College, Waterville, ME 04961, USA}\\
$^{14}${Center for the Exploration of the Origin of the Universe, Seoul National University, Seoul, Korea}\\
$^{15}${Goddard Space Flight Center, Code 665, Greenbelt, MD 20771, USA}\\
$^{16}${Department of Physics and Astronomy, The Johns Hopkins University, Baltimore, MD 21218, USA}\\
$^{17}${University of California, Berkeley}\\
$^{18}${Department of Astronomy, The University of Texas at Austin, Austin, TX 78712, USA}\\
$^{19}${INAF-Osservatorio Astronomico di Roma, Roma, Italy}\\
$^{20}${Max Planck Institute fur Extraterrestriche Physik, 85741 Garching bei Munchen, DE}\\
$^{21}${Department of Physics \& Astronomy, University of California, Irvine, CA 92697, USA}\\
$^{22}${Departamento de Astrof\'isica, Facultad de CC. F\'isicas, Universidad Complutense de Madrid, Madrid, Spain}\\
$^{23}${Aix Marseille Université, CNRS, LAM (Laboratoire d'Astrophysique de Marseille) UMR 7326, 13388, Marseille, France}\\
$^{24}${Scientific Support Office, ESA/ESTEC, Noordwijk, The Netherlands}\\
$^{25}${Leiden Observatory, Leiden University, Leiden, Netherlands}

\bibliographystyle{mnras} 
\bibliography{Mantha01_ClosePairs_CANDELSpaper_second_submission_without_bold}
\end{document}